\documentclass[bibyear]{aa}

\usepackage{mpgmpar}
\usepackage{float}
\usepackage{subfig}
\usepackage{mathrsfs}
\usepackage{boxhandler}

\usepackage[labelfont=bf]{caption}
\usepackage[labelsep=period]{caption}

\usepackage[normalem]{ulem}

\usepackage{mwe}

\usepackage[export]{adjustbox}

\usepackage{color}

\usepackage{graphicx}

\usepackage{natbib}
\bibpunct{(}{)}{;}{a}{}{,}

\usepackage{txfonts}
\begin{document}

   
   
   
   
   \title{Gravity drives the evolution of infrared dark hubs:\\
   JVLA observations of SDC13}

   \author{G. M. Williams
          \inst{1}
          \and
          N. Peretto
          \inst{1}
          \and
          A. Avison\inst{2}\fnmsep\inst{3}
          \and
          A. Duarte-Cabral\inst{1}
          \and
          G. A. Fuller\inst{2}
          }

   \institute{School of Physics and Astronomy, Cardiff University, Queens Buildings, The Parade, Cardiff CF24 3AA, UK \\ email: \texttt{williamsgwen12@gmail.com}
         \and
         Jodrell Bank Centre for Astrophysics, School of Physics and Astronomy, University of Manchester, Manchester, M13 9PL, UK
         \and
         UK ALMA Regional Centre Node
         }

   \date{Received Month Day, Year; accepted Month Day, Year}


  \abstract
   {Converging networks of interstellar filaments, i.e. hubs, have been recently linked to the formation of stellar clusters and massive stars. Understanding the relationship between the evolution of these systems and the formation of cores/stars inside them is at the heart of current star formation research.}
   {The goal is to study the kinematic and density structure of the SDC13 prototypical hub at high angular resolution to determine what drives its evolution and fragmentation.}
   {We have mapped SDC13, a $\sim$1000\,M$_{\odot}$ infrared dark hub, 
    in NH$_{3}$(1,1) and NH$_{3}$(2,2) emission lines, with both the Jansky Very Large Array and Green Bank Telescope. The high angular resolution achieved in the combined dataset allowed us to probe scales down to 0.07pc. After fitting the ammonia lines, we computed the integrated intensities, centroid velocities and line widths, along with gas temperatures and H$_2$ column densities.
    }
   {The mass-per-unit-lengths of all four hub filaments are thermally super-critical, consistent with the presence of tens of gravitationally bound cores identified along them. These cores exhibit a regular separation of $\sim0.37\pm 0.16$\,pc suggesting gravitational instabilities running along these super-critical filaments are responsible for their fragmentation. The observed local increase of the dense gas velocity dispersion towards starless cores is believed to be a consequence of such fragmentation process. Using energy conservation arguments, we estimate that the gravitational to kinetic energy conversion efficiency in the SDC13 cores is $\sim35\%$. We see velocity gradient peaks towards $\sim63\%$ of the cores as expected during the early stages of filament fragmentation. Another clear observational signature is the presence of the most massive cores at the filaments' junction, where the velocity dispersion is the largest. We interpret this as the result of the hub morphology generating the largest acceleration gradients near the hub centre.  
   }
   {We propose a scenario for the evolution of the SDC13 hub in which filaments first form as post-shock structures in a supersonic turbulent flow. As a result of the turbulent energy dissipation in the shock, the dense gas within the filaments is initially mostly sub-sonic. Then gravity takes over and starts shaping the evolution of the hub, both fragmenting filaments and pulling the gas towards the centre of the gravitational well. By doing so, gravitational energy is converted into kinetic energy in both local (cores) and global (hub centre) potential well minima.  Furthermore, the generation of larger gravitational acceleration gradients at the filament junctions promotes the formation of more massive cores.  
   }


   \maketitle


\section{Introduction} \label{intro}

The star formation process can be perceived as the mechanism leading to the concentration of diffuse interstellar clouds into compact, nuclear burning, balls of gas. While the importance of interstellar filaments had been recognised already in the seventies \citep{schneider_elmegreen79}, observations of the Galactic interstellar medium with {\it Herschel} have revealed that they are a key intermediate stage towards the formation of stars \citep{andre10, molinari10, arz11}. Indeed, the majority of prestellar and protostellar cores are embedded in thermally supercritical filaments  \citep{polychroni13, konyves15, marsh16}, i.e. filaments whose mass-per-unit-length $M_{\rm{line}}$ is larger than the theoretical upper limit for an isothermal, infinitely long cylinder to maintain hydrostatic equilibrium \citep{ostriker64}. Understanding the connection between filament evolution and core formation has become one of the main goals of star formation research over the past decade.  

While a lot of effort has focused on the properties of individual filaments in nearby star-forming regions \citep{arz11, peretto12, palmeirim13, panopoulou14, salji15}, here we focus on hub filament systems (HFS), i.e. a small network of spatially converging interstellar filaments \citep{myers09}. The converging nature of such systems is suggestive of the role played by gravity in shaping them. Follow-up observations have shown that hubs are likely collapsing on parsec scales, gathering matter at their centre as a result of the collapse \citep{kirk13, peretto13, peretto14, liu13}. One of the most massive protostellar cores ever observed in the Galaxy has been found lying at the centre of such hubs \citep{peretto13}, indicating that they might play a key role in the formation of massive stars. \citet{schneider12} also showed that stellar protoclusters in the Rosette molecular cloud tend to form at the junction of filamentary structures. Understanding how hubs form and evolve can tell us what physical mechanisms are behind the fragmentation of filaments and how their interaction influence the formation of more massive objects. High-angular resolution observations of infrared dark clouds (IRDCs) provides  us with the opportunity to study such systems very early on in their evolution.

IRDCs are cold (T\,$<$\,20K) and dense (n\,$>$\,10$^{3}$\,cm$^{-3}$) reservoirs of gas seen in extinction against the bright mid-infrared emission of the Galactic plane background \citep{perault96, egan98, simon06, peretto09, butler09, peretto10}. They are considered to be mostly undisturbed by stellar feedback due to the lack of a significant embedded population of stars, and exhibit a wide range of morphologies, masses, and sizes \citep{peretto10}. The most massive of these IRDCs are thought to contain the initial conditions for massive star formation  \citep{rathjack06, pillai06, beuther07}.

\begin{figure}[!t]
\centering
\includegraphics[trim={5.3cm 0 3.7cm 0},clip,scale=.33,left]{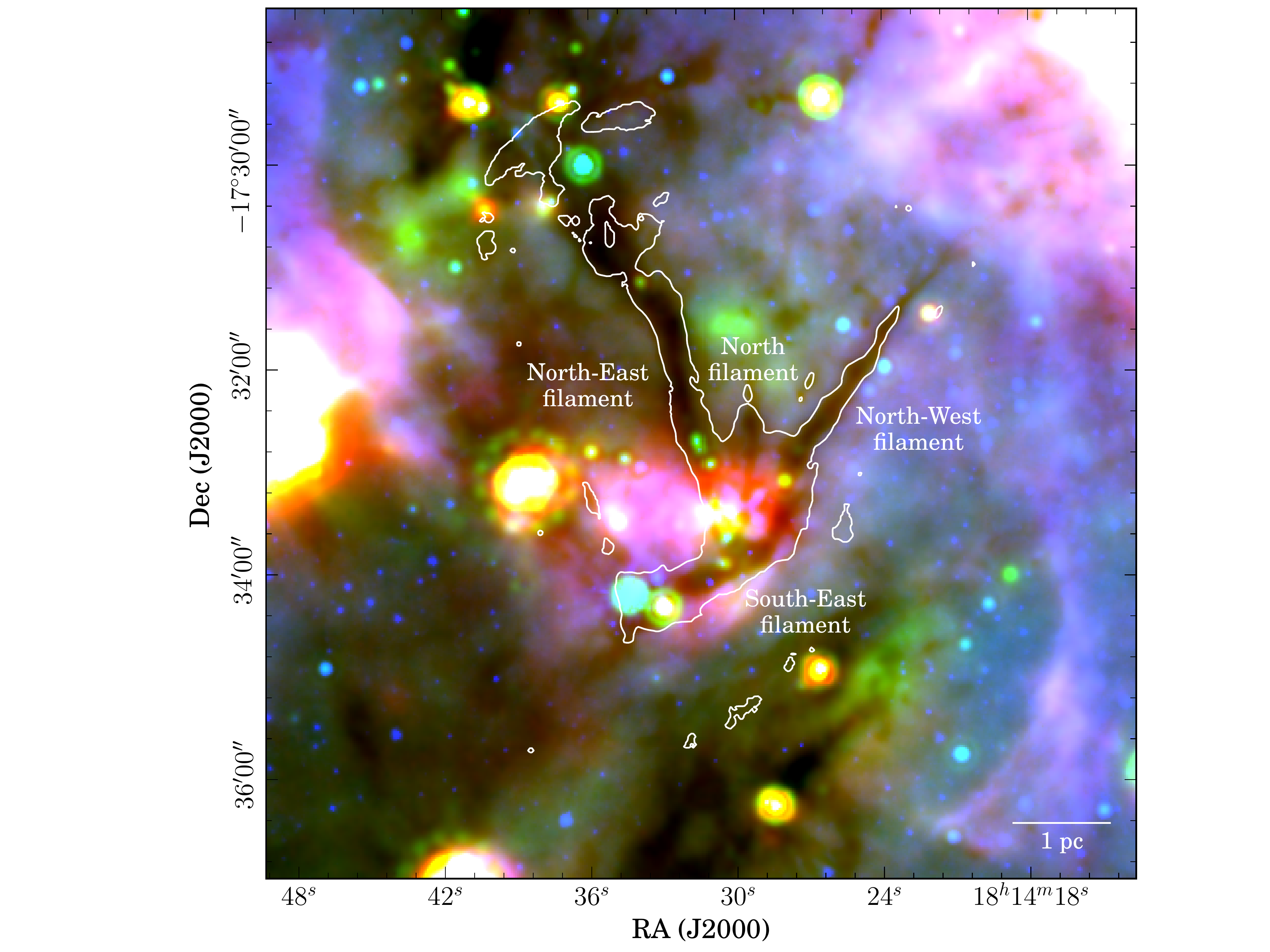}    
\caption{Three colour image of SDC13. R, G and B bands correspond to 70$\mu$m HIGAL \citep{molinari10}, 24$\mu$m Spitzer MIPSGAL \citep{carey09} and 8$\mu$m Spitzer GLIMPSE \citep{churchwell09} maps respectively. The overplotted contour is from the H$_{2}$ column density map at 2\,$\times$\,10$^{22}$\,cm$^{-2}$ (see Section \ref{sec:coldensmap} and the middle panel of Figure \ref{fig:n}). A 1\,pc scale bar is plotted in the bottom right corner.  The filament names are labelled following those used by \cite{peretto14}}         
\label{fig:tri}
\end{figure}

SDC13 (Figure~\ref{fig:tri}) is a remarkable filamentary hub IRDC system that lies 3.6\,$\pm$\,0.4\,kpc away in the Galactic plane, and contains $\sim$\,1000\,M$_{\odot}$ of material \citep{peretto14}.  Each of its four well-defined parsec-long filaments \citep[including SDC13.174-0.07, SDC13.158-0.073 and SDC13.194-0.073,][subsequently named following their on-sky orientation]{peretto09} converges on a central hub.  With the analysis of  1.2mm dust continuum data from the MAMBO bolometer array on the IRAM 30m telescope (at 12$''$ resolution) 18 compact sources were extracted and their starless or protostellar nature identified from 8\,$\mu$m and 24\,$\mu$m \textit{Spitzer} images.  The two most massive cores (named MM1 and MM2) are located right at the junction of the hub filaments, the substructure of which was studied using high resolution ($<3 ''$) 1.3\,mm SMA dust continuum emission \citep{mcguire16}.  MM1 was found to contain two subfragments, the centremost being brighter than the MM2 fragment.  Modelling the cores with RADMC-3D reveals that MM2 requires a steeper density profile than MM1, suggesting that it may be most likely to form a single massive central object \citep{girichidis11}.
Tracing the dense gas in N$_{2}$H$^{+}$(1\,--\,0) with the IRAM 30m telescope (at 27$''$ resolution) \cite{peretto14} identified velocity gradients along each of the filamentary arms of SDC13. Infall velocities larger at the filament ends anti-correlate with the velocity dispersion gradients which reach their maximum at the hub centre. This was interpreted as the consequence of the collapse of the gas along the filaments, generating an increase of kinetic pressure at the centre of the hub and providing the necessary conditions for the formation of super-Jeans cores, i.e. cores with masses that are several times larger than the local Jeans mass.

In this paper we present new high-resolution ammonia observations of SDC13 obtained with the Jansky Very Large Array (JVLA) and the Green Bank Telescope (GBT). These observations allow us to resolve, for the first time, the density structure and kinematics of the SDC13 filaments, expanding on the analysis of previous work.  In Section \ref{obs} we present the observations, and discuss the process of combining the two data sets.  Section \ref{sec:analysis} presents our line fitting and analysis of the observed kinematics of the system. In Section \ref{sec:structures} we discuss the identification of filament and core structures.  We discuss the fragmentation and dynamics of the system in Section \ref{sec:discussion}, and finally, we summarise our conclusions in Section \ref{sec:conc}.

\section{Observations} \label{obs}

We observed  ammonia (NH$_{3}$) at the position of SDC13 with both the JVLA and the GBT. NH$_{3}$ is rather abundant in star forming regions \citep{cheung68}, with an abundance of [NH$_{3}$]$/$[H$_{2}$]$\sim 3\times10^{-8}$, \citep{harju93}.  The (J,K) = (1,1) and (2,2) rotation inversion transitions with their relatively large critical density \citep[$n_{crit} = 10^{3}$\,cm$^{-3}$, ][]{HoTownes83,shirley15} are particularly good tracers of the cold and dense gas during the early stages of star formation as they are excited at the low temperatures ($<$\,20\,K) of molecular clouds and IRDCs, and exhibit very little, if any, depletion.  Studies of the hyperfine splitting of the (1,1) and (2,2) transitions \citep[e.g.][]{gunther, kuko} show that they have 18 and 24 hyperfine components, emitting in the radio regime at 23.69 and 23.72\,GHz respectively \citep[e.g.][]{HoTownes83}. The detection of both these lower metastable states allows for the direct calculation of the opacity, temperature and column density of the gas \citep[e.g.][]{HoTownes83,ungerechts86}.

\begin{table*}[t]
\centering
\caption{Summary of the observational properties of the JVLA, GBT and combined data.}
\label{tab:details}
\begin{tabular}{l|cc|cc|cc}
\hline\hline
\multicolumn{1}{c|}{Telescope} & \multicolumn{2}{c|}{JVLA}     & \multicolumn{2}{c|}{GBT}          & \multicolumn{2}{c}{Combined data}        \\ \hline
Parameters                     & NH$_{3}$(1,1) & NH$_{3}$(2,2) & NH$_{3}$(1,1) & NH$_{3}$(2,2)     & NH$_{3}$(1,1) & NH$_{3}$(2,2)           \\ \hline
Channel width (kHz)            & 3.906         & 3.906         & 2.86          & 2.86              & 7.812     & 7.812               \\
Velocity resolution (km/s)     & 0.049         & 0.049         & 0.036         & 0.036             & 0.098     & 0.098               \\
RMS noise (mJy/beam)           & 3.53          & 3.63          & 170.0         & 180.0             & 3.71      & 3.82                \\
Beam major axis ($''$)         & 3.87          & 3.77          & 31.32         & 31.32             & 4.35      & 4.30                \\
Beam minor axis ($''$)         & 2.81          & 2.87          & 31.32         & 31.32             & 3.44      & 3.48                \\
Position angle ($^{\circ}$)    & 30.95         & 30.56         & 0.0           & 0.0               & 30.95     & 30.56               \\ \hline
\end{tabular}
\tablefoot{For the purpose of hyperfine structure fitting, velocity resolutions were smoothed to 0.098\,km/s for the JVLA, and 0.072km/s for the GBT.  RMS noise was estimated from emission free areas.  The combined data were smoothed by a Gaussian kernel of 2.4$''$, and have the same spectral resolution as the smoothed JVLA data.  The quoted RMS noise of the GBT data includes the applied calibration factor}.
\end{table*}

\subsection{JVLA observations} \label{sec:dr}

The NH$_{3}$(1,1) and NH$_{3}$(2,2) transitions were observed at the position of the SDC13 infrared dark cloud (J2000 18:14:30.0 -17:32:50.0) at 23.7GHz using the JVLA in the DnC configuration, using the 1GHz band with 4MHz sub-bands.  We observed the NH$_{3}$(3,3) transition, but it was not significantly detected.
The angular resolution achieved of 3.3$''$ is an 8-fold improvement on the 27$''$ resolution of the N$_{2}$H$^{+}$(1--0) molecular line IRAM 30m data \citep{peretto14}, probing 0.07pc spatial scales.  A mosaic across the full 5$'$x5$'$ extent of the cloud was performed in eight, half-beam spaced pointings, collected over 8 observing sessions between 18$^{\mathrm{th}}$ and 20$^{\mathrm{th}}$ of May 2013, with 102.4 minutes integration time per position.
Further information on these data are provided in Table \ref{tab:details}.

Data reduction and calibration was completed using CASA\footnote{The Common Astronomy Software Applications package \texttt{http://casa.nrao.edu}} \citep{mcmullin}.  Phase and flux calibration was completed using the two bright quasars J1832-1035 and 3C286, respectively.  The phase calibrator was chosen to sit within 6 degrees to the target on the sky to ensure similar atmospheric conditions.  Bandpass calibration was completed with the J1743-0350 quasar.  We flagged any bad data using the \texttt{flagcmd()} and \texttt{flagdata()} tasks within CASA, and imaged both NH$_{3}$ transitions using the deconvolution task, \texttt{clean()}.
Our imaging implements the Natural weighting scheme, which puts less emphasis on the smallest scale coverage of the data.  This was done as our $u$-$v$ coverage was slightly stretched in this regime due to the low declination of the source.

\subsection{GBT observations} \label{sec:gbtobs}

We used the 7-beam K-band Focal Plane Array  at the GBT with the VEGAS (Versatile GBT Astronomical Spectrometer) backend to observe the NH$_{3}$(1,1) and (2,2) transitions employing a position switched scheme, where the Off source reference position (18:13:59.3 -17:35:02.0) was devoid of NH$_{3}$ emission.  Two maps were scanned in right ascension, while another two were scanned in declination so as to mitigate any artefacts that could arise during imaging had only one mapping scheme been used.  All four maps were observed on 14$^{th}$ Nov 2015, each with an integration time of 17.9 minutes (see Table \ref{tab:details} for data properties).  We ran the GBT pipeline \citep{masters} on each map scan separately to calibrate, reduce and subtract the continuum emission.  All scans were then combined and imaged in AIPS using the aipspy ParselTongue scripts included with the pipeline.  Possible issues with the noise diode voltages at the time of observation skewed our flux calibration but had no effect on the quality of the data. This resulted in the measured flux of the flux calibrator (J1833-2103, a quasar) to be lower than expected  from published calibration tables.  We applied an appropriate calibration factor (a ratio of the expected to the measured flux) to our data to correct for this.

\begin{figure*}[!htbp]
\centering
\includegraphics[scale=.715,left]{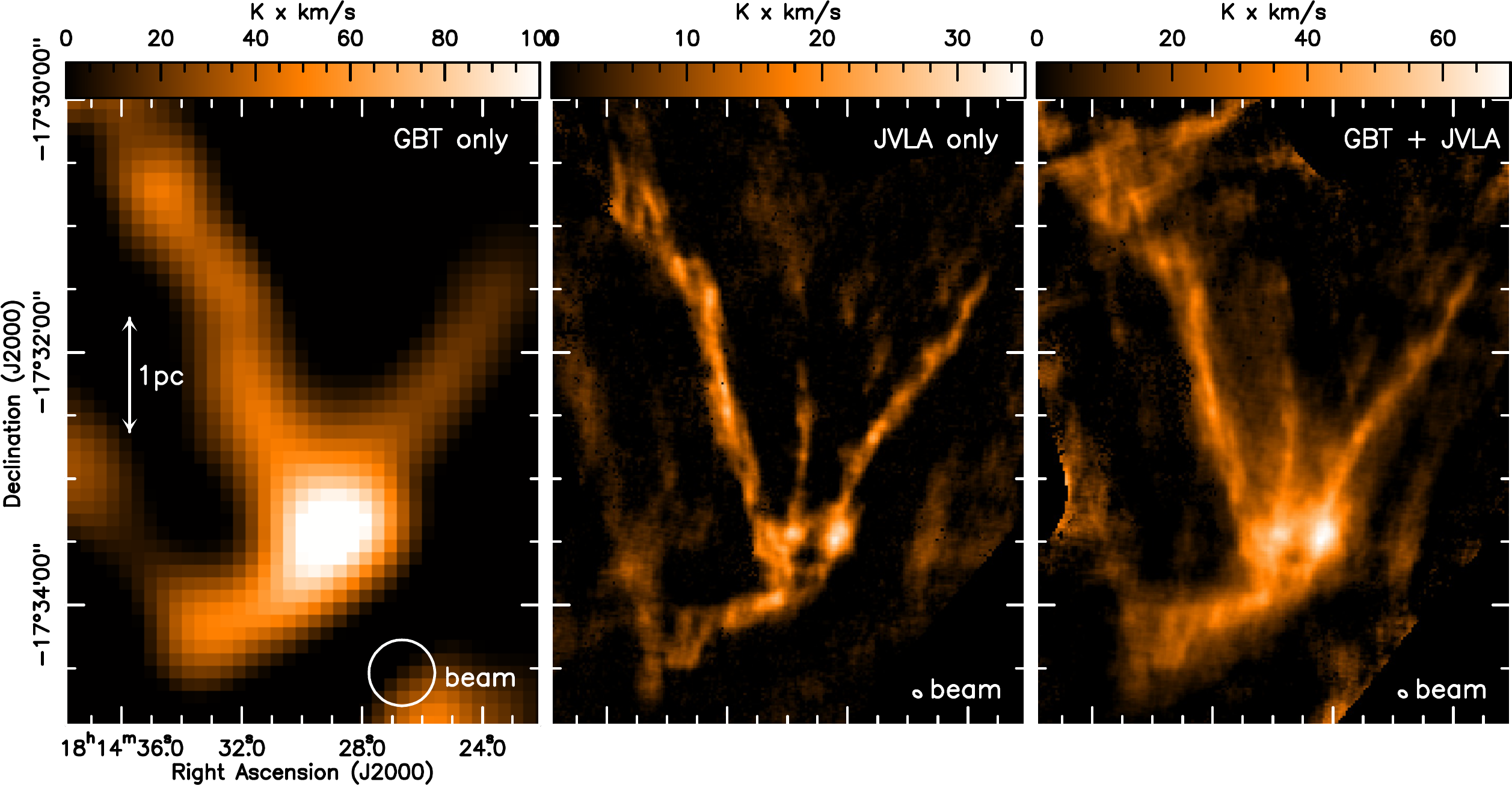}
\caption{Comparison between the integrated intensity maps of the NH$_{3}$(1,1) emission from the GBT (left), the JVLA (middle) and the combined GBT and JVLA (right) data sets. The GBT map displayed here has already had the calibration factor applied (see Section \ref{sec:gbtobs}). The beam sizes are plotted in the bottom right corner of each panel. A parsec scale at the distance of the cloud is plotted in the first panel.}
\label{fig:data-compare}
\end{figure*}

\subsection{Combination of JVLA and GBT data} \label{sec:combined}

Interferometers only probe a limited range of spatial frequencies of the source brightness spatial distribution, the largest frequency being determined by the shortest distance between any pair of the interferometer antennas. As a consequence, interferometers filter out extended emission. In order to recover this large-scale emission, and therefore be able to analyse the source structure on all spatial scales, one needs to combine interferometer data with single dish data. Hence, we combined our JVLA data with the GBT observations.

As a preliminary combination step, we re-clean our JVLA visibilities in the same way as described in Section \ref{sec:dr}, whilst using the GBT cube as a starting model \citep{dirienzo15}.  This method assists in initializing the clean algorithm by using the GBT primary beam to determine the flux scale.  Moreover, as the JVLA is missing information at the shortest baselines, this approach allows the clean algorithm to extrapolate the GBT data and converge on a better solution in this $u$-$v$ regime. Note that this is not the same as performing a joint deconvolution.

Combination is achieved with the use of the CASA task, \texttt{feather}.  With the high and low resolution images as inputs, \texttt{feather} Fourier transforms the two cubes, applies a low-pass filter to the low resolution data and a high-pass filter to the high resolution data, sums the two Fourier and filtered cubes in $u$-$v$ space, and reconvolves the final combined cube.  The low and high pass filters (weights calculated from the input clean beams) are applied to recover the larger scale emission of the GBT data at low $u$-$v$ distances which the JVLA naturally filters out, whilst retaining the fine scale structure probed by the JVLA at larger $u$-$v$ distances which the GBT cannot be sensitive to.  The primary beam response of the JVLA was applied to the GBT data prior to feathering, so that only emission within the JVLA mosaic area was considered.

The \texttt{casafeather} visual interface provides the tools to inspect slices of the $u$-$v$ plane of both the original and weighted deconvolved input images.  This is especially useful while setting the \texttt{sdfactor} parameter, which applies a flux scaling factor to the low resolution image.  For the conservation of flux, one would expect the fluxes of the weighted deconvolved high and low resolution images to be roughly equal.  
One can achieve this by altering the \texttt{sdfactor} until the casafeather plots demonstrate this equality.  Given the uncertainty in the flux calibration of the GBT, we expected a factor would be required, however, we find a satisfactory result with the default \texttt{sdfactor} of 1. This demonstrates that the calibration factor applied to the GBT data in the first instance was sound and appropriate.


Figure \ref{fig:data-compare} shows a comparison of the GBT, JVLA and combined integrated intensity maps of NH$_{3}$(1,1).  It is clear that the combined data recovers more extended flux than the JVLA--only, resulting in a doubling of the JVLA flux on the whole, whilst retaining the small scale detail and RMS noise of the interferometer.  We can clearly see that the negative bowl features, a consequence of missing flux, that surrounds the cloud structure in the JVLA--only map are mostly eradicated in the combined map and filled by the recovered extended emission. A check was made on the conservation of flux, performed by smoothing the feathered image to the resolution of the GBT, converting the units to Jy per GBT beam, and dividing by the original GBT map. In doing this, we find that flux is conserved. In the filament regions, we find a ratio of unity on the whole, whilst in the regions of significant emission such as the hub the combined flux is typically $\sim$20\% less than the GBT map \citep[a similar result to that reported by][]{dirienzo15}. 

In order to gain in signal to noise ratio we convolved the combined datacube with a Gaussian kernel of 2.4$''$, degrading the angular resolution to $\sim4''$.


\begin{figure*}[!htbp]
\centering
\includegraphics[trim={1cm 0cm 1cm 0cm},clip,scale=.5,center]{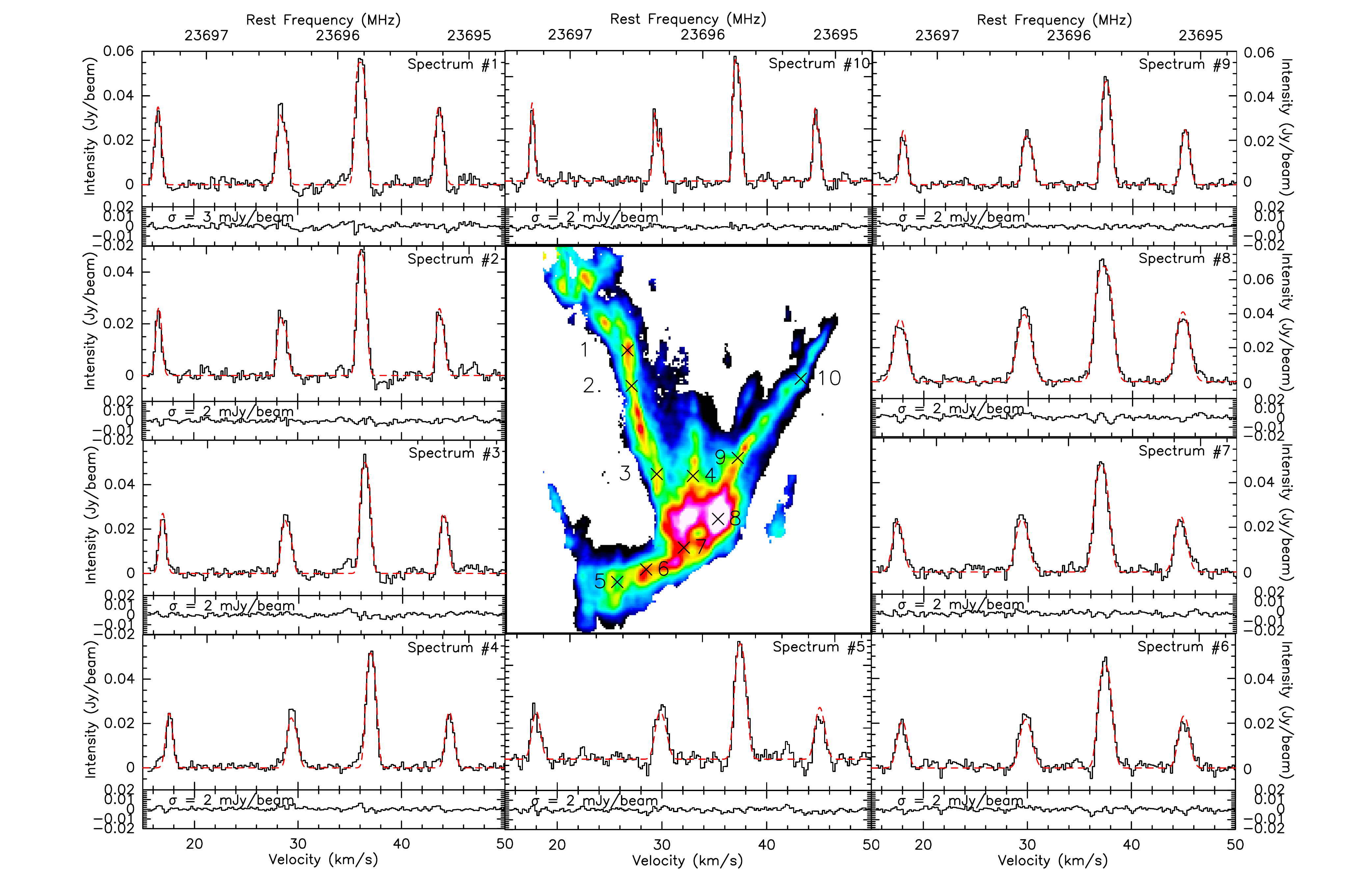}
\caption{Examples of NH$_{3}$(1,1) spectra at ten positions in the cloud, plotted surrounding the H$_{2}$ column density derived from the NH$_{3}$ emission (discussed later in Section \ref{sec:coldensmap}).  The model fit is plotted over each spectrum in red, whilst the residual of the two is plotted in the bottom panel with the standard deviation quoted in mJy/beam.  Regions of specific interest include those where the velocity width is seen to increase (spectra 1 and 6), positions where the centroid velocity changes as a filament meets the central hub (spectra 3, 4 and 7) and the large starless core MM2 (spectrum 8). It is clear that there is only a single velocity component everywhere in the cloud, with some lines being narrow enough to start resolving further hyperfine components (spectrum 10), and some showing the tell-tale signs of the cloud identified by \cite{peretto14} at a $V_{sys} = 54$\,kms$^{-1}$ that likely overlaps in projection with the North-East filament (spectrum 1 and 2).}
\label{fig:spectra}
\end{figure*}

\section{Results} \label{sec:analysis}

Here we present the results obtained from fitting the NH$_{3}$ hyperfine structure, and from deriving a H$_{2}$ column density map from the NH$_{3}$ emission.

\begin{figure*}[!htbp]
\centering
\begin{minipage}{.345\linewidth}
\subfloat{\label{fig:integ11}\includegraphics[trim={0.5cm 0 0 0},scale=.39,left]{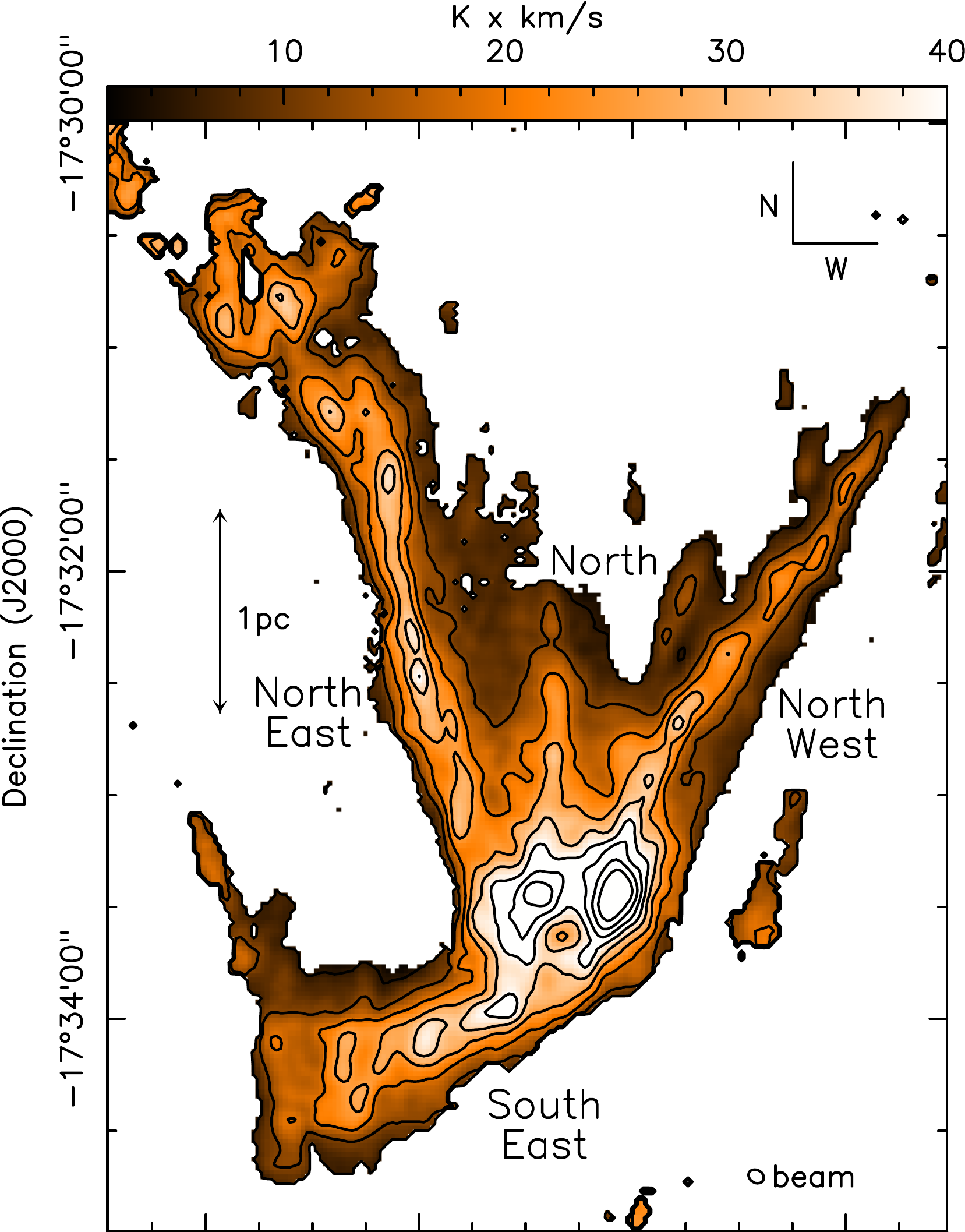}}
\end{minipage}%
\begin{minipage}{.315\linewidth}
\subfloat{\label{fig:velo11}\includegraphics[scale=.39,center]{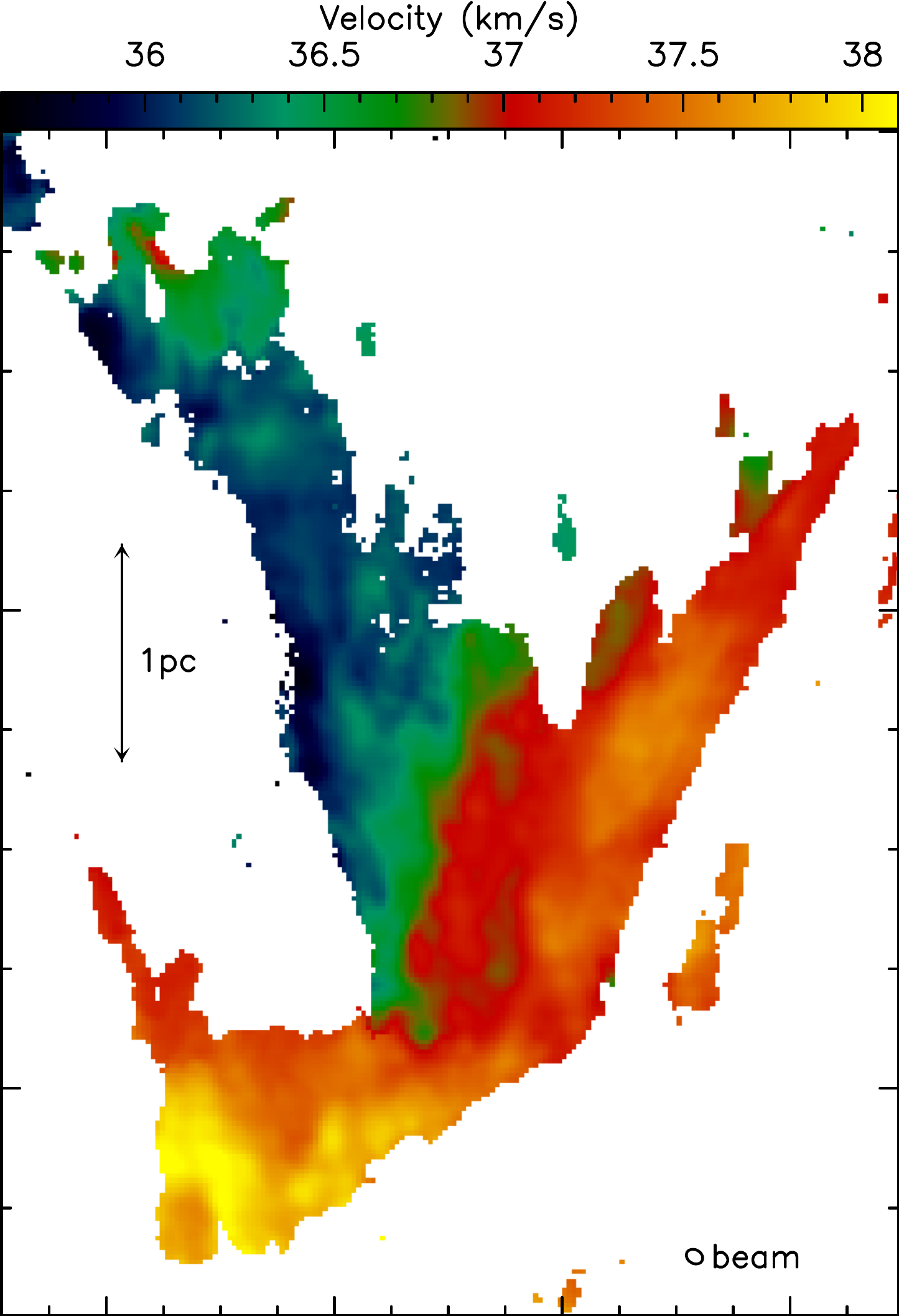}}
\end{minipage}%
\centering
\begin{minipage}{.315\linewidth}
\subfloat{\label{fig:width11}\includegraphics[trim={0 0 0.5cm 0},scale=.39,right]{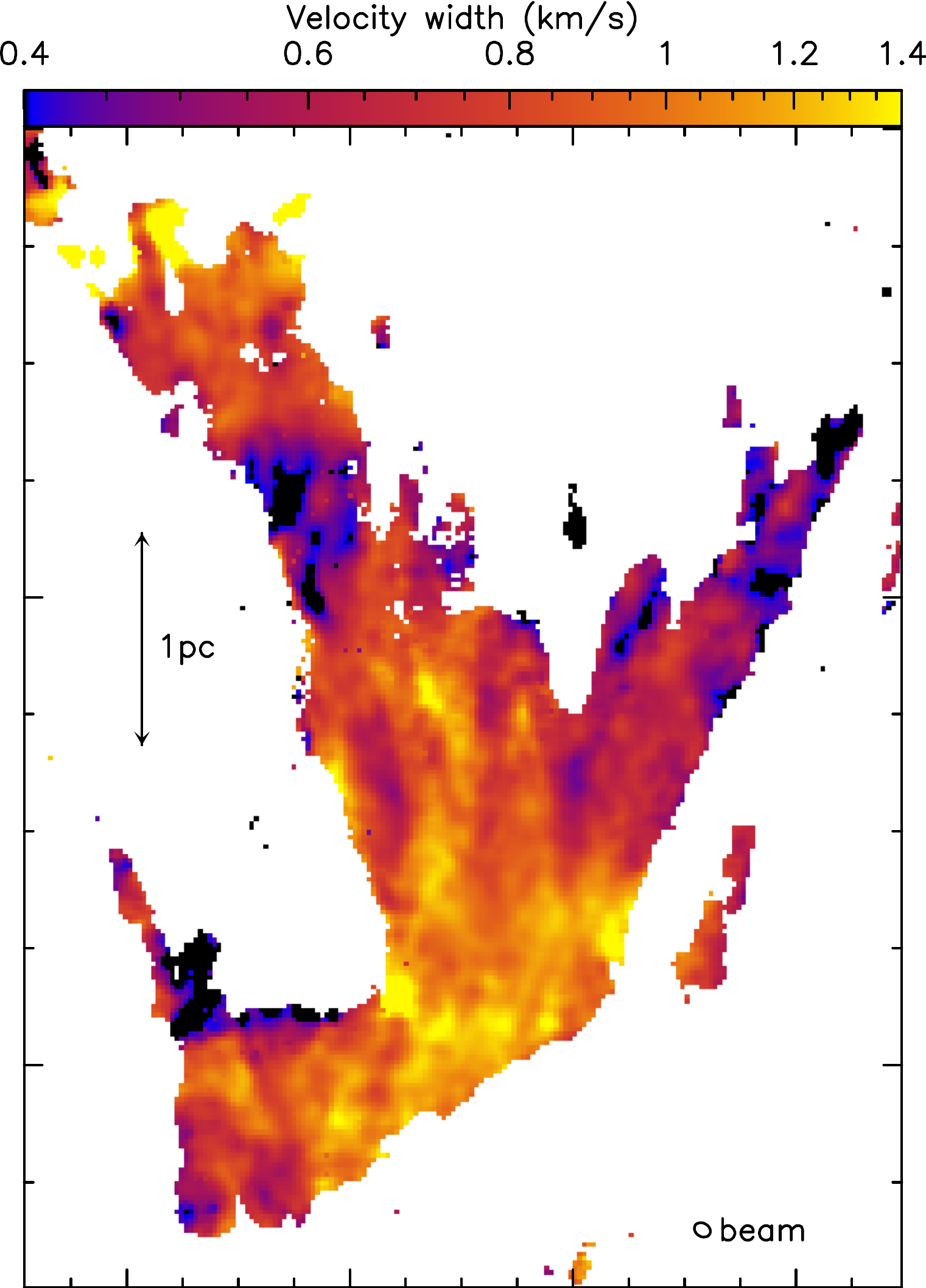}}
\end{minipage}\par\bigskip
\begin{minipage}{.345\linewidth}
\subfloat{\label{fig:integ22}\includegraphics[trim={0.5cm 0 0 0},scale=.39,left]{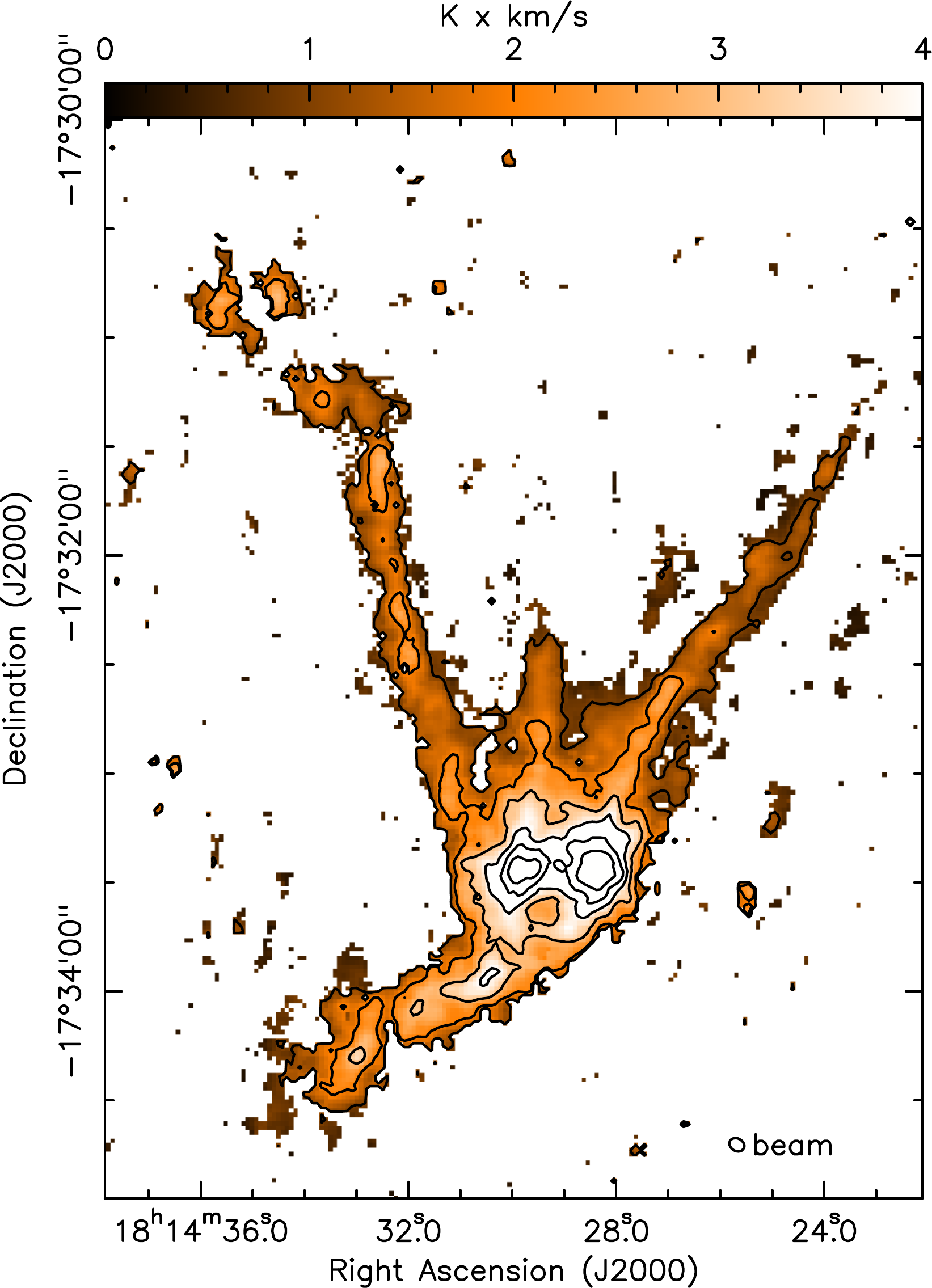}}
\end{minipage}%
\centering
\begin{minipage}{.315\linewidth}
\subfloat{\label{fig:velo22}\includegraphics[scale=.39,center]{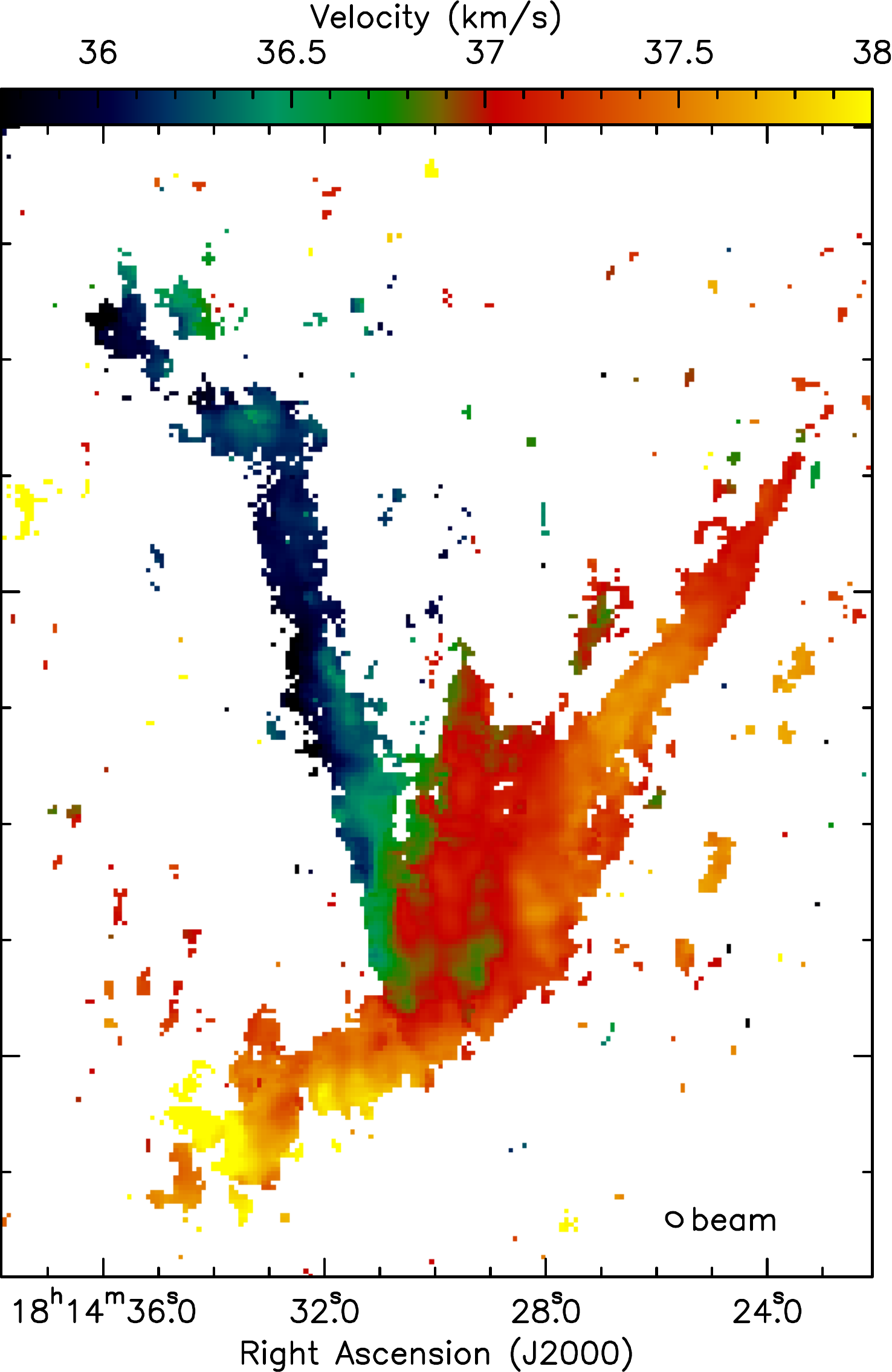}}
\end{minipage}%
\begin{minipage}{.315\linewidth}
\subfloat{\label{fig:width22}\includegraphics[trim={0 0 0.5cm 0},scale=.39,right]{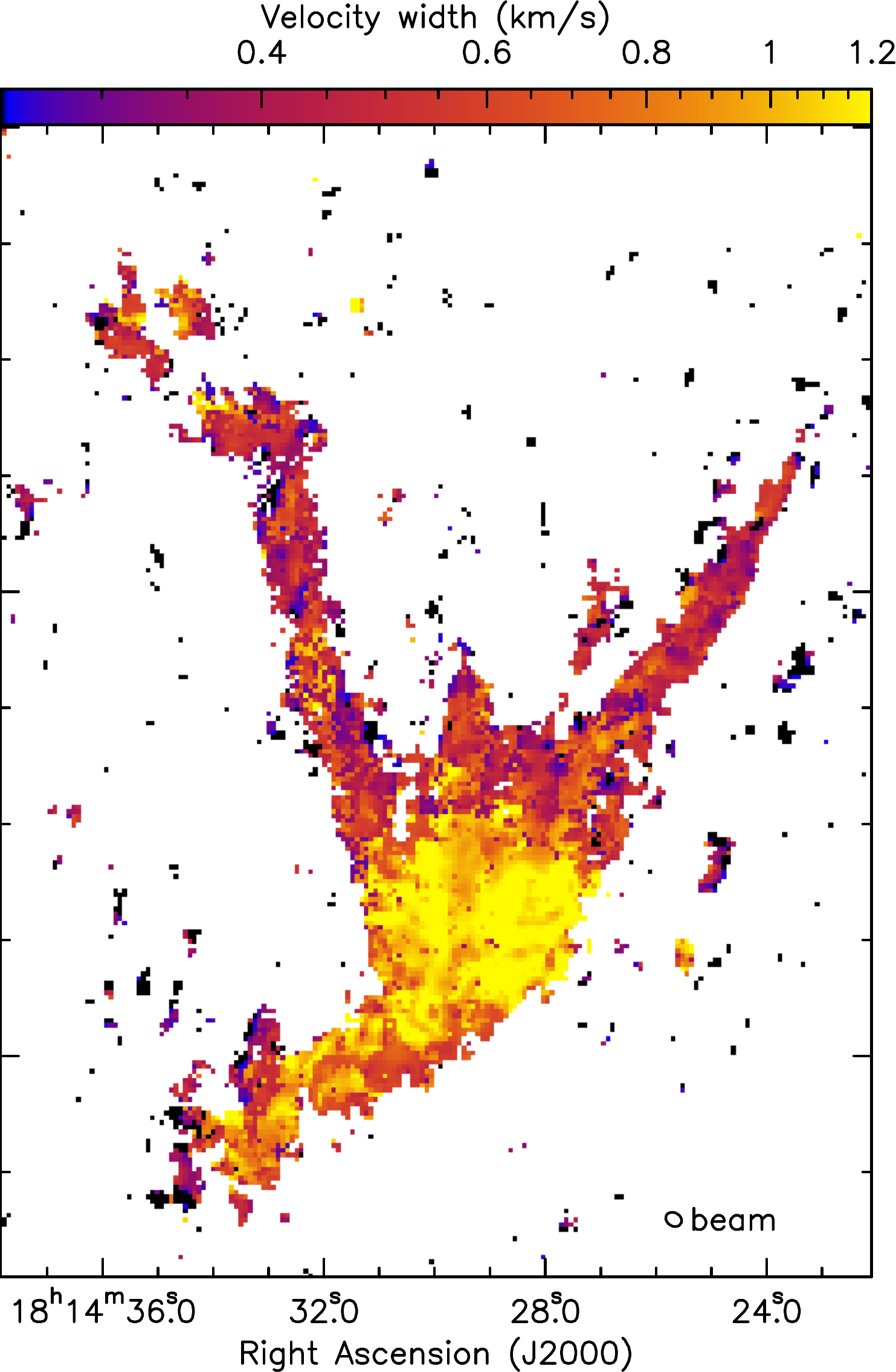}}
\end{minipage}
\caption{Integrated intensity (first column), centroid velocity (middle column) and velocity width (right column) of the NH$_{3}$(1,1) transition (top row) and  NH$_{3}$(2,2) transition (bottom row).  The NH$_{3}$(1,1) data was masked to 5\,$\sigma$, while the weaker NH$_{3}$(2,2) data was masked to 3\,$\sigma$.  Filaments are named in the top left panel.  Contours in the top left panel are in steps of 7\,K\,$\times$\,km/s, from 5\,K\,$\times$\,km/s to 61\,K\,$\times$\,km/s, while contours in the bottom left panel are in steps of 1\,K\,$\times$\,km/s, from 1\,K\,$\times$\,km/s to 6\,K\,$\times$\,km/s. Beam information is plotted in the bottom right corner of each panel, while the scale of 1\,pc is plotted in each panel of the top row.}
\label{fig:maps}
\end{figure*}

\subsection{Line fitting}


We performed hyperfine structure fitting of the combined JVLA and GBT data using the NH$_{3}$(1,1) and NH$_{3}$(2,2) fitting schemes within the CLASS\footnote{The fitting procedure of hyperfine structure is described at: https://www.iram.fr/IRAMFR/GILDAS/doc/pdf/class.pdf} 
software, where the spectra are fitted at every pixel over all channels.    Figure \ref{fig:spectra} shows examples of ten spectra and their hyperfine structure fits from positions distributed across the cloud.  After visual inspection, we do not find multiple velocity components, even at the filament junctions in the hub centre, therefore we proceed to fit a single velocity component everywhere.  Using the \texttt{result} command, a model fit to the data was created, where opacity-corrected Gaussian profiles are fitted to each hyperfine transition.  The channel maps of the resulting cube are shown in Fig.~\ref{fig:channel} in Appendix \ref{appendix-channelmap}.  We use these model fit cubes for all analysis conducted in the rest of the paper.

Due to a miss-alignment of the centre of the bandwidth with respect to the main hyperfine component at the cloud velocity (v$_{sys}$ = 37\,km\,s$^{-1}$) we do not detect the fifth hyperfine satellite of the NH$_{3}$(1,1) line at 57\,km\,s$^{-1}$.  
To correct for this missing flux, in the construction of integrated intensity maps we assumed that the missing component's intensity was identical to that of its symmetric pair at 17\,km\,s$^{-1}$. 
Using the GBT data, which does cover all hyperfine components, we estimate that this correction leads to an uncertainty on the total integrated intensity of $\le5\%$.

\begin{figure*}[!t]
\centering
\begin{minipage}{.33\linewidth}
\subfloat{\label{fig:8micron}\includegraphics[trim={0.5cm 0.15cm 0cm 0cm},scale=.38,left]{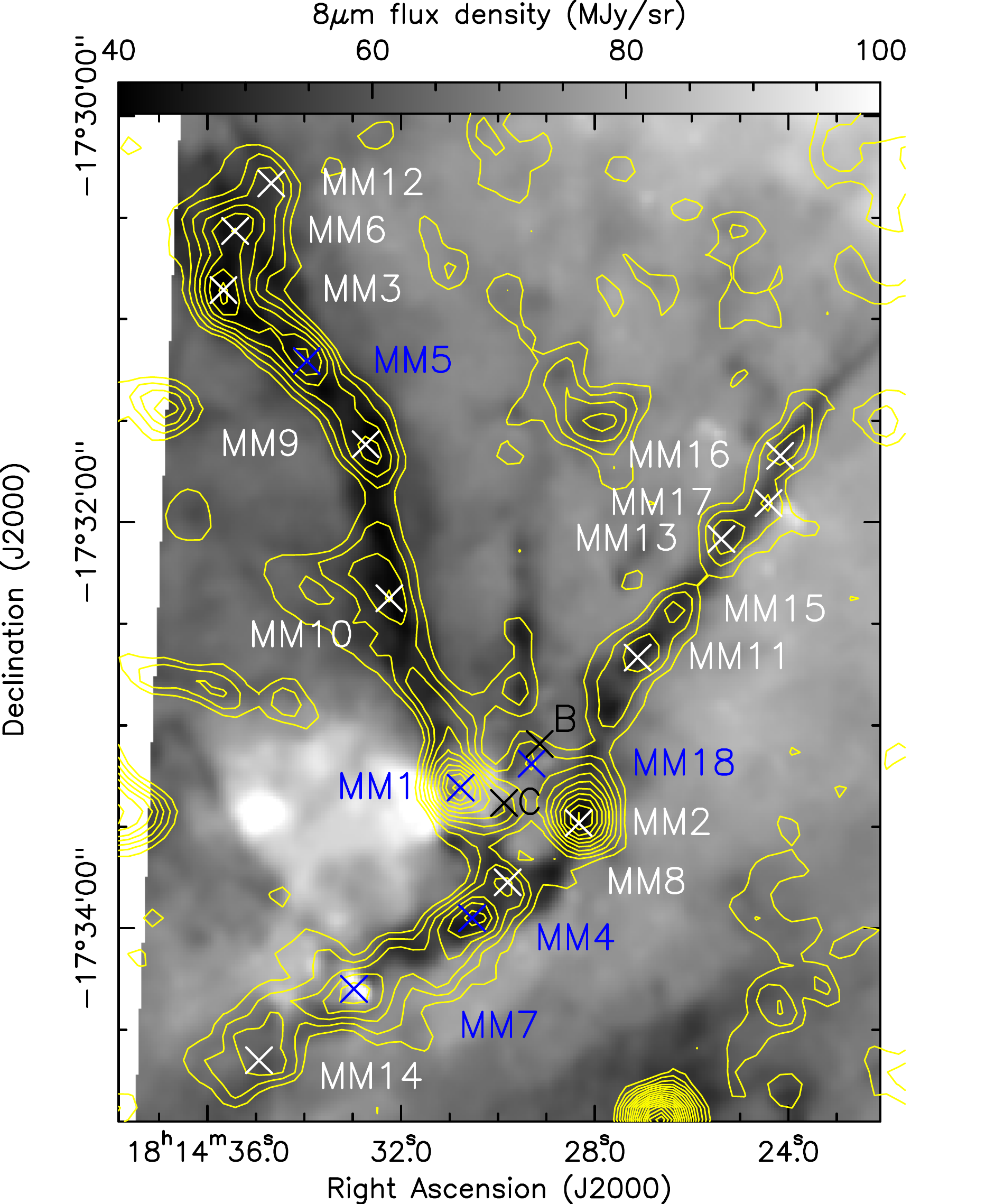}}
\end{minipage}%
\begin{minipage}{.33\linewidth}
\subfloat{\label{fig:n:nh2}\includegraphics[trim={1.5cm 0cm 0 0},clip,scale=.38,center]{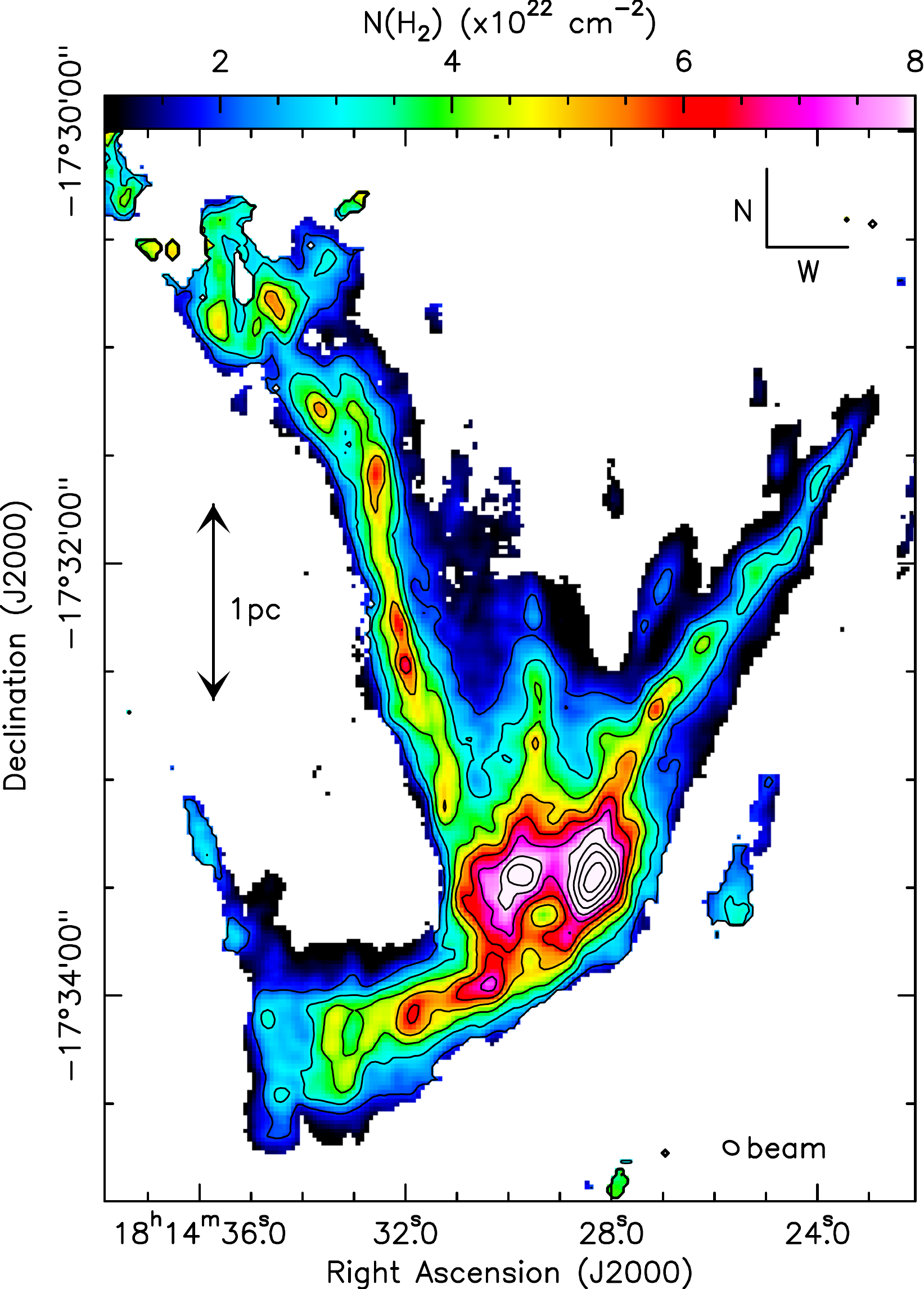}}
\end{minipage}%
\begin{minipage}{.33\linewidth}
\subfloat{\label{fig:skelcore}\includegraphics[trim={0.2cm 2.35cm 0 0},scale=.38,left]{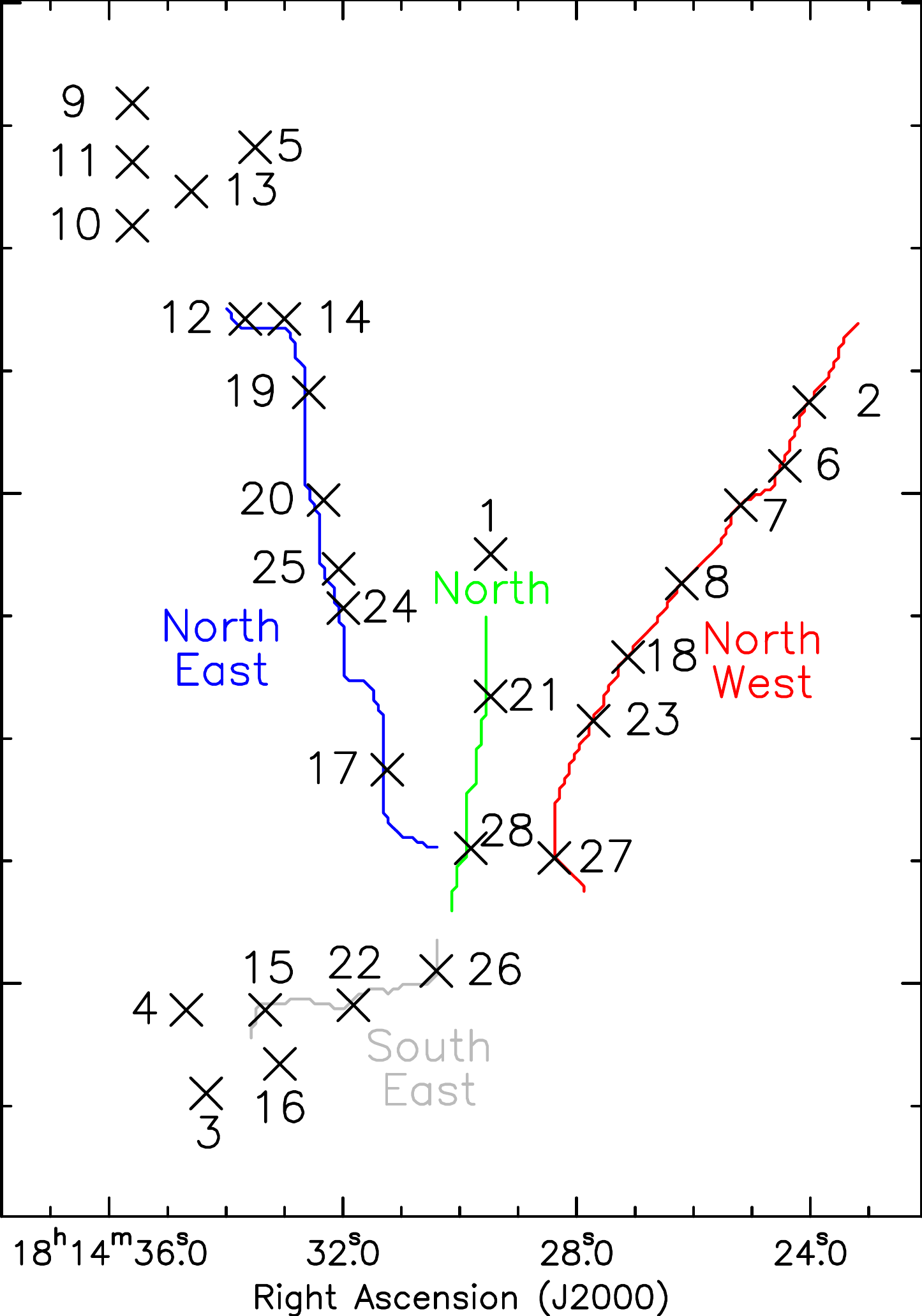}}
\end{minipage}
\caption{\textit{Left}: Spitzer 8\,$\mu$m flux density in units of MJy/sr, overlaid with IRAM 30m MAMBO 1.2mm dust continuum contours in steps of 5mJy/beam, from 3mJy/beam to 88mJy/beam. Crosses denote the positions of identified 1.2mm MAMBO compact sources \citep[white for starless and blue for protostellar,][]{peretto14} and two 1.3mm SMA continuum sources \citep[in black crosses, ][]{mcguire16}. \textit{Middle}: H$_{2}$ column density map in units of $10^{22}$\,cm$^{-2}$, derived from the NH$_{3}$ integrated intensity.  Overlaid contours are placed in 1\,$\times10^{22}$\,cm$^{-2}$ steps, from 2\,$\times10^{22}$\,cm$^{-2}$ to 12\,$\times10^{22}$\,cm$^{-2}$. \textit{Right}: Plot showing the extracted cores (black crosses, numbered according to Table \ref{tab:cores}) and filaments (labelled coloured lines). The extraction of these structures are discussed in Section \ref{sec:structures}. Extra cores at the outskirts of the North-East and South-East arms are in the more diffuse regions of the cloud where the spines did not extend through. Spine colours match those in subsequent figures.}
\label{fig:n}
\end{figure*}

\subsubsection{Integrated intensity} \label{sec:structure}


Figure~\ref{fig:maps} (left) shows the integrated intensity of the NH$_3$(1,1)  (top) and NH$_3$(2,2) (bottom) hyperfine structure lines. The structures in both maps are very similar even though, being typically 6 times weaker, the NH$_3$(2,2) emission is less extended and noisier than the NH$_3$(1,1) emission. When comparing the integrated intensity maps (or column density map - see Section~\ref{sec:coldensmap}) to both the 1.2mm dust continuum and 8\,$\mu$m maps (see Figure~\ref{fig:n}) we find excellent agreement between the structures seen in the different images.  All four filamentary arms seen in dust extinction/emission are well resolved in our NH$_{3}$ maps.
This demonstrates that ammonia is an excellent tracer of cold dense gas as numerous previous works had already theoretically  \citep{bergin97} or observationally  \citep{lu14} shown. There is, however, one exception to this excellent match between ammonia and dust emission. Indeed, we do not detect MM1 \citep[74.8\,$\pm$\,27.1M$_{\odot}$ within 0.26pc,][labelled in Figure \ref{fig:n}]{peretto14}, the largest protostellar source detected in the 1.2mm dust continuum map. This is a rather striking result since ammonia has consistently shown to be a good tracer of both cold and warm dense gas in star-forming clouds \citep{pillai11, ragan11}. Because of the protostellar nature of MM1, the lack of ammonia in the gas phase points towards a destruction mechanism rather than depletion. Chemical modelling of MM1 would be necessary to better constrain the physical origin of the lack of ammonia, but this goes beyond the scope of this paper.  
Also, the cavity in column density (centred on 18:14:29.3 -17:33:37.4) is also seen in the 8\,$\mu$m \textit{Spitzer} data as diffuse emission, with no evidence of any related compact source that may have caused clearing of surrounding material.

\subsubsection{Centroid velocity} \label{sec:velo}

The channel map of the main NH$_{3}$(1,1) transition in this model cube is shown in Figure \ref{fig:channel}. It shows large-scale velocity structures, with filaments appearing at different velocities, with a total velocity span of $\sim7$~km/s, consistent with the global velocity structure observed in N$_2$H$^+$(1--0) by \cite{peretto14}. Figure \ref{fig:maps} (middle columns) display the line of sight centroid velocity from both the NH$_{3}$(1,1) and (2,2) transitions. As a result of the increased angular resolution (almost by an order of magnitude), we observe a far more complex velocity field in SDC13 than previously revealed by the IRAM 30m data. In particular, we resolve both longitudinal and radial velocity gradients across all filaments \citep[where filaments are named following the same convention as][and their extraction is discussed in Section \ref{sec:fil}]{peretto14}.  This is most evident across the bottom half of the North-East oriented filament, and across the entire South-East oriented filament (labelled in Figures \ref{fig:maps} and \ref{fig:n}). At face value, it is not clear that our ammonia velocity map is consistent with the  N$_2$H$^+$(1-0) data published by \cite{peretto14}.  In order to check for consistency, we fitted the hyperfine components of the NH$_{3}$(1,1) and NH$_{3}$(2,2) emission from the GBT data  only (see Appendix \ref{appendix-gbt}), at a similar angular resolution to the IRAM 30m data.  The GBT velocity map  reveals gradients along each and every filamentary arm towards the centre of the system,  identical to those observed in the N$_{2}$H$^{+}$(1--0) IRAM 30m data.  The large number of NH$_{3}$(1,1) hyperfine components makes the best fit centroid velocity and velocity width both very accurate. We estimate that the related uncertainty is less than half a channel width, i.e. $\sim0.05$\,km\,s$^{-1}$.  


\subsubsection{Velocity width} \label{sec:wid}

Figure \ref{fig:maps} (right column) shows the FWHM velocity width of the dense gas. We report an increased velocity width at the filament hub junction, peaking at $\sim1.6$~km/s, consistent with the broadening reported by \cite{peretto14}. Interestingly, we also observe local increases of the velocity width sprinkled along each filamentary arms.  
As discussed later in Section \ref{sec:structures}, these are correlated to peaks in the H$_{2}$ column density.  
Note that opacity broadening is accounted for as a consequence of the fitting procedure used. However, because of the relatively large peak NH$_{3}$(1,1) main line opacity of 4.7, we independently performed a Gaussian fit of the optically thinner of the hyperfine components. By doing so, we observe the same increase of the line width. We conclude that the line widths plotted in Figure~\ref{fig:maps} represent the true velocity dispersion of the dense gas.
Furthermore, we also notice the velocity width increases around some localised regions at the edges of filaments, most evident on the Eastern edge of the lower portion of the North-East filament, and the Western edge of the lower portion of the North-West filament.


\subsection{H$_2$ column density map} \label{sec:coldensmap}

Ammonia is often called the thermometer of molecular clouds \citep[e.g.][]{maret09}, as the temperature of the gas can be directly calculated from the ratio of the intensities of the two main lower metastable transitions, NH$_{3}$(1,1) and NH$_{3}$(2,2). In turn, the ammonia column density map can be calculated using the temperature, opacity, and brightness temperature maps. The derivations of each of these quantities can be found in Appendix~\ref{appendix-A}.      
Using a constant abundance of ammonia with respect to H$_2$ of $3\times10^{-8}$ \citep{harju93} one can then compute the H$_2$ column density map of SDC13.    This calculation is however limited by the weaker emission of the NH$_{3}$(2,2) transition, which in our case traces 42\% less of the cloud area than the NH$_{3}$(1,1) transition.  To overcome this, we take a median value of the temperature across the cloud of 12.7\,K (the full temperature map and histogram are shown in Appendix~\ref{appendix-A}).  The corresponding column density map is displayed in Figure~\ref{fig:n}.  Comparing this column density map to that calculated with a non-uniform temperature, we are satisfied that assuming a constant temperature across the entire cloud does not introduce any significant changes to the morphology of features in the map, whilst extending the coverage of the map \citep[a result also reported by][]{morgan13}.  Increasing the temperature by a standard deviation (1.8\,K) decreases the column density by 12\%.  It is important to note that given our non-detection of the MM1 protostellar core, our assumption of constant abundance across the entire cloud (whilst a fair assumption given the excellent correlation between the NH$_{3}$(1,1) integrated intensity and the 8$\mu$m Spitzer emission) does not hold entirely for such portions of the cloud as the MM1 region.  Deriving the H$_{2}$ column density from the 1.2\,mm dust continuum emission \citep[presented in][]{peretto14} and comparing it to the NH$_{3}$ column density shows that the assumed abundance is consistent with the data, and has a dispersion of only $\sim3\%$. Overall, we take a conservative estimate of the uncertainty on the column density of 20\%.   



\begin{table*}[]
\centering
\caption{Observed and calculated properties of extracted cores.}
\label{tab:cores}
\begin{tabular}{cccccccccccccc}
\hline\hline
ID & \begin{tabular}[c]{@{}c@{}}RA\\ (J2000)\end{tabular} & \begin{tabular}[c]{@{}c@{}}Dec\\ (J2000)\end{tabular} & \begin{tabular}[c]{@{}c@{}}Min.\\ ($''$)\end{tabular} & \begin{tabular}[c]{@{}c@{}}Maj.\\ ($''$)\end{tabular} & \begin{tabular}[c]{@{}c@{}}P.A.\\ ($_{\circ}$)\end{tabular} & \begin{tabular}[c]{@{}c@{}}Aspect\\ ratio\end{tabular} & \begin{tabular}[c]{@{}c@{}}N(H$_{2}$)\\ (10$^{22}$cm$^{-2}$)\end{tabular} & \begin{tabular}[c]{@{}c@{}}Radius\\ ($''$)\end{tabular} & \begin{tabular}[c]{@{}c@{}}Radius\\ (pc)\end{tabular} & \begin{tabular}[c]{@{}c@{}}Mass$_{low}$\\ (M$_{\odot}$)\end{tabular} & \begin{tabular}[c]{@{}c@{}}Mass$_{up}$\\ (M$_{\odot}$)\end{tabular} & $\alpha_{vir}$ & \begin{tabular}[c]{@{}c@{}}MM\\ assoc.\end{tabular} \\ \hline
1  & 18:14:29.5 & -17:32:14.9 & 2.3 & 5.0  & 10.3  & 2.2 & 2.22 & 3.0 & 0.07 & 0.5  & 5.4   & 1.47 & -- \\
2  & 18:14:24.0 & -17:31:37.7 & 2.6 & 9.9  & -32.6 & 3.8 & 2.90 & 4.5 & 0.09 & 1.9  & 14.4  & 0.54 & MM16  \\
3  & 18:14:34.3 & -17:34:26.9 & 3.0 & 3.7  & 8.4   & 1.2 & 2.95 & 3.0 & 0.07 & 0.5  & 7.2   & 1.43 & -- \\
4  & 18:14:34.7 & -17:34:06.5 & 3.5 & 6.5  & -7.2  & 1.8 & 2.94 & 3.7 & 0.08 & 0.6  & 10.3  & 1.49 & -- \\
5  & 18:14:33.5 & -17:30:35.3 & 2.7 & 5.7  & -35.3 & 2.1 & 3.02 & 3.4 & 0.08 & 0.7  & 9.1   & 2.00 & -- \\
6  & 18:14:24.4 & -17:31:53.3 & 1.7 & 5.7  & -26.5 & 3.3 & 3.30 & 2.7 & 0.07 & 0.4  & 7.1   & 0.56 & MM17 \\
7  & 18:14:25.2 & -17:32:02.9 & 2.9 & 8.8  & -26.8 & 3.0 & 3.37 & 4.4 & 0.09 & 1.2  & 15.8  & 0.54 & MM13 \\
8  & 18:14:26.2 & -17:32:22.1 & 2.8 & 6.6  & -25.7 & 2.3 & 3.84 & 4.0 & 0.08 & 1.1  & 15.0  & 0.65 & MM15 \\
9  & 18:14:35.6 & -17:30:24.5 & 1.9 & 2.8  & -85.3 & 1.5 & 3.80 & 1.5 & 0.05 & 0.2  & 4.1   & 4.15 & -- \\
10 & 18:14:35.6 & -17:30:54.5 & 3.7 & 6.8  & 5.3   & 1.8 & 4.22 & 4.6 & 0.10 & 2.6  & 21.8  & 0.61 & MM3  \\
11 & 18:14:35.6 & -17:30:38.9 & 1.6 & 3.3  & 17.9  & 2.1 & 3.87 & 1.7 & 0.05 & 0.2  & 4.5   & 1.50 & -- \\
12 & 18:14:33.7 & -17:31:17.3 & 3.9 & 7.4  & 33.4  & 1.9 & 4.35 & 5.0 & 0.10 & 3.5  & 25.9  & 0.45 & MM5 \\
13 & 18:14:34.6 & -17:30:46.1 & 4.7 & 6.9  & 28.1  & 1.5 & 4.64 & 5.3 & 0.11 & 4.2  & 30.8  & 0.64 & -- \\
14 & 18:14:33.0 & -17:31:17.3 & 1.2 & 4.9  & 34.5  & 4.1 & 4.20 & 1.7 & 0.05 & 0.1  & 4.9   & 0.84 & -- \\
15 & 18:14:33.3 & -17:34:06.5 & 2.4 & 7.8  & -5.5  & 3.2 & 4.43 & 3.9 & 0.08 & 0.9  & 16.9  & 0.87 & -- \\
16 & 18:14:33.1 & -17:34:19.7 & 2.2 & 4.4  & -18.6 & 2.0 & 4.35 & 2.6 & 0.07 & 0.3  & 8.9   & 0.88 & MM7  \\
17 & 18:14:31.2 & -17:33:07.7 & 2.7 & 11.9 & 0.7   & 4.4 & 4.59 & 5.2 & 0.11 & 2.2  & 29.6  & 0.43 & -- \\
18 & 18:14:27.1 & -17:32:40.1 & 2.7 & 7.8  & -32.9 & 2.9 & 4.86 & 4.2 & 0.09 & 2.3  & 20.9  & 0.47 & MM11 \\
19 & 18:14:32.6 & -17:31:35.3 & 3.0 & 11.7 & 1.3   & 3.9 & 5.05 & 5.5 & 0.11 & 3.5  & 36.0  & 0.25 & MM9  \\
20 & 18:14:32.3 & -17:32:01.7 & 1.1 & 6.5  & 21.2  & 5.9 & 4.73 & 1.8 & 0.06 & 0.1  & 6.0   & 0.73 & -- \\
21 & 18:14:29.5 & -17:32:49.7 & 1.7 & 6.3  & -4.4  & 3.8 & 4.83 & 2.4 & 0.06 & 0.3  & 9.0   & 0.92 & -- \\
22 & 18:14:31.8 & -17:34:05.3 & 3.7 & 6.8  & -38.4 & 1.8 & 5.68 & 4.6 & 0.10 & 2.0  & 29.4  & 0.83 & -- \\
23 & 18:14:27.7 & -17:32:55.7 & 1.6 & 2.3  & -7.1  & 1.5 & 5.44 & 1.0 & 0.05 & 0.1  & 4.2   & 1.05 & -- \\
24 & 18:14:32.0 & -17:32:28.1 & 2.2 & 4.7  & 10.9  & 2.2 & 5.87 & 2.7 & 0.07 & 1.0  & 12.6  & 0.75 & MM10 \\
25 & 18:14:32.1 & -17:32:18.5 & 1.5 & 4.5  & 13.2  & 2.9 & 5.71 & 2.0 & 0.06 & 0.5  & 8.4   & 0.94 & -- \\
26 & 18:14:30.4 & -17:33:56.9 & 4.6 & 8.6  & -41.9 & 1.9 & 6.48 & 5.4 & 0.11 & 3.5  & 44.3  & 0.85 & MM4  \\
27 & 18:14:28.4 & -17:33:29.3 & 8.1 & 12.6 & -7.4  & 1.6 & 9.11 & 9.9 & 0.18 & 36.7 & 195.7 & 0.28 & MM2 \\
28 & 18:14:29.8 & -17:33:26.9 & 6.4 & 11.4 & -37.6 & 1.8 & 8.20 & 8.1 & 0.15 & 11.6 & 119.2 & 0.30 & --  \\ \hline
\end{tabular}
\tablefoot{Column 1: core ID number; Cols. 2 and 3: right ascension and declination of the core peak emission; Cols. 4 and 5: core major and minor axes; Col. 6; position angle; Col. 7: major-to-minor axes ratio; Col. 8: mean value of H$_{2}$ column density within core boundary; Col. 9: deconvolved core radius in arcseconds; Col. 10: deconvolved core radius in parsec; Col. 11: lower limit on the mass; Col. 12: upper limit on mass; Col. 13: Virial ratio, calculated using equation \ref{eq:vir} where the upper limit on the mass was used; Col. 14: Core associations to the MM identifications of previous work \citep{peretto14}.  The detected protostellar sources are MM4, MM5 and MM7. Systematic error from the kinematical distance to SDC13 ($3.6\pm0.4$\,kpc) is associated to all size parameters. Uncertainty on the column density and masses originate from the assumption of constant [NH$_{3}$]$/$[H$_{2}$]$\sim$ $3\times10^{-8}$ abundance \citep{harju93} and temperature (12.7\,K) in the cloud.
}
\end{table*}

\section{Analysis: Structure extraction} \label{sec:structures}

In this section, we discuss the identification of cores and filaments within SDC13.

\subsection{Cores} \label{sec:core}

Dendrograms are a useful tool for the understanding of hierarchical structure, and are invaluable for understanding the fragmentation of molecular clouds \citep[e.g.][]{rosolowsky08dendro}.  We used the dendrogram code developed by \cite{peretto09} to extract all clumps from the NH$_{3}$ derived H$_{2}$ column density map at regular isocontours.  A minimum isocontour value is set at N(H$_{2}$) = $1\times10^{22}$\,cm$^{-2}$ defining the ``trunk" of the dendrogram tree, with a regular isocontour spacings at 1\,$\sigma$ of 0.04\,$\times\,10^{22}$\,cm$^{-2}$ throughout, and a 5\,$\sigma$ detection threshold.  A pixel limit of 7 defines the smallest size of structure considered resolved, as it roughly matches the beam shape.  Cores were identified as the highest levels in the dendrogram hierarchy (termed ``leaves") meaning that they contain only themselves and no other sub-structures.  All observed core properties are listed in Table \ref{tab:cores}, such as the central core coordinates, major and minor axes, position angle, aspect ratio, radius and mean H$_{2}$ column density.  Core radii were calculated by considering each isocontour core boundary a disc with an equivalent area.  The smallest deconvolved core radius can be seen to match half of the beam width, a direct effect of matching the pixel limit of the extraction code to the beam. 



We identified 28 cores, of which seven lie along each of the North-West and North-East filaments, three lie along the South-East filament and two along the North filament (plotted in the right panel of Figure~\ref{fig:n}). The remaining cores mostly lie at the end of the North-East filament. Their starless (or protostellar) nature was identified by the lack (or presence) of Spitzer $8\mu$m and $24\mu$m sources.  Only three protostellar sources were detected, which were already identified by \cite{peretto14} (listed in Table \ref{tab:cores}).  Core masses were calculated by taking the mean H$_{2}$ column density (middle panel of Figure \ref{fig:n}, calculated from the NH$_{3}$ column density) within the core boundary, and setting the average molecular weight, $\mu = 2.8$. Two mass estimates were calculated for each core, one excluding the outer column density shell \citep[effectively excluding the ``background", equivalent to the \textit{clipping} scheme of][]{rosolowsky08dendro} regarded as the lower limit of the mass, and one that incorporates the background column density \citep[equivalent to the \textit{bijection} scheme of][]{rosolowsky08dendro}, regarded an upper limit of the mass. Small column density peaks on top of a large background column density will generate very different mass estimates. The cores we identify here range from low- to high-mass, with the most massive ones being located near the filament junctions (core IDs 26, 27 and 28 - See Figure~\ref{fig:n}). Both mass ranges, as well as the virial ratio (see Section~\ref{sec:virial}) are listed in Table \ref{tab:cores}. The association of the newly identified cores with those published in \cite{peretto14} is also provided in the table.  The dominant source of uncertainties in the estimate of core properties are the distance (11\%), the abundance (within a factor of 2), and the way cores are defined (bijection versus clipping - see Table 2). Note that uncertainties related to distance and abundance are systematic and therefore will not change observed trends.

\begin{figure}[!t]
\centering
\includegraphics[scale=.5,left]{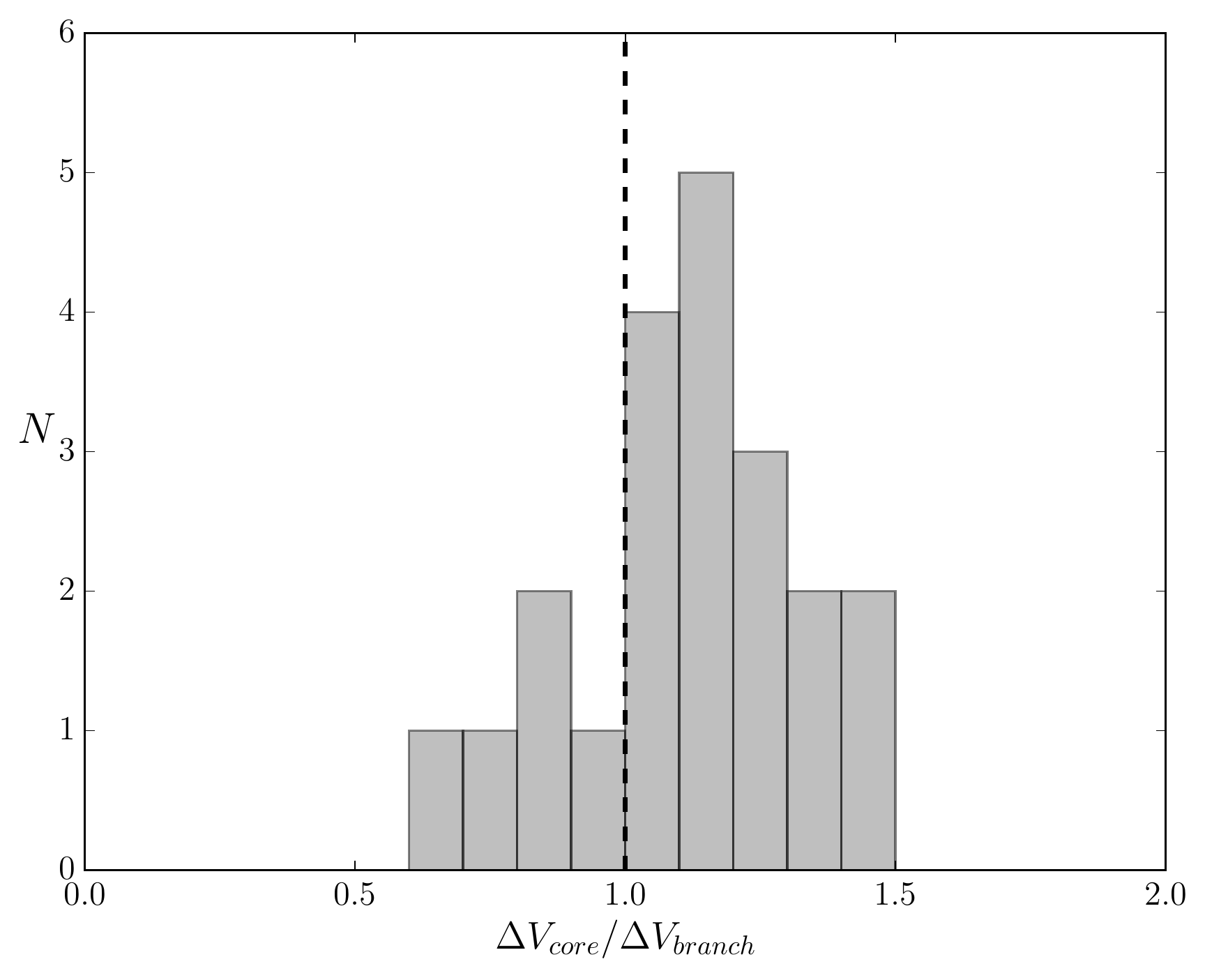} 
\caption{Histogram of the ratio of the average velocity width within a core region ($\Delta V_{core}$) to that of the underlying branch structure ($\Delta V_{branch}$) in the dendrogram tree. The vertical dashed line indicates where $\Delta V_{core} = \Delta V_{branch}$. }
\label{fig:core_branch}
\end{figure}

\begin{figure}[!t]
\centering
\includegraphics[scale=.4,left]{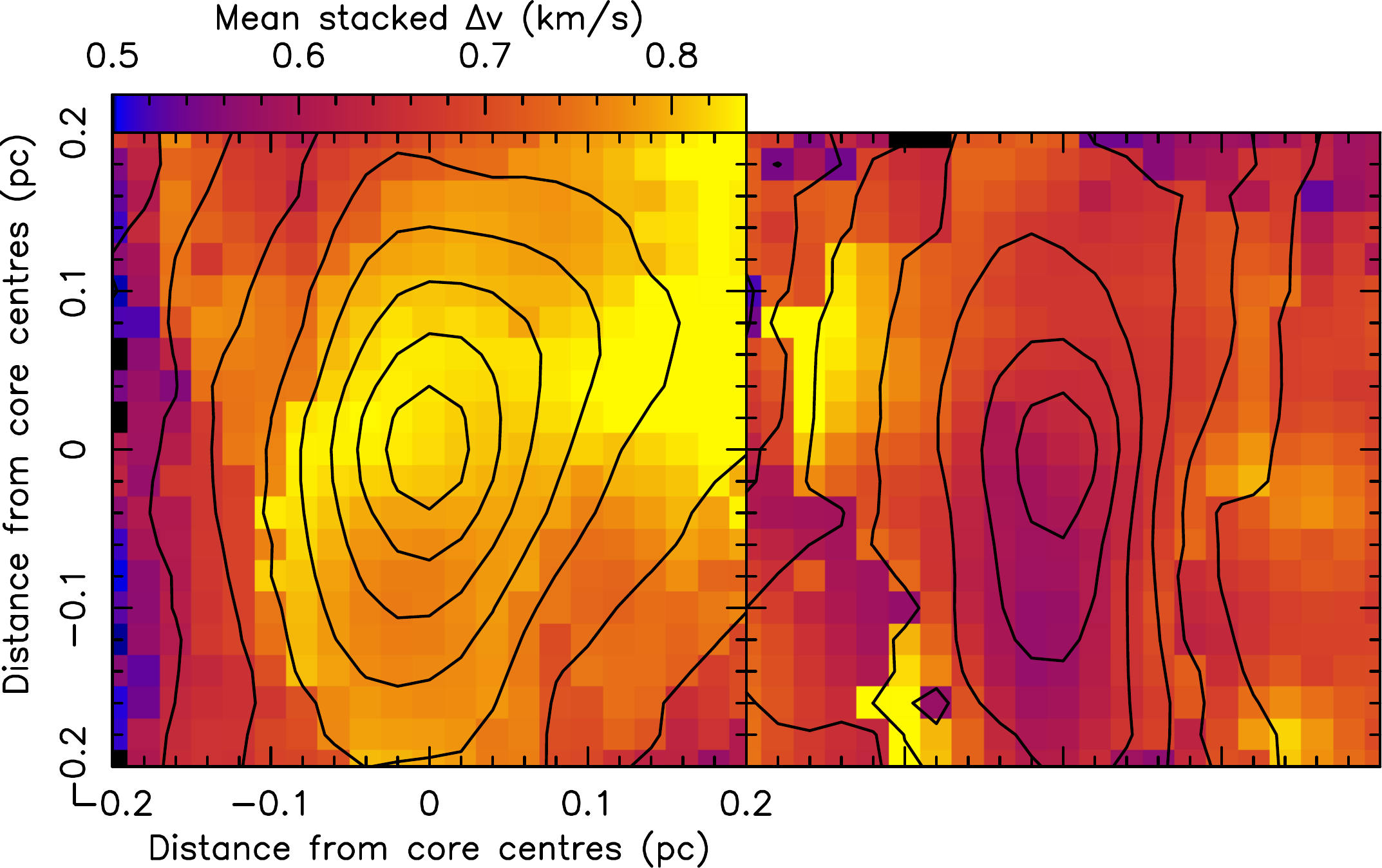}
\caption{The average stacked velocity width of cores identified with $\Delta V_{core} / \Delta V_{branch} > 1$ (left) and those with $\Delta V_{core} / \Delta V_{branch} < 1$ (right). Included in the construction of the left and right panels are 12 and 5 cores respectively. Only starless cores within the filamentary arms are included, excluding both protostellar and central hub region sources. Contours are of the respective average stacked column density in 0.5$\times 10^{22}$\,cm$^{-2}$ steps from 0 to 6$\times 10^{22}$\,cm$^{-2}$ on the left, and in 0.4$\times 10^{22}$\,cm$^{-2}$ steps from 0.2 to 3.8$\times 10^{22}$\,cm$^{-2}$ on the right.}
\label{fig:width-stacked}
\end{figure}

As already mentioned in Section \ref{sec:wid}, we do observe a correlation between the increase of velocity width and the presence of cores. In order to quantify the number of cores displaying such behaviour, we systematically computed the average velocity width within the extracted core area at its position in the dendrogram tree where it is first identified as a leaf ($\Delta V_{core}$).  Doing so we find that 73\% of the cores show an increase in their mean velocity width of $\geq$\,10\% when compared to the velocity width of the underlying  ``branch'' structure of the dendrogram tree ($\Delta V_{branch}$, as shown in Figure \ref{fig:core_branch}).  Furthermore, 87.5\% of these belong to the starless core population (according to the lack of mid-infrared point source).  
To visually demonstrate this increase even further, we split the cores into two sub-samples.  The first contains cores with $\Delta V_{core} / \Delta V_{branch} > 1$, whilst the second contains cores with $\Delta V_{core} / \Delta V_{branch} \leq 1$. 
In Figure~\ref{fig:width-stacked} we stack each of the sub-samples separately and take the mean velocity width (for starless cores only).  The overplotted contours are of the average H$_{2}$ column density. This figure clearly shows that the velocity dispersion is increased over the entire extent of the cores, with a peak towards their centres.

\subsection{Filaments} \label{sec:fil}

The orientation and morphology of all filamentary structures were constrained by identifying the spines of the filaments. As the name suggests, the spine is considered the backbone of a filament, tracing the pixels where the column density exhibits a local maximum in at least one direction.  This was quantified by computing the 2$^{\rm{nd}}$ derivative matrix  \citep[e.g.][]{schisano14}, i.e., the Hessian matrix, for each pixel in the column density map.  Then, by diagonalising the matrix, and selecting areas where at least one of the eigenvalues is negative, we can identify the filament areas. To reduce this to a spine, we used the \texttt{THIN} IDL function. The local orientation of the filaments at each pixel along the spine is provided by the angle of the first eigenvector and the $x$ axis of the image. Figure~\ref{fig:n} shows the spines of each of the four identified filaments, and Table \ref{tab:fil-properties} compiles their properties.  The masses of all filaments (including the portions of the spines that intersect the central hub) total to $\sim$\,1000\,M$_{\odot}$, equal the mass quoted by \cite{peretto14}.  Separating the central hub region from the filaments places half the total mass in the hub region and half in the filaments.

\begin{table*}[!htbp]
\centering
\caption{Summary of filament properties.}
\label{tab:fil-properties}
\begin{tabular}{l|cccccccccc}
\hline\hline
Filament & \begin{tabular}[c]{@{}c@{}}Length\\ (pc)\end{tabular} & \begin{tabular}[c]{@{}c@{}}Orientation\\ ($^{\circ}$)\end{tabular} & \begin{tabular}[c]{@{}c@{}}$\lambda_{core}$\\ (pc)\end{tabular} & \begin{tabular}[c]{@{}c@{}}Width\\ (pc)\end{tabular} & \begin{tabular}[c]{@{}c@{}}M$_{line}$\\ (M$_{\odot}$/pc)\end{tabular} & \begin{tabular}[c]{@{}c@{}}M$_{\mathrm{fil}}$\\ (M$_{\odot}$)\end{tabular} & \begin{tabular}[c]{@{}c@{}}$\tau_{crit}$\\(Myr)\end{tabular} & \begin{tabular}[c]{@{}c@{}}$\dot{M}$\\ (M$_{\odot}/$pc$/$Myr)\end{tabular} & \begin{tabular}[c]{@{}c@{}}$\tau_{age}$\\(Myr)\end{tabular} & \begin{tabular}[c]{@{}c@{}}$|\nabla V_{r}|$\\ (km$/$s$/$pc)\end{tabular} \\ \hline
North (N)       & 1.32 & -8.1   & --                & 0.27 & 270 & 259 & --   & --   & --  & 0.3 \\
North-West (NW) & 3.06 & -25.6  & 0.34\,$\pm$\,0.06 & 0.16 & 147 & 393 & 0.64 & 48.9 & 3.0 & 0.2 \\
South-East (SE) & 1.17 & +145.2 & 0.41\,$\pm$\,0.02 & 0.37 & 374 & 223 & 0.78 & 40.0 & 9.4 & 1.1 \\
North-East (NE) & 2.89 & +18.0  & 0.33\,$\pm$\,0.21 & 0.25 & 264 & 401 & 0.62 & 50.5 & 5.3 & 1.5 \\ \hline
\end{tabular}
\tablefoot{Column 1: Filament name. Col. 2: The length of the filament spine in parsec. Col. 3: The mean orientations from the minimisation of the eigenvalues of the Hessian matrices, in degrees. The origin at 0$^{\circ}$ was defined by the North direction, and the positive values are anticlockwise. Col. 4: median core separation, excluding hub cores, quoted with their standard deviation.  Given we have excluded hub centre cores, this leaves only one core in the North filament..  Column 5: Mean filament width in parsec (from 2.35 times the standard deviation of each radial column density) evaluated from the regions of the filament outside the hub centre only.  Col. 6: Mass per unit length, evaluated from the integration of radial column density profiles outside the cloud hub. Col. 7: the total mass of the filament, including the regions that pass through the cloud hub centre. Col. 8: The lower limit on the age of the filament (equation \ref{eq:age}) (in-evaluable for the North filament given its single core.). Col. 9: The accretion rate of material onto the filament during its formation. Col. 10: the time elapsed since the filament became critical. Col. 11: mean radial velocity gradient across the inner 0.2\,pc filament width.  Systematic errors from the kinematical distance to SDC13 ($3.6\pm0.4$\,kpc) and unknown filament inclination are associated to the filament lengths and mean core separations. Inclination angle affects the timescales quoted, as discussed in Section~\ref{sec:coresep}.}
\end{table*}

\begin{figure*}[!htbp]
\centering
\includegraphics[trim={0.5cm 1.1cm 0.6cm 1.85cm},clip,scale=.495,center]{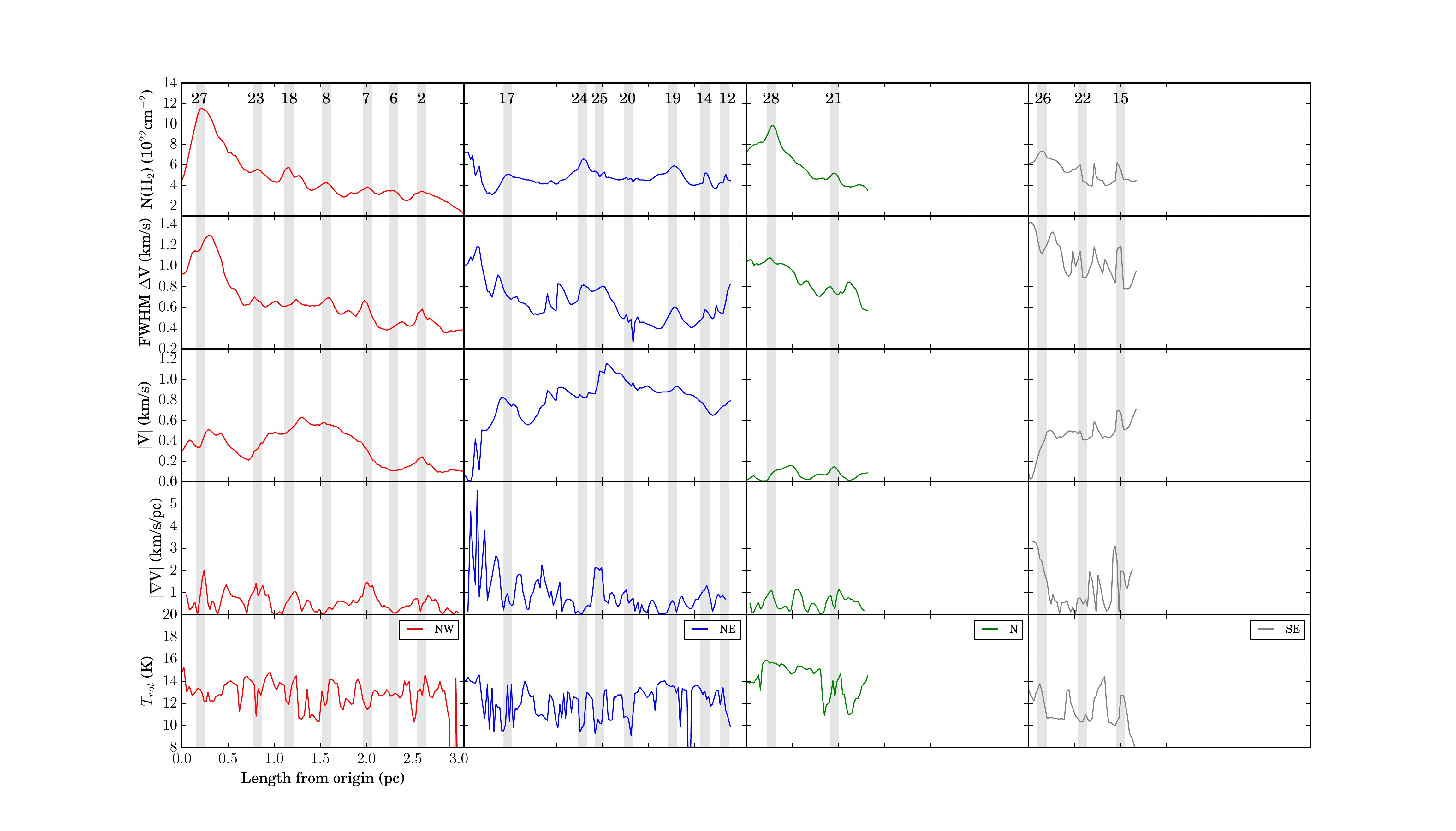}
\caption{Profiles along the spines, where each column denotes a different filament (North-West in red, North-East in blue, North in green and South-East in grey). The origin of each spine was centred at the hub region. The first row plots the column density of H$_{2}$ calculated from the fitted NH$_{3}$ integrated intensity. The second row plots the FWHM velocity width, while the third row plots the absolute centroid line-of-sight velocity (offset from the cloud systemic velocity of 37\,km\,s$^{-1}$), both also from the hyperfine structure fitting. The fourth row plots the absolute velocity gradient evaluated over the mean core size of $\sim$0.1\,pc.  The fifth row plots the rotational temperature derived from the NH$_{3}$ emission (see Appendix \ref{sec:trot}) which has a standard deviation of 1.8\,K. The drops of temperature below 8~K are artificial and due to missing NH$_{3}$(2,2) detections at these particular locations. The vertical shaded regions in each panel correspond to the peak N(H$_{2}$) positions of the cores along the spine, and are $\sim$0.1\,pc wide.}
\label{fig:along}
\end{figure*}

\subsubsection{Longitudinal filament profiles} \label{sec:filalong}

\begin{figure}[!t]
\centering
\includegraphics[scale=.505,right]{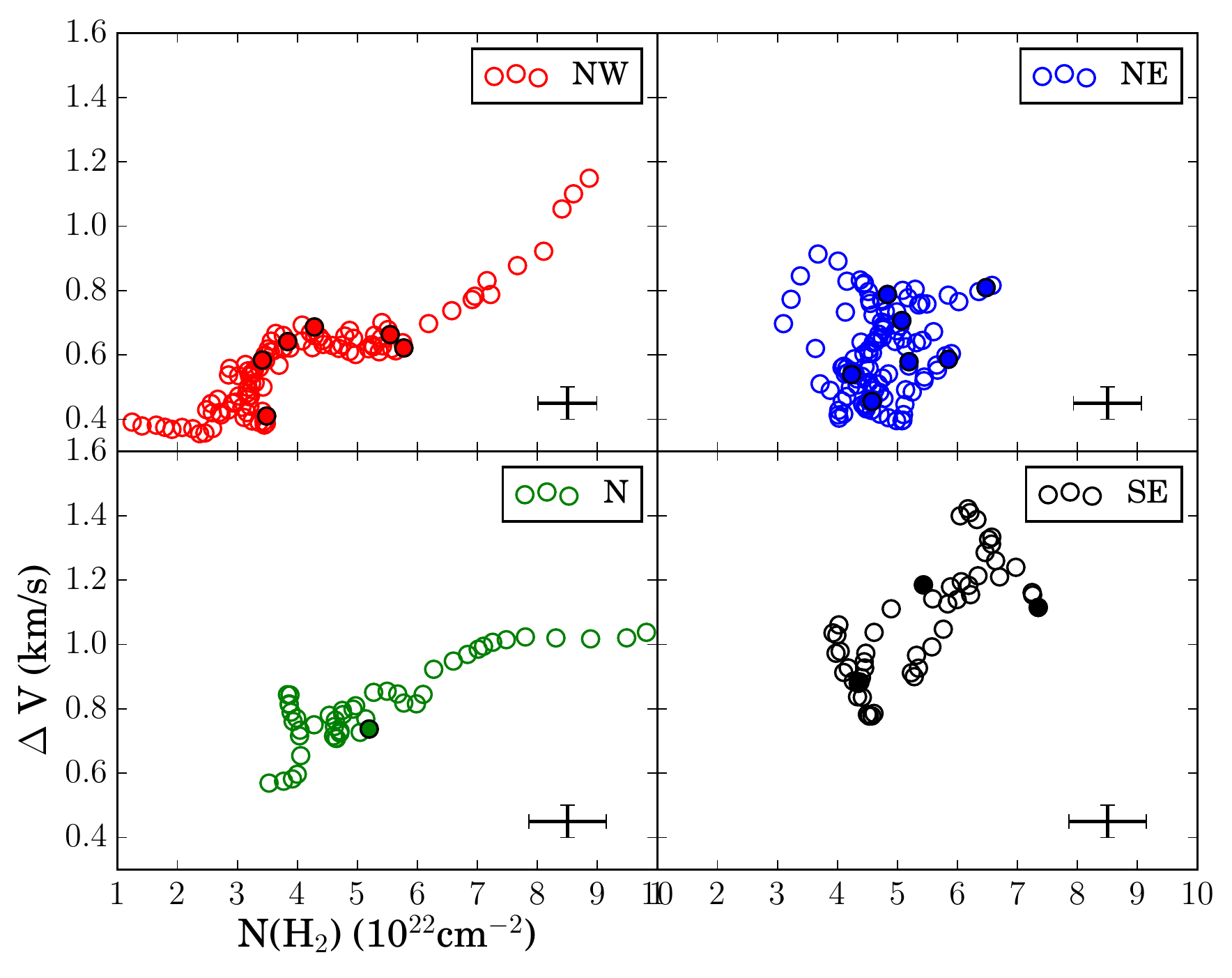} 
\caption{The correlations of velocity width to H$_{2}$ column density along the filament spine directions for the North-West (top left) North-East (top right), North (bottom left) and South-East (bottom right) filaments, excluding the portion of the spines that intersect the central hub regions (identified as where the filament can no longer be distinguished from the hub). Filled circles denote the spinal pixel position of the identified cores along the filaments.  A mean error is plotted in the bottom right of each panel, equal to $\pm20\%$ on the mean H$_{2}$ column density, and $\pm0.05$\,km\,s$^{-1}$ on $\Delta V$.}
\label{fig:nh2_vs_width}
\end{figure}

We first study the filaments along their spines.  Figure \ref{fig:along} plots the line profiles of all four filaments in H$_{2}$ column density (first row), velocity width (second row) and absolute centroid line-of-sight velocity (third row) where the origin of the filaments was defined to reside at the centre of the hub. Filaments show variations in all quantities, and on a range of spatial scales. 
First, we notice that these variations are correlated in all three quantities, being particularly obvious for the North-East and North-West filaments.  More specifically, core positions (marked by shaded regions on all panels) correlate with the velocity width, but also in centroid velocity where local velocity gradients are developed.  This is shown in the fourth row of Figure~\ref{fig:along} where we plot the absolute velocity gradient along the spine, evaluated over the mean core size of 0.1\,pc. We note that a significant fraction of the cores ($\sim63\%$) are located at a peak of velocity gradient.  In the bottom row of Figure~\ref{fig:along} we plot the rotational temperature derived from the NH$_{3}$ emission.  Although quite variable along the length of the filament, we do not find a correlation between the temperature and core positions, nor the kinematic properties.
We conduct Spearman's rank correlation coefficient test on these correlations for all four filaments separately. We find a definite correlation between the column density and velocity width peaks in every filament (as shown in Figure~\ref{fig:nh2_vs_width}) with vanishingly small p-values in every filament with coefficients ranging from 0.24 in the North-East filament (indicating a moderate correlation), to 0.74--0.91 in the other three filaments (indicating a strong correlation).  We find similarly small p-values in the column density and centroid velocity correlations, showing to us a definite link between all three of these properties.
Furthermore, we notice that there are other common features to all filaments, such as the strongest column density and velocity dispersion peaks located at the origin of the filaments. Note that this is not true for the North-East filament, but would likely be true if the abundance of ammonia towards the MM1 protostellar core \citep[][labelled in the left panel of Figure~\ref{fig:n}]{peretto14} were not decreased (see Section~\ref{sec:structure}). More generally, it is interesting to note that similarities emerge in the large scale morphology of all profiles between pairs of filaments (the North-West and North-East filaments on one side, and the North and South-East filaments on the other). Overall, such similarities are suggestive of the common physical origin of the SDC13 filament evolution, as discussed in detail throughout Section~\ref{sec:discussion}. 

\subsubsection{Radial filament profiles} \label{sec:filacross}

To construct a radial view of the filaments, 
we interpolate along an 0.4pc slice perpendicularly oriented to each spinal pixel using a Taylor expansion method (where the interpolation step is equivalent to half a pixel).
Every slice is used to construct radial position-velocity (PV) profiles from the NH$_{3}$ hyperfine structure fitted cube.  Any prevailing velocity gradient oriented parallel to the filaments however causes each radial PV slice to be centred differently in velocity, effectively smearing the mean radial PV profile. This was corrected by aligning each PV slice to the central velocity of the slice.   


\begin{figure}[!htbp]
\centering
\includegraphics[scale=.43,center]{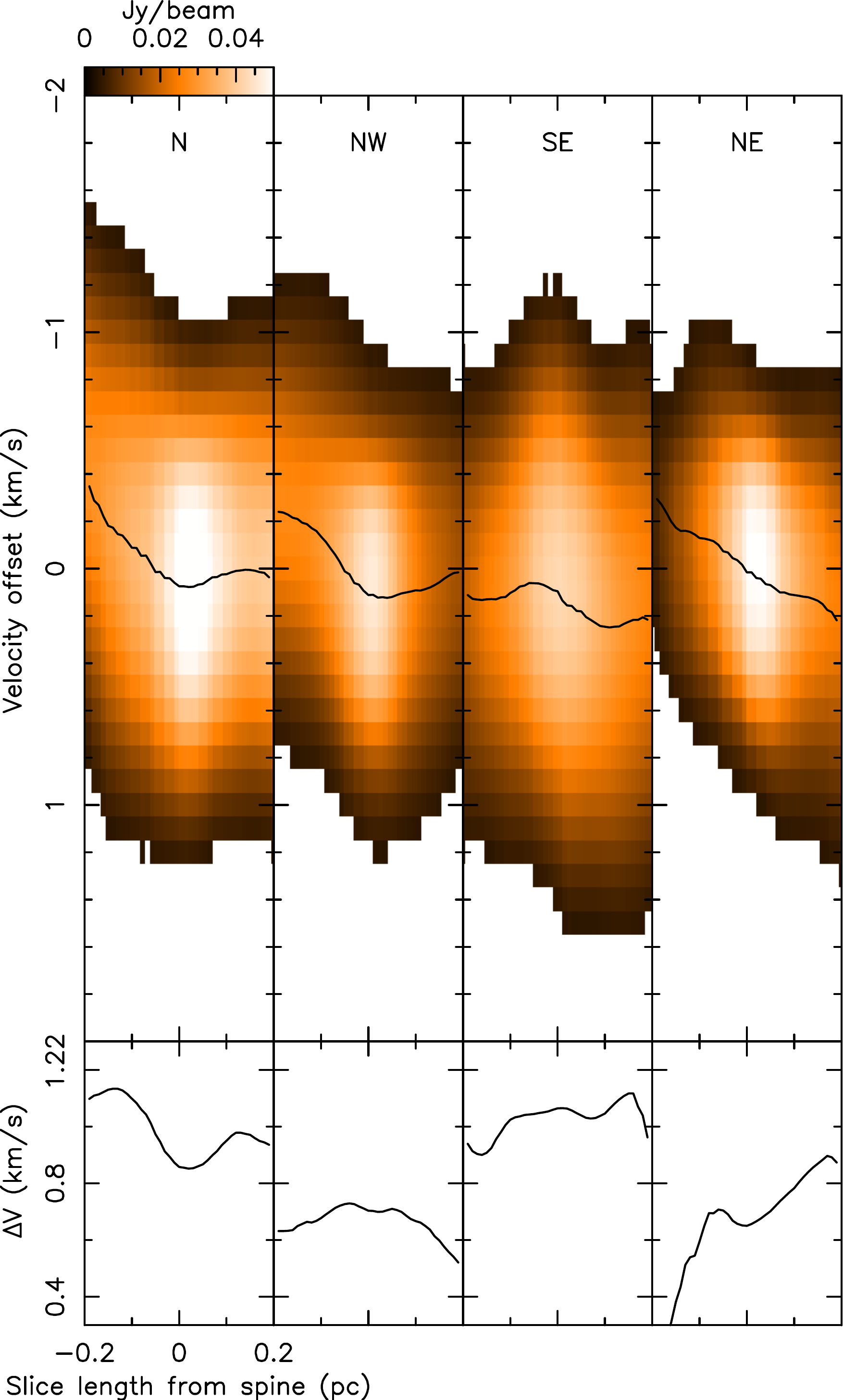}
\caption{\textit{Top}: The mean position-velocity diagrams across the North, North-West, South-East and North-East filaments (from left to right), where position denotes the length of a perpendicular slice relative to the central spine pixel in parsec, and the velocity are relative to the peak velocity of each slice. The overplotted black line shows the trend of the intensity-weighted mean offset velocity at each position along the radial slice. Mean radial gradients were computed within the inner 0.2pc width around the spinal pixel. \textit{Bottom}: The mean variation in the velocity width in the radial direction, in km$/$s.}
\label{fig:pvacross}
\end{figure}


It is clear from the on-axis concentric nature of the highest intensity emission in the North and North-West filaments in Figure \ref{fig:pvacross} that any radial velocity gradient does not dominate the velocity field.  The intensity-weighted mean of these profiles does appear flat at their centremost regions around the spinal pixel, with relatively low radial velocity gradients of 0.3\,km\,s$^{-1}$\,pc$^{-1}$ and 0.2\,km\,s$^{-1}$\,pc$^{-1}$ respectively.  In contrast, the South-East and North-East filaments have rather significant mean radial velocity components across their entire width, corresponding to 1.1\,km\,s$^{-1}$\,pc$^{-1}$ and 1.5\,km\,s$^{-1}$\,pc$^{-1}$ respectively.  Although significant in this context, these gradients are not as strong as others observed by \cite{fernandez-lopez14} or \cite{beuther15} for example, who observe radial gradients an order of magnitude larger than longitudinal velocity gradients in Serpens South and IRDC 18223, respectively. 

The radial variations of the velocity width are plotted in the bottom panels of Figure~\ref{fig:pvacross}. These profiles show that the longitudinally averaged velocity dispersion varies across all filaments, with, in some cases, local minima along the filament spine, and local maxima at the edges. However, as discussed in Section 5, these trends result from the average of a number of different processes that are mixed-up together in such profiles.

\subsection{JVLA/GBT combined versus JVLA--only datasets}

\begin{figure*}[!t]
\centering
  \includegraphics[width=1.0\textwidth]{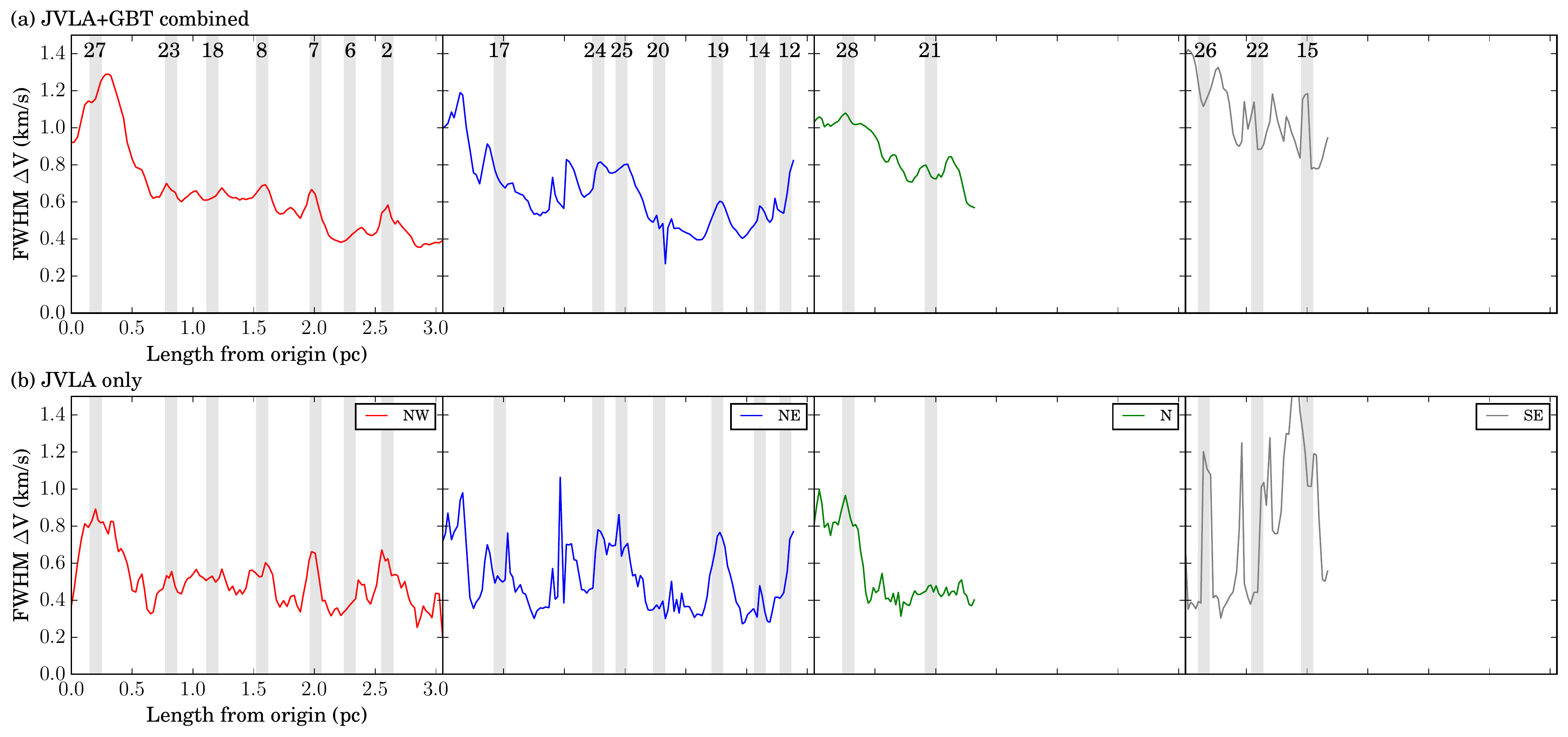}%
\caption{Velocity width profiles of the North-West (red), North-East (blue), North (green) and South-East (grey) filaments in the combined data (top row, as presented in the second row of Figure~\ref{fig:along}) and the JVLA--only data (bottom row), evaluated over the same filament spines as identified in Section~\ref{sec:fil}.}
\label{fig:width_profiles_jvla}
\end{figure*}

\begin{figure}[!t]
\centering
  \includegraphics[width=0.47\textwidth]{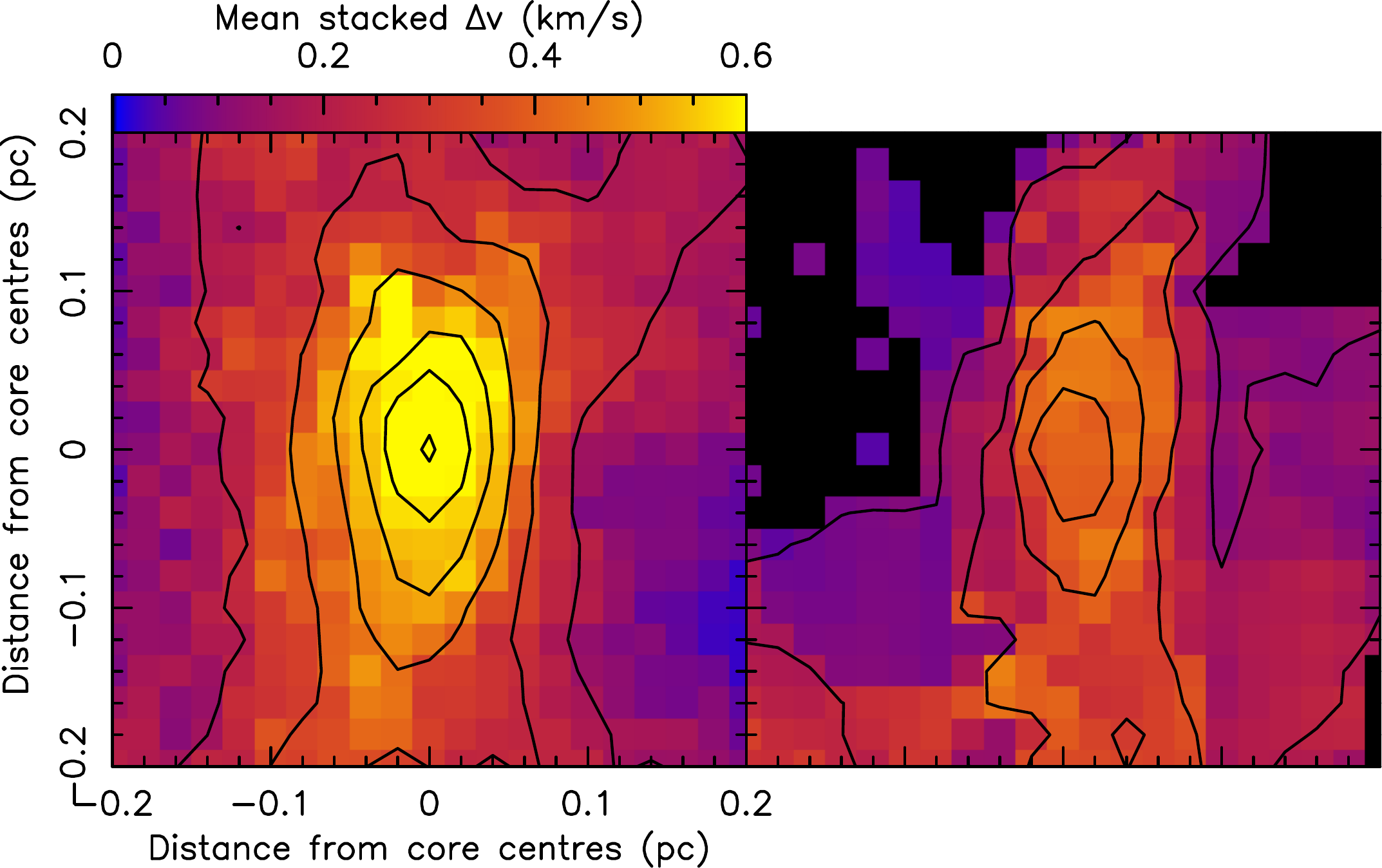}%
\caption{Reproduction of Figure~\ref{fig:width-stacked}, but for the JVLA--only data. The average stacked velocity width of cores identified with velocity width peaks (left) and those without peaks (right).  Included in the construction of the left and right panels are 13 and 4 cores respectively.  Plot details are the same as those detailed in the caption of Figure~\ref{fig:width-stacked}.}
\label{fig:width_stacked_jvla}
\end{figure}

In previous sections, results obtained from the combined JVLA/GBT dataset were presented. As already mentioned in Section~\ref{sec:combined}, such a combined dataset is optimal to recover emission at all spatial scales. However, when trying to characterise the gas properties of the densest parts of the filaments (i.e. cores and filament spines) as we have done above, results can be affected by the contamination of the larger scale, more diffuse, foreground/background emission. The effect of this contamination would be to dilute the observational signatures one is trying to identify.

For that reason we redid some of the analysis presented in earlier sections using the JVLA--only dataset, which traces only the most compact/densest regions of the cloud. We particularly focus on the velocity dispersion variations along the filament spines and towards the cores. By fitting the hyperfine components of the JVLA--only NH$_{3}$ emission, we find that the increase of the velocity width at the core positions is even larger than in the combined data set. This is illustrated in Figures~\ref{fig:width_profiles_jvla} and \ref{fig:width_stacked_jvla}, reproductions of Figures~\ref{fig:width-stacked} and \ref{fig:along} but of the JVLA--only data.  From these figures, one can see that the JVLA--only velocity width towards the core centres is 1.5--2 times larger than that of the surroundings, compared to a 1.1--1.4 factor in the JVLA/GBT combined data.  We also notice that in the JVLA--only data the remaining starless cores also display an increase of velocity width. This analysis strengthens our conclusions about the local increase of velocity dispersion towards the SDC13 starless cores.

\section{Discussion} \label{sec:discussion}

\begin{figure*}
    \centering
    \begin{minipage}{.5\linewidth}
    \subfloat{\label{fig:deprojnw}\includegraphics[scale=.68,center]{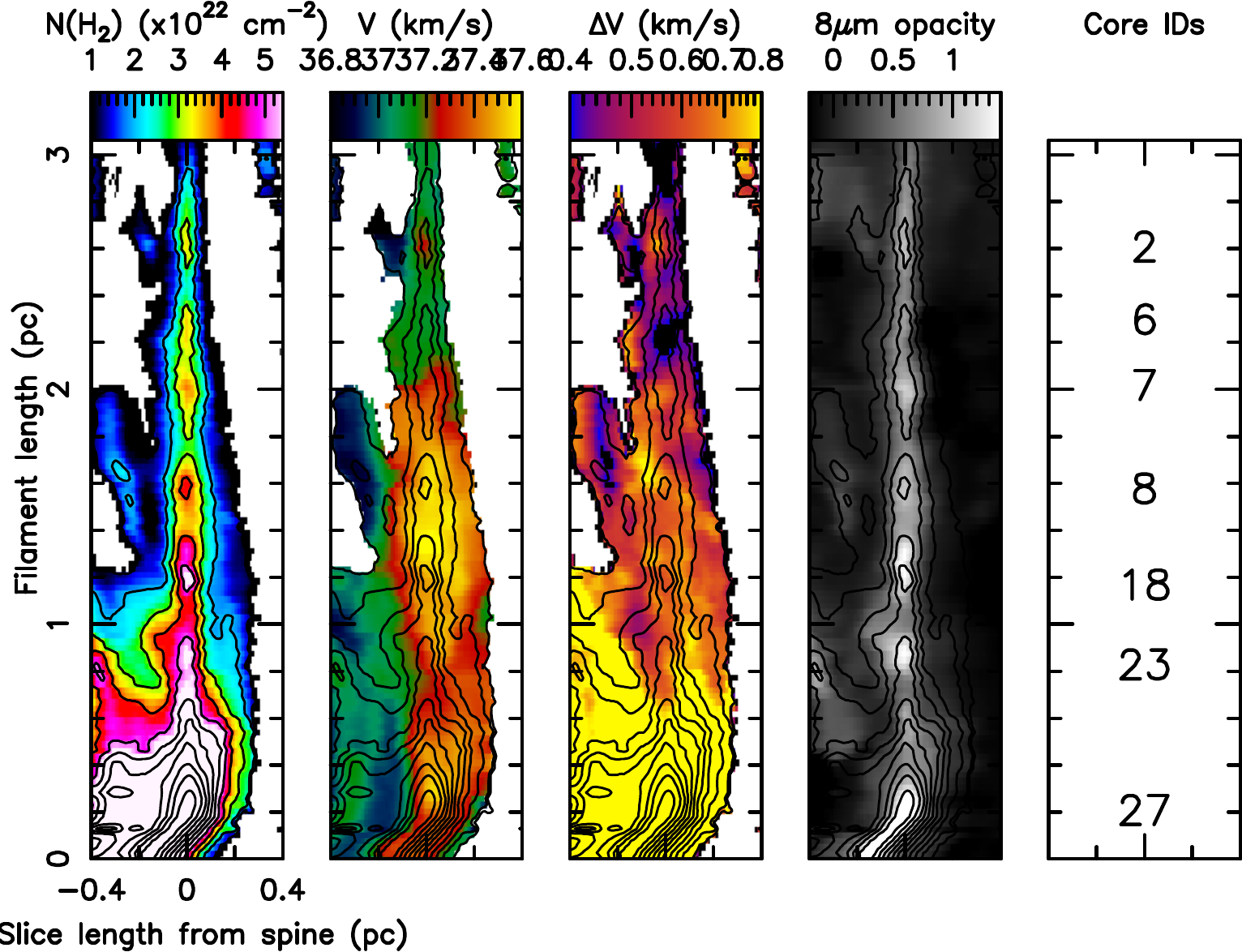}}
\end{minipage}\par\medskip
\begin{minipage}{.5\linewidth}
    \subfloat{\label{fig:deprojne}\includegraphics[scale=.68,center]{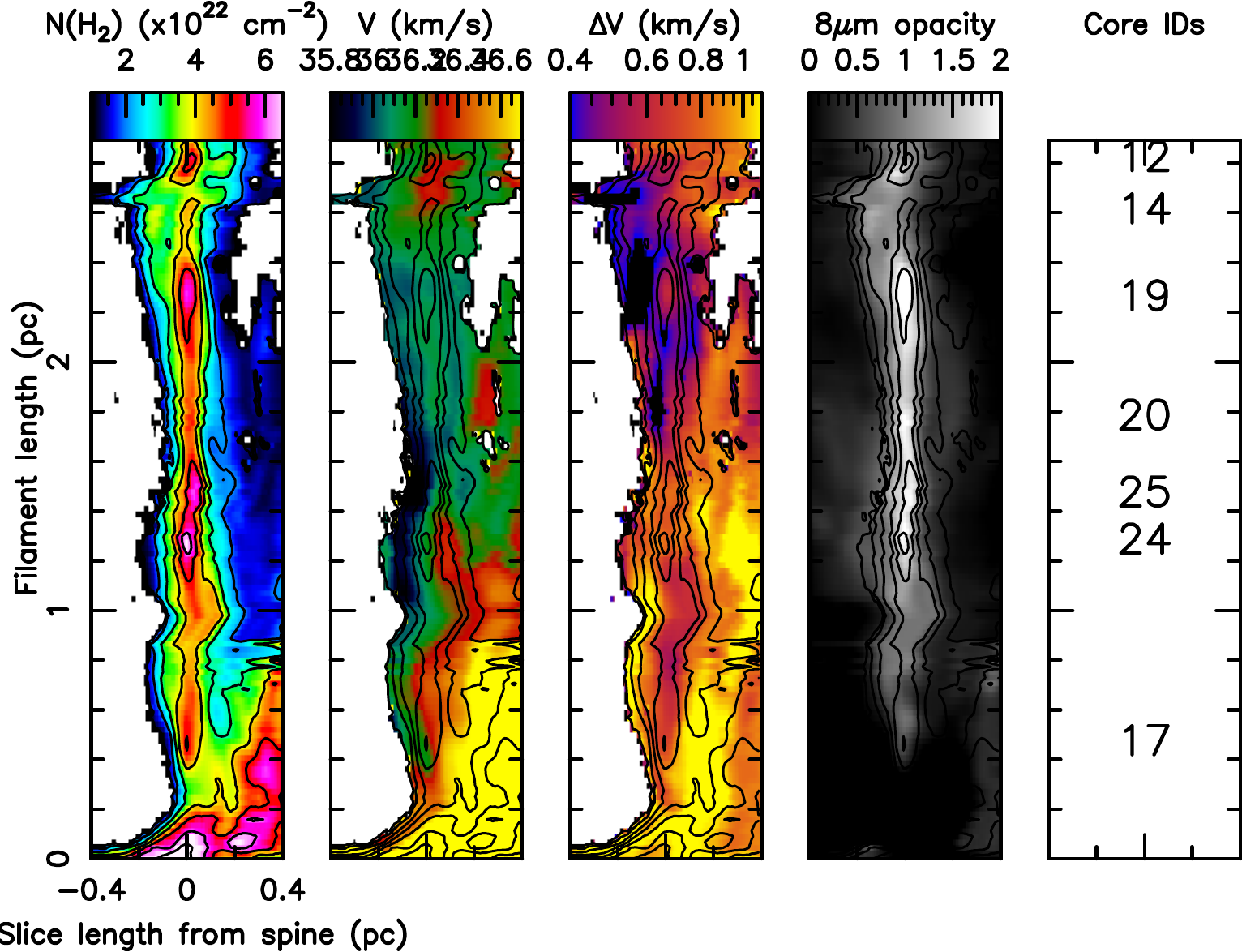}}
\end{minipage}
\caption{Deprojected views of the North-West (top) and North-East (bottom) filaments. The filament length is plotted on the y-axis, while the length of the radial slices from the central spine pixel in plotted on the x-axis. In both sub-figures, the panels show (from left to right) the H$_{2}$ column density derived from the NH$_{3}$ emission, the line-of-sight velocity, the velocity width, the opacity derived from the 8$\mu$m Spitzer emission, and finally the core ID as listed in Table \ref{tab:cores}. Contours in the first four panels are of the column density, from $1\times10^{22}$\,cm$^{-2}$ to $11\times10^{22}$\,cm$^{-2}$, spaced by $1\times10^{22}$\,cm$^{-2}$. The plots for the South-East and North filaments are in Appendix \ref{appendix-B}.}
\label{fig:deprojected}
\end{figure*}

In this section, we discuss the various observational properties of the filaments in SDC13 in terms of filament evolution theories. To facilitate the visualisation of the filament properties, the radial interpolation carried out in Section \ref{sec:filacross} was carried out on the column density, centroid velocity, velocity width and Spitzer 8\,$\mu$m opacity for all filaments. Aligning each slice to the pixel that lies along the filament spines gives us a unique deprojected perspective of the filaments (Figure~\ref{fig:deprojected}) i.e. a view of the filament independent of their on-sky projection, making some of the filament features discussed below stand out.

\subsection{Supercritical filaments} \label{sec:supercritical}

\begin{figure*}[!t]
\centering
  \includegraphics[scale=.58,center]{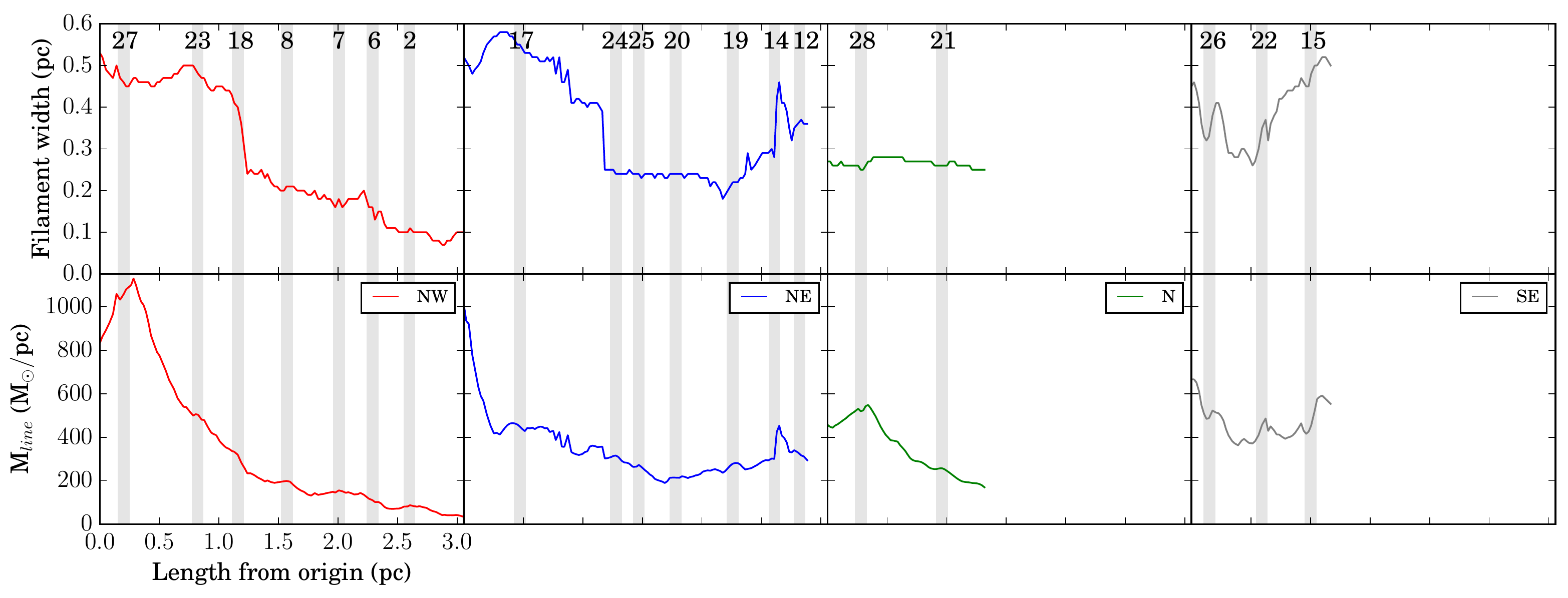} 
\caption{Caption as for Figure~\ref{fig:along}.  The first row plots the filament width calculated from 2.35 times the standard deviation of the radial column density distributions.  The jump in width seen at roughly 1.2\,pc in the North-West and North-East filaments is due to the exit of the filament spine from the hub. The mean filament width was evaluated beyond this point.  The second row plots the mass per unit length, evaluated from conducting a Trapezoidal integration under all radial column density distributions.}
\label{fig:width_mline}
\end{figure*}

One important parameter of interstellar filaments regarding their stability is their mass-per-unit-length, M$_{\mathrm{\mathbf{line}}}$. The first step towards measuring this quantity is to determine their radial column density profiles.
In our attempt to compute these profiles, neither a Gaussian distribution nor a Plummer profile \citep{whitworth01} accurately reflects the complexity of the radial column density distributions \citep[unlike][for example, see Appendix \ref{app:radial_nh2}]{arz11}.  Therefore, we instead integrate all radial column density distributions using the Trapezoidal rule at every position along the filaments spines to calculate the mass per unit length. A rough estimation of the filament widths were obtained by taking $\sqrt{8\ln2}$ times the standard deviation of the radial column density profiles. In calculating the mean M$_{\mathrm{line}}$ and filament widths (listed in Table~\ref{tab:fil-properties}), we excluded the regions that intersected the cloud hub, (identified as belonging to the radial column density slices where the extended nature of the hub emission makes the determination of the width impractical). Figure~\ref{fig:width_mline} plots the evolution of the filament width and M$_{\mathrm{line}}$ along the filament lengths respectively.  

The critical mass per unit length, M$_{\mathrm{line,crit}} = 2a_{0}^{2}/G$ (where $a_{0}$ is the isothermal sound speed and G is the Gravitational constant) is the critical value above which an interstellar filament becomes gravitationally unstable to radial contraction and fragmentation \citep{ostriker64}.  For a typical 10\,K filament,  M$_{\mathrm{line,crit}} = 16$\,M$_{\odot}$\,pc$^{-1} [\frac{T}{10K}]$.  At the mean rotational temperature of SDC13 of 12.7\,K, $a_0=$\,0.21\,km\,s$^{-1}$ and the critical mass per unit length is 20.4\,M$_{\odot}$\,pc$^{-1}$.  All SDC13 filaments can be classed as thermally supercritical (see Table \ref{tab:fil-properties}) therefore prone to radial gravitational contraction and fragmentation along their lengths. 
When considering the effective sound speed, $a_{eff,0} = \sqrt{a_{o}^{2} + \sigma_{NT,0}^{2}} = 0.26\pm0.02$\,km\,s$^{-1}$ (where $\sigma_{NT,0} = 0.15\pm0.02$\,km\,s$^{-1}$, taken from Figure~\ref{fig:mach} as the most representative value of the filament dense gas prior to fragmentation - see Section~\ref{sec:probe_infall}) the critical mass per unit length becomes 31.3\,M$_{\odot}$\,pc$^{-1}$, a factor of 1.5 larger than the previously calculated thermal M$_{\mathrm{line,crit}}$, but still a factor of 4 to 10 lower than the measured filament line masses.


\subsection{Core separation and age estimates} \label{sec:coresep}

\begin{figure}
    \centering
    \begin{minipage}{.5\linewidth}
    \subfloat{\label{fig:hist}\includegraphics[trim={1cm 0cm 1.5cm 1cm},clip,scale=.5,center]{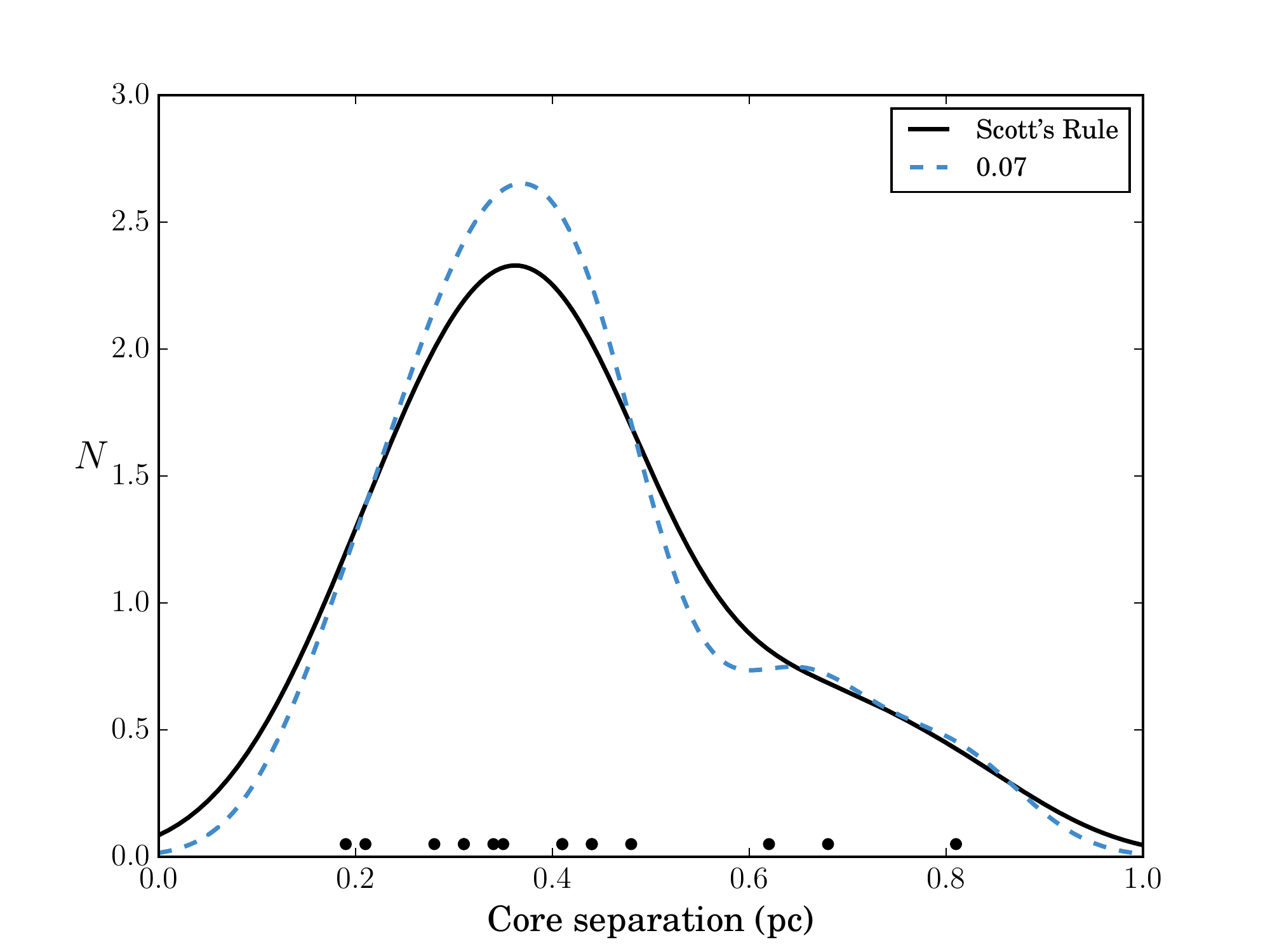}}
\end{minipage}\par\medskip
\begin{minipage}{.5\linewidth}
    \subfloat{\label{fig:cumulative}\includegraphics[trim={1cm 0 1.5cm 1cm},clip,scale=.5,center]{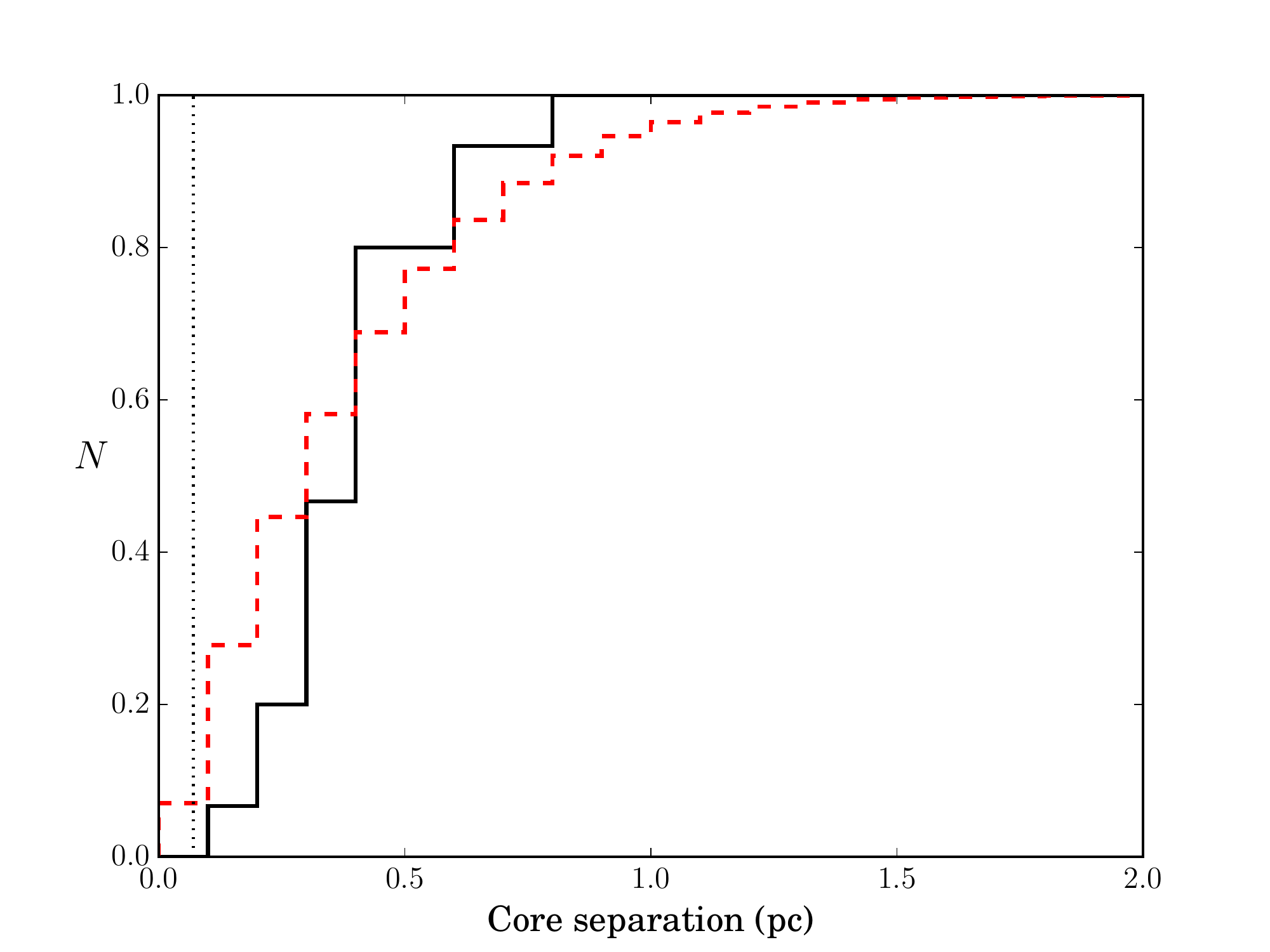}} 
\end{minipage}
\caption{\textit{Top}: Gaussian Kernel Density Estimation (KDE) plot of the separation of each consecutive core pair in each of the four filaments. Each line denotes a different bandwidth Gaussian, 0.07\,pc (equivalent to the beam size, in dashed blue) and 0.1\,pc (derived by the standard Scott's Rule, and equivalent to the mean core radius, in black).  Black circles at $N = 0.05$ are overplotted to show the positions of the individual spacings in the sample.  The standard deviation is 0.16\,pc. \textit{Bottom}: Normalised, cumulative distribution of the observed core separation (black line) and the 100,000 simulated core separations  (dashed red line). The bin size in both plots was set to be slightly larger than the beam at 0.1\,pc in size.  The vertical dotted line in both denotes the data spatial resolution of 0.07\,pc, used as a cut-off of the allowed simulated spacings.}
\label{fig:hist-sp}
\end{figure}

The separation of cores along filaments can provide indication on the physical mechanism driving the fragmentation. The top panel of Figure~\ref{fig:hist-sp} shows the Gaussian Kernel Density Estimation (KDE) plot of core separations in all filaments (excluding hub cores) where the two plotted distributions each use a different bandwidth Gaussian.  The core separations were calculated on the original on-sky projected filament spines, and not from the deprojected plots in Fig.~\ref{fig:deprojected}, although both provide identical results.
There is clearly a peak in each distribution at a core separation $\lambda_{core} \simeq 0.37\pm0.16$\,pc.
Regular spacings in IRDCs with core separation ranging between $\sim0.2$ to 0.4\,pc have already been reported \citep{beuther15, henshaw16, zhang09}.  For comparison, we simulated the four filaments by randomly placing the same number of cores as observed along each of the four filament lengths and calculated the separation of each consecutive core pair \citep[e.g.][]{teixeira16}, and repeated the process 100,000 times.
In doing this, we stipulated that the smallest core separation in these random core distributions must exceed the beam size, equivalent to 0.07\,pc spatial scale. We can clearly see that the observed cumulative histogram (black) is steeper than the simulated cumulative histogram (red) in the bottom panel of Figure~\ref{fig:hist-sp}. To quantify these differences, we used the Kolmogorov-Smirnov test to estimate the probability that these two histograms originate from the same parent distribution. We find that there is a 15\% chance that this is the case, exceeding the significant test limit of 5\%.  However, coupled with the relatively large K-S statistic of 0.28 (suggesting a large maximum distance between the two cumulative distributions), we reject the null hypothesis and consider that these two histograms are significantly different, demonstrating that the core spacing in the SDC13 filaments are not randomly distributed.  The median value of $\lambda_{core}$ in each filament is given in Table~\ref{tab:fil-properties}. Note that looking at the deprojected view of the filaments in Figure~\ref{fig:deprojected}, the spacing of the North-West filament seems to be more regular than the others.  This is supported by the lower core spacing standard deviation of the North-West filament, i.e. 0.06\,pc (whilst excluding core number 27 in the hub centre) versus 0.21\,pc for the North-East filament.

\cite{inutsuka92} and \cite{miyama94} showed that filaments in hydrostatic equilibrium fragment under density perturbations whose wavelengths are four times the filament diameter. Looking at the relevant values given in Table~\ref{tab:fil-properties} we see that the ratio between the core separation of the SDC13 filaments and their widths varies between 1.1 to 2.1 which, given the unknown inclination, is compatible with a ratio of 4. However, in a turbulent medium, filaments in equilibrium are likely to be rare objects. Interestingly, \cite{clarke16, clarke17} studied the fragmentation of non-equilibrium, accreting filaments and showed that the fastest growing mode of density perturbations in such systems, $\lambda_{core}$, is a function of the time it takes to build-up a critical filament through accretion, $\tau_{crit}$, and the effective sound speed, $a_{eff,0}$: 
 
     \begin{equation}
        \lambda_{core} =2a_{eff,0}\tau_{crit} \, ,
        \label{eq:age}
     \end{equation}

\noindent By measuring the core separation and the gas temperature, it then becomes possible to derive $\tau_{crit}$. One can also relate $\tau_{crit}$ to the accretion rate onto the filament $\dot{M}$ by:

    \begin{equation}
        \tau_{crit}=\frac{M_{\mathrm{line,crit}}}{\dot{M}}
    \end{equation}
    
\noindent where M$_{\mathrm{line,crit}}$ is the critical mass per unit length. Values for $\tau_{crit}$ and $\dot{M}$ (derived using the previously calculated $a_{eff,0}$\,=\,0.26\,km\,s$^{-1}$ and M$_{\mathrm{line,crit}}$\,=\,31.3\,M$_{\odot}$\,pc$^{-1}$) are given in Table \ref{tab:fil-properties} for each filament. According to this model, we see that it takes on average $\sim$\,0.68\,Myrs to form the SDC13 filaments up to the critical mass per unit length, with an average accretion rate $\sim$\,46.5\,M$_{\odot}$\,pc$^{-1}$Myr$^{-1}$. Interestingly, assuming that the derived accretion rate remained constant over the entire filament lifetime, one can derive the age, $\tau_{age}$, through:

    \begin{equation}
        \tau_{age}=\frac{M_{line}}{\dot{M}}
    \end{equation}

As shown in Table~\ref{tab:fil-properties} we obtain an average value of $\tau_{age}\sim$\,5.9\,Myr, which implies that the elapsed time since the filaments became critical is $\sim$\,5.2\,Myr. Based on the observed longitudinal velocity gradients and assuming that the gas is free-falling, \cite{peretto14} estimated that the SDC13 filaments had been collapsing for $\sim1$ to 4\,Myr. This is consistent with the estimate based on the \cite{clarke16, clarke17} model. However, in the calculation of $\tau_{crit}$ we have not taken into account projection effects in the estimate of $\lambda_{core}$. Doing so will increase $\tau_{crit}$ by a factor $1/\cos(\theta)$ where $\theta$ is the angle between the direction of the main axis of the filament and the plane of the sky, which for an average angle of $67^{\circ}$ \citep{peretto14} is equal to 1.7\,Myr. This brings the estimated age of the filament to 15.1\,Myr, more than a factor of 3 larger than the dynamical timescale estimated by \cite{peretto14}. 

\subsection{Collapse timescales}

\begin{figure}[!t]
\centering
  \includegraphics[width=0.45\textwidth]{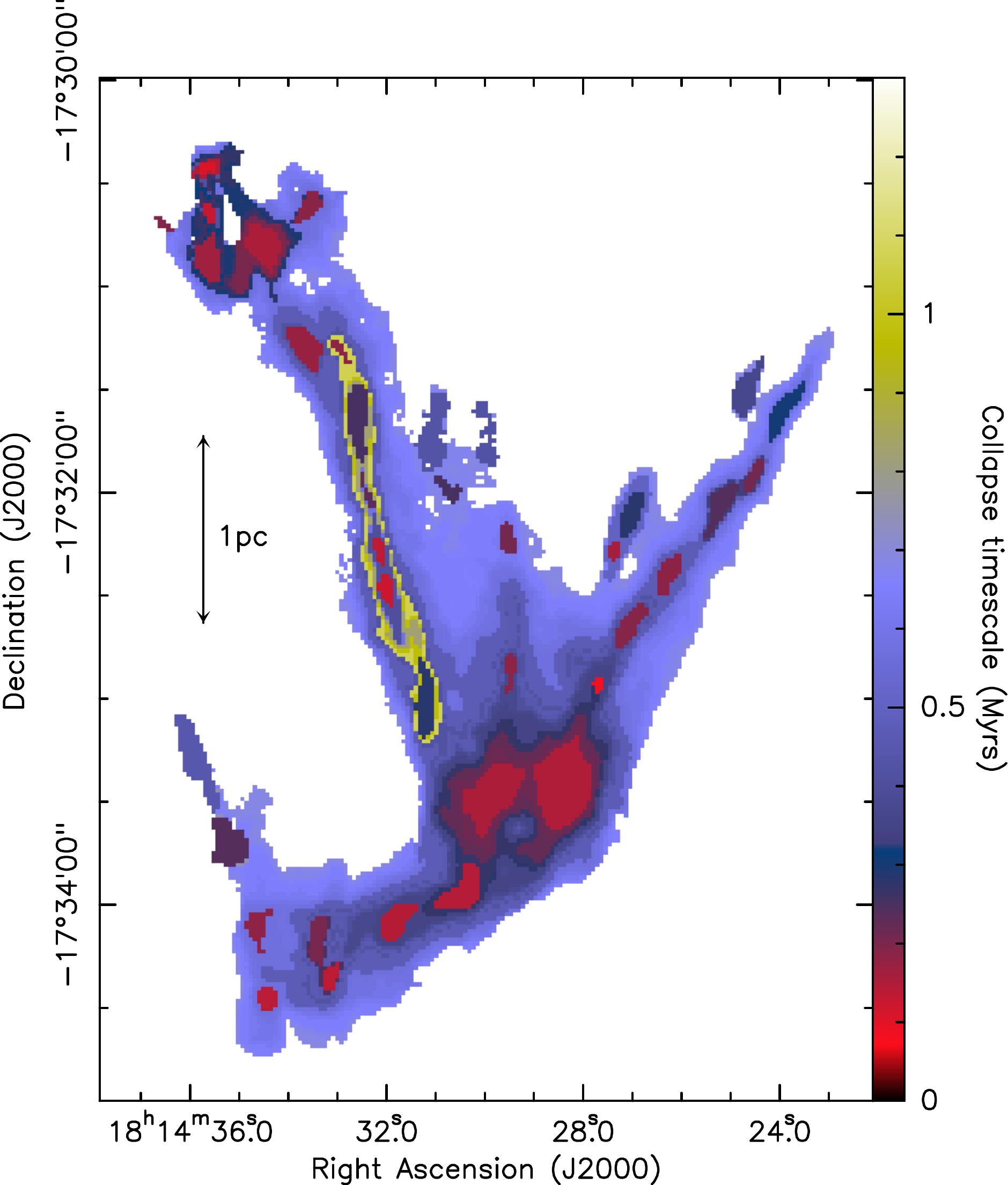}%
\caption{Map where the outline of all extracted structures are coloured by their collapse timescale t$_{col}$ in Myrs.}
\label{fig:collapse}
\end{figure}

With all the hierarchical structure in SDC13 extracted using the dendrogram method, we calculate their radii, aspect ratios and mean densities.  As the free-fall collapse time of a structure depends only on the density, we calculate t$_{ff}$ for all extracted structures, from the base of the dendrogram tree up to the leaves using:
    \begin{equation}
        t_{ff} = \left (\frac{3\pi }{32\mathrm{G}\rho } \right )^{1/2} \, ,
        \label{eq:tff}
    \end{equation}
\noindent where G is the gravitational constant, and $\rho$ is the density of the core (for which we use the upper limit on the mass in column 12 of Table \ref{tab:cores}).  However, this is not an appropriate approach for non-spherical objects like the high-aspect ratio filamentary structures within SDC13 \citep{pon12,toala12}. \cite{clarke15} derive a collapse timescale (t$_{col}$) valid for both filamentary and near spherical structures:
    \begin{equation}
        t_{col} = (0.49 + 0.26\mathrm{A_{o}}) (\mathrm{G}\rho)^{-1/2} \, ,
        \label{eq:tcol}
    \end{equation}
\noindent where A$_{\mathrm{o}}$ is the aspect ratio, valid down to values of A$_{o}$ $\gtrsim$ 2. The filamentary nature of SDC13 from large to small scale (i.e. 61$\%$ of cores having $A_0>2$ - Table~\ref{tab:cores}) justifies the need to use equation \ref{eq:tcol}. Figure~\ref{fig:collapse} shows the collapse time $t_{col}$ for all identified structures in the SDC13 dendrogram.  Overall we see a decrease of the collapse time, from $\sim$\,0.7\,Myr to 0.1\,Myr as we go from large to small structures as a result of the structures' decreasing aspect ratios and increasing densities. A consequence of this hierarchical collapse time is that cores will collapse well before the filaments. This is the basic idea behind the hierarchical star formation models of \cite{vazquez-semadeni09, vazquez-semadeni17}.  On the same figure, we see one structure in the North-East filament that departs from the rest of the SDC13 structures, with a longer collapse time than the structure in which it is embedded in. This is due to the particularly long aspect ratio of that particular structure. We also notice that the same structure exhibits the strongest radial velocity gradient (see Section 5.2) and the smallest core separation (see Table~\ref{tab:fil-properties}).  As speculated later in the paper, we argue that this is a direct consequence of the compression of the pre-existing filament by the feedback of a nearby star formation event. 

\subsection{Linewidth as a probe of infall} \label{sec:probe_infall}

\begin{figure}[!t]
\centering
\includegraphics[trim={0.3cm 0.3cm 0 0},clip,scale=.52,center]{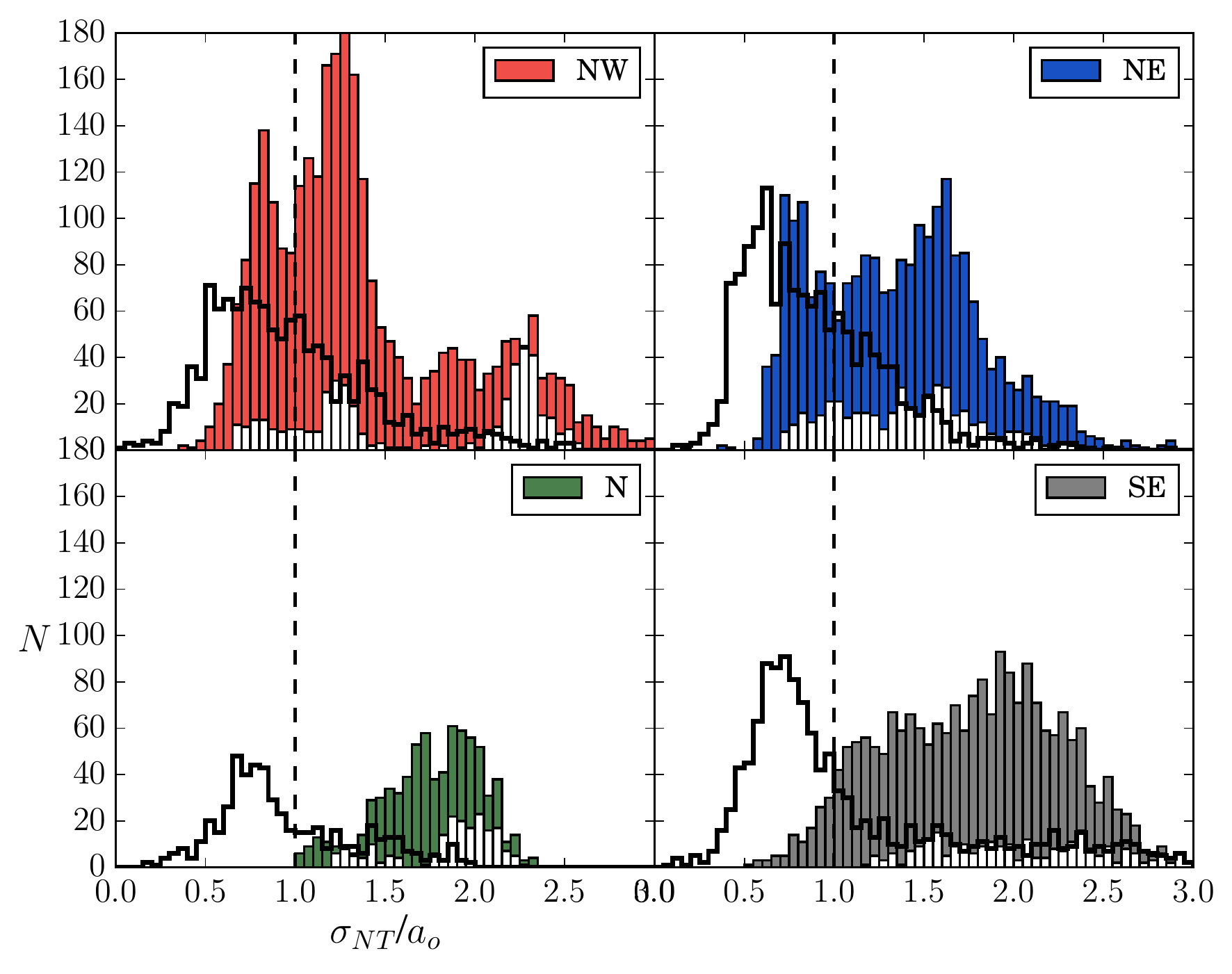} 
\caption{Histogram of the non-thermal contribution to the velocity dispersions, divided by the isothermal sound speed at the temperature of the NH$_{3}$ gas of 12.7\,K i.e. the Mach number. The red histogram is the North-West filament, blue is the North-East, green is the North while grey is the South-East.  The vertical dashed line denotes the transition from subsonic to supersonic motions.  Overplotted in white in each panel is the Mach number of the core regions only from the combined data. The overplotted black line is the distribution of the JVLA--only velocity dispersion results.}
\label{fig:mach}
\end{figure}

The increase of velocity width within the cores can be quantitatively compared to the velocity dispersion within the filaments. In that respect, we calculated the non-thermal contribution of each pixel within SDC13 following \citep{myers83}:
    \begin{equation}
        \sigma_{NT} = \sqrt{\frac{\Delta V_{obs}^{2}}{8\ln2}-\frac{\mathrm{k_{b}T}}{m_{\mathrm{NH_{3}}}}}
        \label{eq:sigmant}
    \end{equation}
\noindent where $\Delta V_{obs}$ is the fitted FWHM velocity width, $k_{b}$ is the Boltzmann constant, and $m_{\mathrm{NH_{3}}}$ is the mass of an NH$_{3}$ molecule.   To determine whether the filaments are subsonic or supersonic in nature, we compare this non-thermal sigma velocity dispersion to the isothermal sound speed $a_{0}$ at 12.7\,K of 0.21\,km\,s$^{-1}$. Figure~\ref{fig:mach} shows the histograms of $\sigma_{NT} / a_{0}$ for all four filaments, where pixels within core regions only are overplotted in white.  Overplotted in black are the corresponding distributions taken from the JVLA--only data. We see that in all filaments the gas in the combined data is predominantly supersonic, however, the North-West and North-East filaments also peaks significantly in the sub/transonic regime.  This second peak is coincident with that of the JVLA data which only exhibits a minor supersonic tail in its distribution whilst peaking sub/transonically.


\cite{barranco98} and \cite{goodman98} find that $\sigma_{NT}$ remains constant across core regions, and increases once outside the core boundary.  They call this behaviour velocity ``coherence'', suggesting that cores have sizes that are intimately linked to the turbulence properties.  
We observe a different behaviour, with 88\% of pixels within core regions have $\sigma_{NT} / a_{0} \geq 1$. This is consistent with our systematic identification of 73\% of cores exhibiting a peak in velocity width. However, 87\% of pixels within the surrounding filament also have $\sigma_{NT} / a_{0} \geq 1$.  Despite this similarity in the core and filament pixel distributions, conducting a Kolmogorov-Smirnov test of the two show vanishingly small p-values with moderate coefficient values (between 0.2 and 0.3), indicating they are significantly different.  In the JVLA data however, 55\% of pixels within core regions peak supersonically compared to only 36\% of filament pixels.  This further demonstrates that with the addition of the extended background emission in the combined data, the kinematics of the cores have become diluted and similar to that of the surrounding filament. Probing the densest gas with the JVLA--only shows that the kinematics of the cores are more distinct from the surrounding filament.

A decrease of velocity dispersion towards low-mass prestellar cores has been previously observed  \citep{fuller92, goodman98, caselli02, pineda10, pineda15}. More recently, this transition to coherence has also been observed towards entire, parsec-long filaments \citep{hacar11,hacar16}. Transonic cores and filaments are expected to form in a turbulent ISM where supersonic shocks generate stagnation regions where turbulent energy has been dissipated \citep{padoan01, klessen05, federrath16}. In SDC13, with mostly transonic velocity dispersion (where both the combined and JVLA data sets peak in Figure~\ref{fig:mach}, hence best representing the dense gas of the system), everything indicates that the filaments represent such post-shock regions, the increase of the dispersion in some localised region at the edge of the filament being reminiscent of what \cite{klessen05} sees in their turbulent simulation of core formation. In this context, the velocity dispersion increase towards the core would be then purely generated by gravity \citep[e.g.][submitted]{ballesteros-paredes17}.

Another possible explanation for a local increase of velocity dispersion towards the cores could be the presence of embedded protostellar sources. Even though we do not see any mid-infrared sources toward them, such protostars could be embedded enough to remain undetectable with both {\it Spitzer} and {\it Herschel}. These protostars could have associated outflows which disrupt the surrounding gas, contributing to a local increase in the velocity dispersion \citep{duarte-cabral12}. ALMA observations of SDC13 cores at sub-arcsecond resolution could settle this issue.

\subsection{Evolution of the virial ratio from large to small scale} \label{sec:virial}

\begin{figure}[!t]
\centering
\subfloat{%
  \includegraphics[width=0.45\textwidth,center]{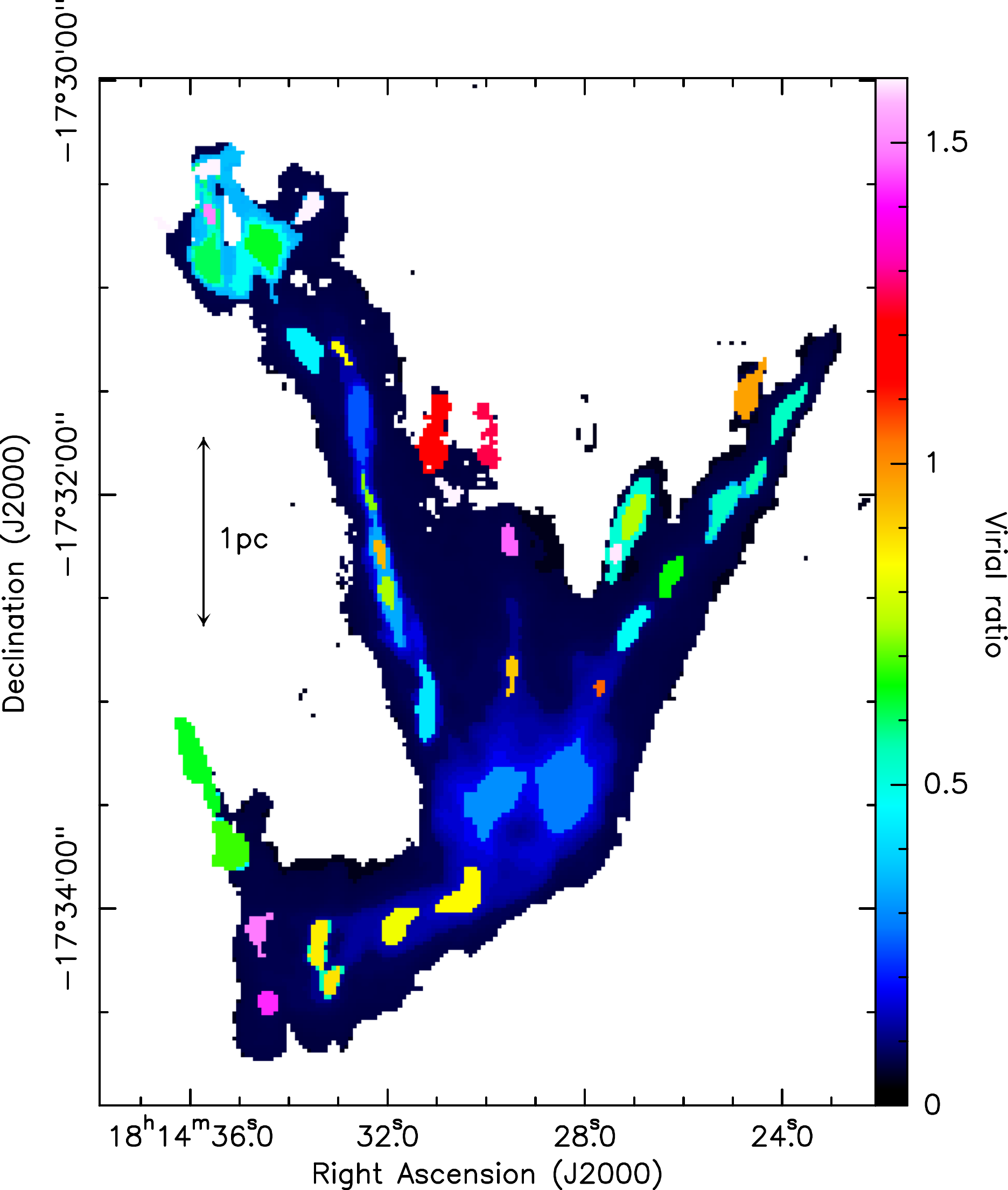}%
  }\par
  \label{fig:virial}
\caption{Virial ratio map of all identified dendrogram structures.
}
\label{fig:treevirial}
\end{figure}

Using the dendrogram tree, we can estimate the virial ratio for all extracted structures. Following \cite{bertoldimckee92}, the virial ratio of a uniform density sphere is given by:
    \begin{equation}
        \alpha_{vir} = \frac{5a_{eff}^2 R}{\mathrm{G}M} \, ,
        \label{eq:vir}
    \end{equation}
\noindent where $R$ is the radius and $M$ is the mass of the cores found by the extraction code respectively, and $a_{eff}$ is the total velocity dispersion considering both thermal and non-thermal contributions from all molecules. We use the upper limit on the mass as it matches the amount of gas responsible for the observed velocity dispersion. In the absence of significant magnetic energy density, structures with  $\alpha_{vir} < 2$ are thought to be gravitationally bound, those with  $\alpha_{vir} \sim1$ are compatible with hydrostatic equilibrium, those with $\alpha_{vir} < 1$ are likely to be gravitationally unstable, while structures with  $\alpha_{vir} > 2$ are unbound and either dispersing or held together by external pressure. Figure \ref{fig:treevirial} shows the map of virial ratios for all SDC13 structures. Based on this map, it is clear that as we go from large to small scale, the virial ratio increases from $\alpha_{vir} \sim 0.1$ towards 1. Such low-values suggests that SDC13 is gravitationally unstable on all scales, and that the transfer of gravitational to kinetic energy could be the process responsible for the increase of the virial ratios on small scales.

\begin{figure}[!t]
\centering
\subfloat{%
  \includegraphics[width=0.5\textwidth,right]{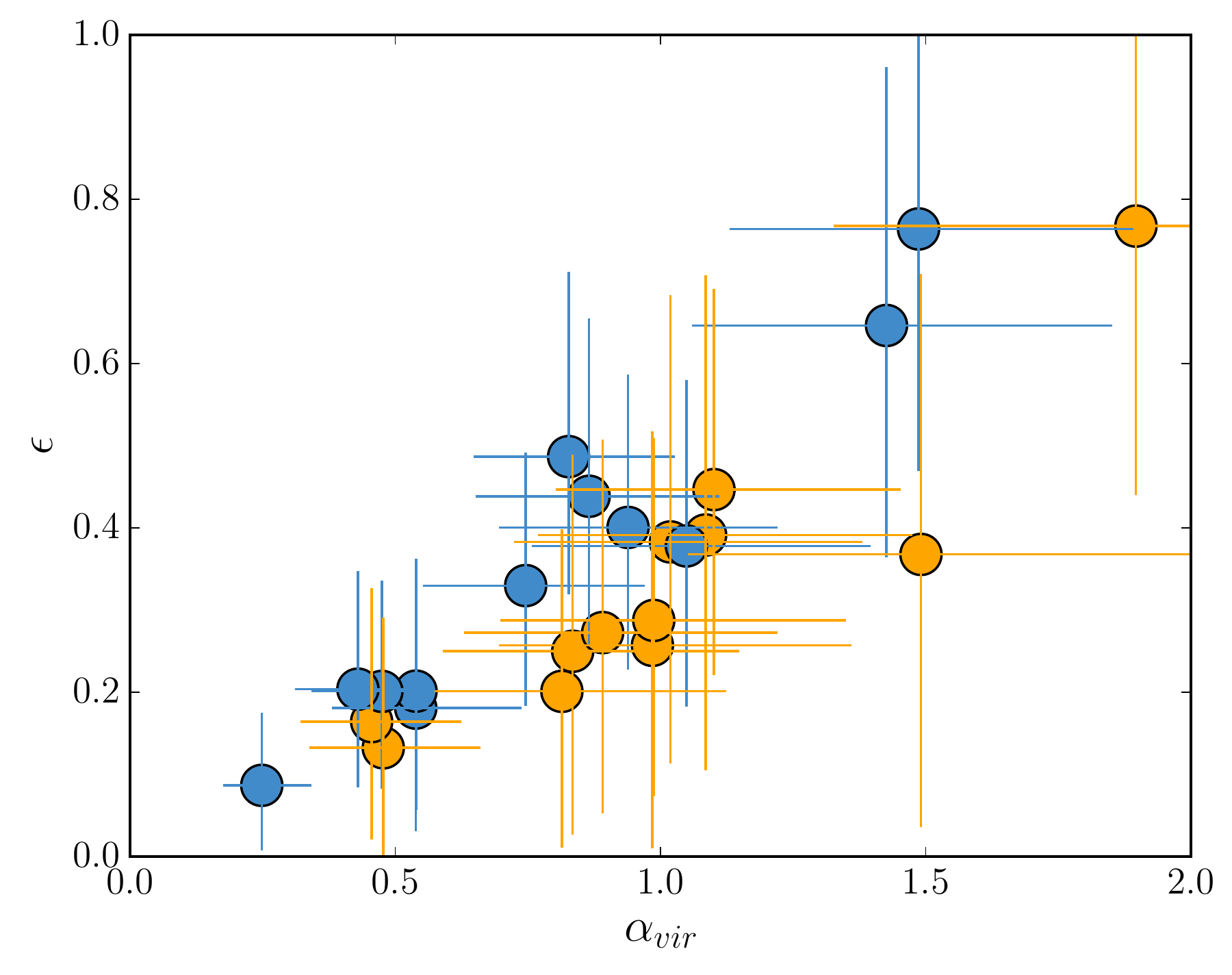} 
  }\par
  \label{fig:virial}
\caption{Observed virial ratio plotted against the energy conversion efficiency of gravitational energy into kinetic energy for the combined JVLA/GBT data (in blue), and the JVLA--only data (in orange). Only starless along the filaments are plotted, as protostellar or hub centre cores are subject to extra energy source that will effect the estimate of $\epsilon$.  Errors were propagated using a Monte Carlo error propagation method.}
\label{fig:epsilon}
\end{figure}

Such an evolution of the viral ratio can be analytically described by the following equation:

    \begin{equation}
        \alpha_{vir}=2\frac{E_{k,0}}{|E_g|}+2\epsilon\left(1-\frac{|E_{g,0}|}{|E_g|}\right)
        \label{eq:vir_1}
    \end{equation}
    
\noindent where $E_k$ and $E_g$ are the kinetic and gravitational energies, respectively, the $0$ index indicates the initial value, i.e. at the start of the collapse/fragmentation, and $\epsilon$ is the fraction of the gravitational energy which is converted into kinetic energy. Note that as $|E_g|$ becomes increasingly larger during the collapse, $\epsilon=1$ translates into $\alpha_{vir} \rightarrow 2$, which is expected if all the gravitational energy were to be converted into kinetic energy.  A mean $a_{eff,0}\sim0.26\pm0.02$\,km\,s$^{-1}$ is evaluated from Figure~\ref{fig:mach} (as discussed in Section~\ref{sec:supercritical}).  By estimating each energy term of equation \ref{eq:vir_1} for every core showing an increase of velocity dispersion one can determine which value of $\epsilon$ is required to reproduce the observed $\alpha_{vir}$ value. Assuming that the cores of initial diameter $\lambda_{core}$ result from the fragmentation of a uniform density transonic filament of initial velocity dispersion $a_{eff,0}$, one can rewrite equation \ref{eq:vir_1} as:

    \begin{equation}
        \alpha_{vir}=\frac{3}{\beta G}\frac{Ra_{eff,0}^2}{M}+2\epsilon\left(1-\frac{2\beta_{0}}{\beta}\frac{R}{\lambda_{core}}\right)
        \label{eq:vir_2}
    \end{equation}

\noindent where $\beta=\frac{3-k_{\rho}}{5-2k_{\rho}}$ is a factor depending on the power law index $k_{\rho}$ of the density profile of the collapsing core, and $\beta_{0}$ is its initial value before the onset of collapse (see Appendix~\ref{app:virial} for a more detailed derivation). For simplicity, we assumed an initially flat density profile (i.e. $\beta_{0}=3/5$), and $k_{\rho} = 2$ once collapse begins (i.e. $\beta = 1$).  Figure~\ref{fig:epsilon} plots the energy conversion efficiency $\epsilon$ against the virial ratio for every core for both the combined JVLA/GBT  data and JVLA--only data. On this plot, only starless cores are included, as protostellar cores, or cores at filament junctions are subject to extra energy sources that will affect the estimate of $\epsilon$.  One can see that a conversion efficiency between 20--50\% can explain most of the observed velocity dispersion increases towards starless cores. The JVLA--only data points (orange circles), believed to be less contaminated by foreground/background emission to the cores, are clustered around a median efficiency of $35\%$.   


\cite{klessen10} proposed a theory according to which accretion is capable of driving turbulent motions in galaxies on all scales, finding an $\epsilon$ value of a few percent when considering both galaxy and molecular cloud sized structures. 
\cite{clarke17} evaluate the efficiency required to reproduce a given velocity dispersion due to accretion driven turbulence in filament structures.  They find an $\epsilon = 5 - 10$\,\% is sufficient to drive a $\sigma_{\mathrm{1D}}\sim$\,0.23\,km\,s$^{-1}$, a consistent efficiency to that found by \cite{heitsch13}.  While these are theoretical estimates, our observationally derived value of $\sim40\%$ is somewhat larger. This could point towards an even more important role of gravity-driven turbulence than previously thought \citep[see also][submitted]{traficante18, ballesteros-paredes17}.




\subsection{Origin of the radial velocity gradients} \label{sec:radvel}

Two of the four SDC13 filaments exhibit strong radial velocity gradients, the largest of which in the combined data set reaches 1.5\,km\,s$^{-1}$\,pc$^{-1}$, an order of magnitude larger than in the remaining two filaments.  In the JVLA--only data, although the morphology of the centroid line-of-sight velocity is the same as in the combined data, the magnitude of the radial velocity gradients double, approaching $\sim$3.0\,km\,s$^{-1}$\,pc$^{-1}$ in both the North-East and South-East filaments. The physical origin of such gradients is unclear as they could be the result of accretion/compression, rotation, shear, or a combination of these. It is important to realise that, given the measured velocity gradients (up to 3.0\,km\,s$^{-1}$\,pc$^{-1}$), the crossing time of the filaments is $\sim0.3-0.4$\,Myr, which is a factor of $\sim3$ lower than the estimated age of the filaments. This implies that, whatever the origin of the gradient is, it cannot be of disruptive nature or the filaments would already have dispersed. This therefore excludes shear motions as a possible origin. Here, we are going to investigate if the magnitude of the gradients are compatible with gravity and/or rotation.

For gravity to be responsible for radial velocity gradients in filaments, large-scale accretion into a plane is the only option as axisymmetric accretion would not produce any velocity gradients. Assuming that the filaments are infinitely long,  then one can estimate the acceleration of a piece of gas at a radius $r$ \citep{palmeirim13}, as:

   \begin{equation}
        \rm{v}_r= 2 \sqrt{G M_{line} \ln\left ( \frac{r_{initial}}{r} \right )} \, .
        \label{eq:vff}
    \end{equation}
    
 Substituting for the mean observed $M_{\mathrm{line}}$ of SDC13 we may calculate the $r_{initial}$ that would produce the observed velocity ($\sim 0.3$~km/s) at a distance of 0.20~pc from the spine of the filament. By doing so we obtain $r_{initial}\simeq0.21$\,pc. The associated dynamical timescale is $3.3\times10^4$~yr, nearly two orders of magnitude shorter than our age estimates for the filaments. This suggests that gravity alone is not the origin of the observed radial velocity gradients in SDC13.  As proposed by \cite{arz13}, turbulence generated by the accretion of matter onto filaments can provide an additional kinetic pressure that can eventually counteract the pull of gravity, and slow down the filament radial contraction, even though such a process might not be efficient enough \citep{seifried15}. Magnetic fields can also contribute to slow down radial contraction, either when they run parallel to the main axis of filaments \citep{seifried15}, or when in a helical configuration \citep{fiege00}. However, B fields are often observed to be perpendicular to the main axis of the star-forming filaments \citep[e.g.][]{palmeirim13, cox16}, therefore leaving the radial contraction unaffected.

We now investigate whether rotation could explain these observed gradients.  \cite{recchi14} simulate the rotation about the main axis of filaments in equilibrium.  They predict the radial velocity gradient required for rotation to stabilize filaments with varying M$_{\mathrm{line}}$ and temperature profiles against gravitational instabilities.  Following their calculations for a mean SDC13 filament mass of 319\,M$_{\odot}$, mean filament length and width of 2.11\,pc and 0.26\,pc respectively (hence central density of 4.8\,$\times10^{-17}$\,kg$/$m$^{3}$), and uniform temperature profile (given we assume a constant temperature everywhere), a radial velocity gradient of 1.42\,km\,s$^{-1}$pc$^{-1}$ is required for rotation to halt the fragmentation of SDC13.  Despite this being larger than the mean radial velocity gradient across the South-East filament (and less than that across the North-East filament) it is clear from the core extraction that there is indeed fragmentation along both filaments.  Hence, we conclude that the observed radial velocity gradient cannot be solely due to rotation as the fragmentation was not halted.  The model also only considers equilibrium filaments, which  seems to be an unrealistic assumption to make for such a dynamic system as SDC13. 


Therefore, we believe that the most likely origin for the observed radial velocity gradients is compression. We speculate that an external site of active star formation, clearly seen in Figure~\ref{fig:tri} within 1\,pc proximity to the North-East filament on the sky (centred on 18:14:33.5 -17:33:30.0) may be compressing the North-East and South-East filaments of SDC13, influencing the radial velocity field and their density structures.  The WISE catalog of H{\sc ii} region sources \citep{anderson14} reveals a H{\sc ii} region they term as being radio-quiet (called G013.170-00.097) coinciding with this external star formation site, however there is no kinematic information available.  Further to this, large scale $^{13}$CO(1--0) and C$^{18}$O(1--0) emission from the IRAM 30\,m telescope seem to show a cavity in emission corresponding to this region (Williams et al. in preparation), consistent with the clearing of material due to feedback effects.  We find no maser emission such as water \citep{walsh08, purcell12} nor methanol \citep{breen12} around the position of the H{\sc ii} region.  Other works have shown the influence of external feedback on the density structure of star-forming filaments \citep{peretto12,tremblin14}.

\subsection{Origin of the longitudinal velocity gradients}

As mentioned in Section~\ref{sec:filalong}, the SDC13 filaments exhibit a complex oscillating longitudinal velocity profile which is at least partially correlated to their column density structure. In their simulations of turbulent cloud evolution, \cite{smith16} find very similar looking filament velocity profiles (see their Figure~6). However, they argue that they do not see any clear correlation between the velocity fluctuations and the location of density peaks, concluding that the velocity fluctuations are the result of transient motions generated by the turbulent field in which the filaments are immerged. While we might be partly witnessing such effect in some of the SDC13 filaments, we do observe that $\sim63\%$ of the SDC13 cores are located at a peak of velocity gradient as expected for cores accreting from their parent filaments.

\begin{figure}[!t]
\centering
\includegraphics[scale=.45,center]{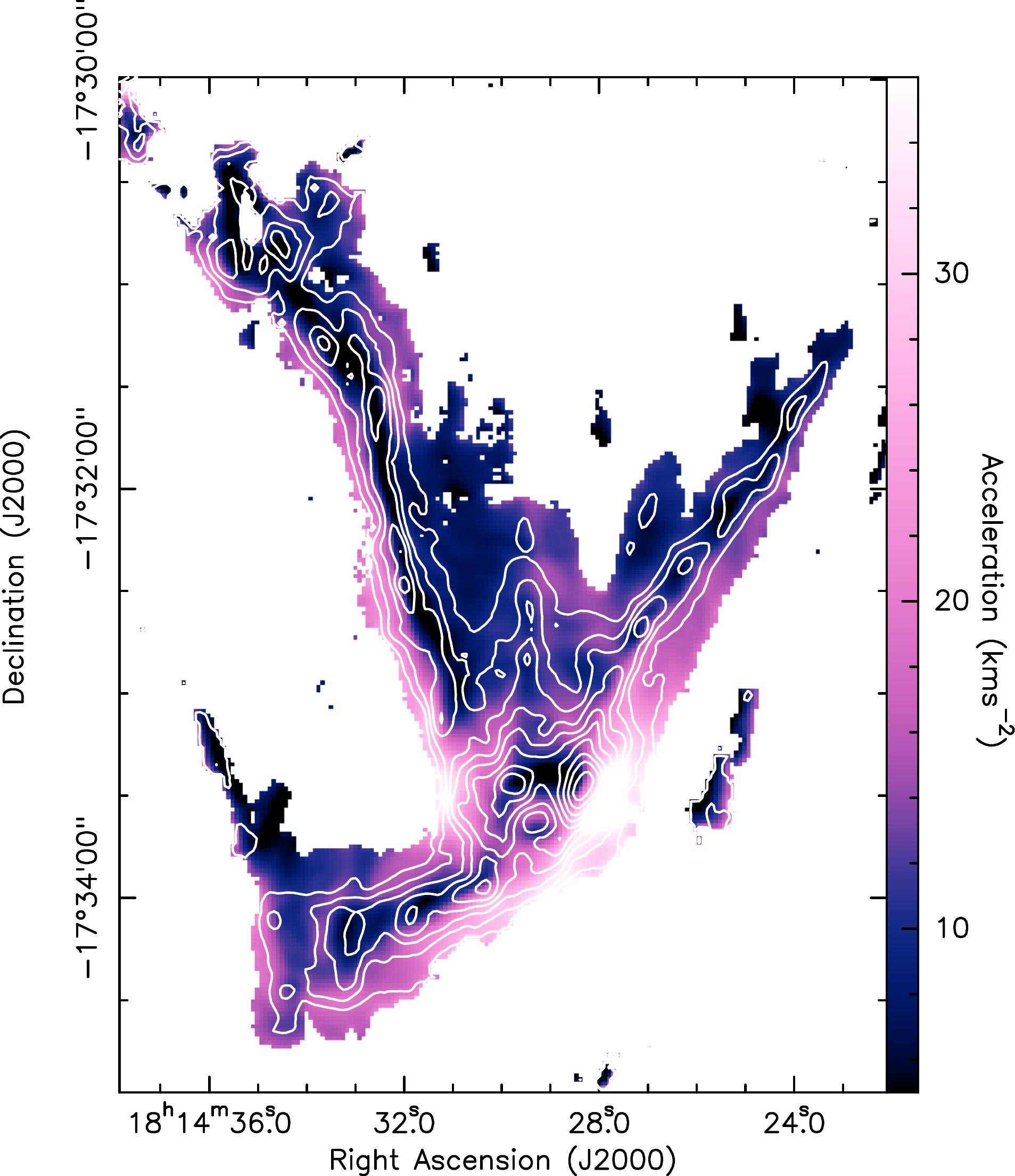}
\caption{Map of gravitational acceleration in SDC13 in units of 1\,$\times$\,10$^{-12}$\,km\,s$^{-2}$. Mass was calculated from the H$_{2}$ column density map in the middle panel of Figure \ref{fig:n}. Contours are of the H$_{2}$ column density in Figure~\ref{fig:n}, plotted in 1\,$\times10^{22}$\,cm$^{-2}$ steps, from 2\,$\times10^{22}$\,cm$^{-2}$ to 12\,$\times10^{22}$\,cm$^{-2}$.}
\label{fig:accel}
\end{figure}

\begin{figure*}[!t]
\centering
\includegraphics[trim={3cm 0.5cm 4cm 1cm},clip,scale=.53,center]{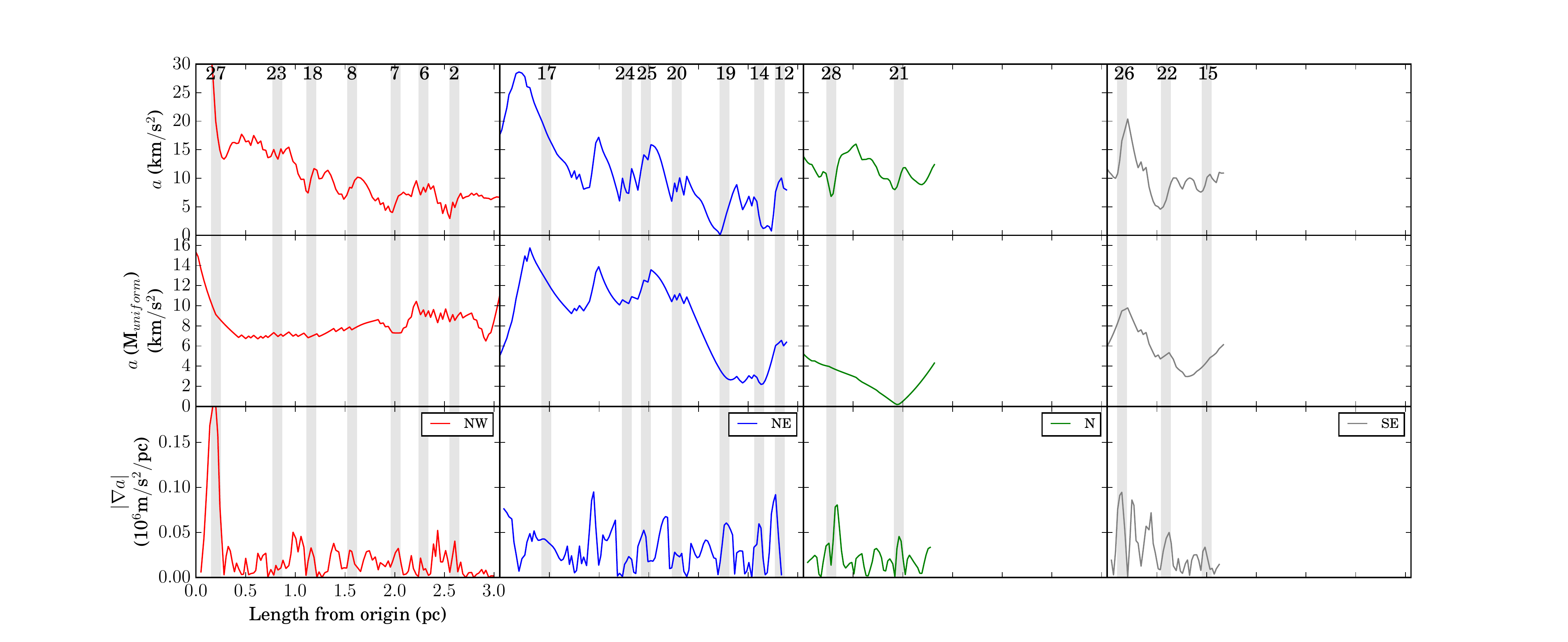} 
\caption{Profiles along the spines, where each column denotes a different filamentary arm, as labelled (North-West in red, North-East in blue, North in green and South-East in grey). The origin of each of the spines was defined to begin at the heart of the hub. The first row plots the acceleration due to gravity in SDC13 where the mass was calculated using the H$_{2}$ column density map, while a uniform, mean cloud mass was used for the acceleration in the second row.  The third row plots the absolute gradient of the acceleration from row three evaluated at each position over the mean core size of $\sim$0.1\,pc. The vertical shaded regions correspond to the positions of the cores along each of the filaments, are $\sim$0.1\,pc wide and overplotted by the core ID number in Table~\ref{tab:cores}.}
\label{fig:gaussprofiles}
\end{figure*}

One specificity of SDC13 is its hub morphology, suggestive of converging gravity-driven flows towards the filament junction. In order to evaluate if the hub morphology has an impact on the density and velocity profiles of SDC13, for each position we computed the amplitude of the acceleration following:

 \begin{equation}
        |\vec{a_j}|= \left|-\sum_{i\neq j}\frac{Gm_i}{r_{i,j}^3}\vec{r_{i,j}}\right| 
        \label{eq:acc}
  \end{equation}

\noindent where indices $i$ and $j$ correspond to given pixels in our SDC13 column density map, $m_i$ is the gas mass at pixel $i$, and $r_{i,j}$ is the distance between pixels $i$ and $j$. Observationally, one only has access to the 2D projected quantities, and acceleration estimated that way can only provide trend indications and order of magnitude estimates. Figure \ref{fig:accel} shows the acceleration map obtained from the SDC13 H$_2$ column density map.  As expected, the accelerations along the filament spine are at a minimum due to the axisymmetric nature of gravity in such systems. The acceleration appears to peak around the filament hub junctions and the edges of some filaments, reminiscent of the behaviour seen in the velocity dispersion map in Figure~\ref{fig:maps}.  More interestingly, we have constructed acceleration profiles along each filament. These are shown in the first row of Figure~\ref{fig:gaussprofiles}. We observe oscillations of the acceleration and an increase of the acceleration towards the hub centre which are reminiscent of the velocity structure. In order to check if this behaviour is due to the presence of cores or to the hub morphology, we constructed another acceleration map with uniform column density, effectively removing the presence of density fluctuations.  The second row of Figure~\ref{fig:gaussprofiles} shows such profiles. One can see that, for the two main filaments (North-East and North-West), the acceleration profiles remain mostly unchanged, demonstrating that the hub morphology is the main driver of the acceleration fluctuation we observed, and therefore the driver of the SDC13 kinematics as a whole.  Another interesting point is given by the last row of Figure~\ref{fig:gaussprofiles} where the gradients of the acceleration is provided. We notice here that the strongest acceleration gradient peak is associated with the most massive core within SDC13. Gradients of acceleration represent accumulation points and are therefore privileged locations for the formation of massive cores. The systematic presence of massive cores near the hub centre could therefore be a direct consequence of large acceleration gradients generated by the hub morphology.

\section{Summary and conclusions} \label{sec:conc}

We have conducted a study of the SDC13 infrared dark hub filament system using the rotation inversion transition of NH$_{3}$(1,1) and NH$_{3}$(2,2). The fitting of the hyperfine structure of these lines allowed the construction of centroid velocity, velocity dispersion, and H$_2$ column density maps. These maps were used to identify both core structures and filament spines, allowing us to characterise their kinematics. We find that all filaments are supercritical, with an average transonic non-thermal velocity dispersion, and that core spacing is typically regular along them ($\sim 0.37\pm0.16$\,pc). Using semi-analytical models of non-equilibrium filaments \citep{clarke16,clarke17} we determine that the filaments are a few Myrs old, consistent with the SDC13 dynamical timescale derived by \cite{peretto14}.  We find that the large radial velocity gradients across two of the four filament cannot be due to gravity nor rotation, but most likely due to the compression caused by a nearby H{\sc ii} region. We also find that the velocity dispersion increases towards 73\% of the identified cores, which we interpret as the result of the accumulation of matter due to the filament fragmentation process itself. We derive an average gravitational to kinetic energy conversion efficiency within these cores of $\epsilon\simeq35\%$, larger than theoretical values published in the literature.  Finally, we propose that the presence of massive cores near the filament junction is due to the large acceleration gradients produced by the hub morphology.

All elements put together, we propose a scenario for the evolution of the SDC13 hub-filament system in which filaments first form as post-shock structures in a supersonic turbulent flow. As a result of the turbulent energy dissipation in the shock, the dense gas within the filaments is initially mostly sub-sonic. Then gravity takes over and starts shaping the evolution of the hub, both fragmenting filaments and pulling the gas towards the centre of the gravitational well. By doing so, gravitational energy is converted into kinetic energy in both local (cores) and global (hub centre) potential well minima, generating more massive cores at the hub centre as a result of larger acceleration gradients.

\begin{acknowledgements}
      GMW gratefully acknowledges the support of an STFC postgraduate studentship. NP wishes to acknowledge support from STFC under grant number ST/N000706/1 and  ST/M000893/1.  We thank the anonymous referee whose report helped improve the quality of this paper.  This research made use of APLpy, an open-source plotting package for Python hosted at http://aplpy.github.com.
\end{acknowledgements}


\bibliographystyle{aa}
\bibliography{sdc13_jvla_gbt}

\begin{thebibliography}{97}
\expandafter\ifx\csname natexlab\endcsname\relax\def\natexlab#1{#1}\fi

\bibitem[{{Anderson} {et~al.}(2014){Anderson}, {Bania}, {Balser}, {Cunningham},
  {Wenger}, {Johnstone}, \& {Armentrout}}]{anderson14}
{Anderson}, L.~D., {Bania}, T.~M., {Balser}, D.~S., {et~al.} 2014, ApJS, 212, 1

\bibitem[{{Andr{\'e}} {et~al.}(2010){Andr{\'e}}, {Men'shchikov}, {Bontemps},
  {K{\"o}nyves}, {Motte}, {Schneider}, {Didelon}, {Minier}, {Saraceno},
  {Ward-Thompson}, {di Francesco}, {White}, {Molinari}, {Testi}, {Abergel},
  {Griffin}, {Henning}, {Royer}, {Mer{\'{\i}}n}, {Vavrek}, {Attard},
  {Arzoumanian}, {Wilson}, {Ade}, {Aussel}, {Baluteau}, {Benedettini},
  {Bernard}, {Blommaert}, {Cambr{\'e}sy}, {Cox}, {di Giorgio}, {Hargrave},
  {Hennemann}, {Huang}, {Kirk}, {Krause}, {Launhardt}, {Leeks}, {Le Pennec},
  {Li}, {Martin}, {Maury}, {Olofsson}, {Omont}, {Peretto}, {Pezzuto}, {Prusti},
  {Roussel}, {Russeil}, {Sauvage}, {Sibthorpe}, {Sicilia-Aguilar}, {Spinoglio},
  {Waelkens}, {Woodcraft}, \& {Zavagno}}]{andre10}
{Andr{\'e}}, P., {Men'shchikov}, A., {Bontemps}, S., {et~al.} 2010, \aap, 518,
  L102

\bibitem[{{Arzoumanian} {et~al.}(2011){Arzoumanian}, {Andr{\'e}}, {Didelon},
  {K{\"o}nyves}, {Schneider}, {Men'shchikov}, {Sousbie}, {Zavagno}, {Bontemps},
  {di Francesco}, {Griffin}, {Hennemann}, {Hill}, {Kirk}, {Martin}, {Minier},
  {Molinari}, {Motte}, {Peretto}, {Pezzuto}, {Spinoglio}, {Ward-Thompson},
  {White}, \& {Wilson}}]{arz11}
{Arzoumanian}, D., {Andr{\'e}}, P., {Didelon}, P., {et~al.} 2011, \aap, 529, L6

\bibitem[{{Arzoumanian} {et~al.}(2013){Arzoumanian}, {Andr{\'e}}, {Peretto}, \&
  {K{\"o}nyves}}]{arz13}
{Arzoumanian}, D., {Andr{\'e}}, P., {Peretto}, N., \& {K{\"o}nyves}, V. 2013,
  \aap, 553, A119

\bibitem[{{Ballesteros-Paredes} {et~al.}(2017){Ballesteros-Paredes},
  {V{\'a}zquez-Semadeni}, {Palau}, \& {Klessen}}]{ballesteros-paredes17}
{Ballesteros-Paredes}, J., {V{\'a}zquez-Semadeni}, E., {Palau}, A., \&
  {Klessen}, R.~S. 2017, arXiv:1710.07384

\bibitem[{{Barranco} \& {Goodman}(1998)}]{barranco98}
{Barranco}, J.~A. \& {Goodman}, A.~A. 1998, ApJ, 504, 207

\bibitem[{{Barrett} {et~al.}(1977){Barrett}, {Ho}, \& {Myers}}]{barrett77}
{Barrett}, A.~H., {Ho}, P.~T.~P., \& {Myers}, P.~C. 1977, ApJL, 211, L39

\bibitem[{{Bergin} \& {Langer}(1997)}]{bergin97}
{Bergin}, E.~A. \& {Langer}, W.~D. 1997, \apj, 486, 316

\bibitem[{{Bertoldi} \& {McKee}(1992)}]{bertoldimckee92}
{Bertoldi}, F. \& {McKee}, C.~F. 1992, \apj, 395, 140

\bibitem[{{Beuther} {et~al.}(2015){Beuther}, {Ragan}, {Johnston}, {Henning},
  {Hacar}, \& {Kainulainen}}]{beuther15}
{Beuther}, H., {Ragan}, S.~E., {Johnston}, K., {et~al.} 2015, \aap, 584, A67

\bibitem[{{Beuther} \& {Steinacker}(2007)}]{beuther07}
{Beuther}, H. \& {Steinacker}, J. 2007, ApJL, 656, L85

\bibitem[{{Breen} {et~al.}(2012){Breen}, {Ellingsen}, {Caswell}, {Green},
  {Voronkov}, {Fuller}, {Quinn}, \& {Avison}}]{breen12}
{Breen}, S.~L., {Ellingsen}, S.~P., {Caswell}, J.~L., {et~al.} 2012, MNRAS,
  426, 2189

\bibitem[{{Butler} \& {Tan}(2009)}]{butler09}
{Butler}, M.~J. \& {Tan}, J.~C. 2009, ApJ, 696, 484

\bibitem[{{Carey} {et~al.}(2009){Carey}, {Noriega-Crespo}, {Mizuno}, {Shenoy},
  {Paladini}, {Kraemer}, {Price}, {Flagey}, {Ryan}, {Ingalls}, {Kuchar},
  {Pinheiro Gon{\c c}alves}, {Indebetouw}, {Billot}, {Marleau}, {Padgett},
  {Rebull}, {Bressert}, {Ali}, {Molinari}, {Martin}, {Berriman}, {Boulanger},
  {Latter}, {Miville-Deschenes}, {Shipman}, \& {Testi}}]{carey09}
{Carey}, S.~J., {Noriega-Crespo}, A., {Mizuno}, D.~R., {et~al.} 2009, PASP,
  121, 76

\bibitem[{{Caselli} {et~al.}(2002){Caselli}, {Benson}, {Myers}, \&
  {Tafalla}}]{caselli02}
{Caselli}, P., {Benson}, P.~J., {Myers}, P.~C., \& {Tafalla}, M. 2002, ApJ,
  572, 238

\bibitem[{Cheung {et~al.}(1968)Cheung, Rank, Townes, Thornton, \&
  Welch}]{cheung68}
Cheung, A.~C., Rank, D.~M., Townes, C.~H., Thornton, D.~D., \& Welch, W.~J.
  1968, Phys. Rev. Lett., 21, 1701

\bibitem[{{Churchwell} {et~al.}(2009){Churchwell}, {Babler}, {Meade},
  {Whitney}, {Benjamin}, {Indebetouw}, {Cyganowski}, {Robitaille}, {Povich},
  {Watson}, \& {Bracker}}]{churchwell09}
{Churchwell}, E., {Babler}, B.~L., {Meade}, M.~R., {et~al.} 2009, PASP, 121,
  213

\bibitem[{{Clarke} \& {Whitworth}(2015)}]{clarke15}
{Clarke}, S.~D. \& {Whitworth}, A.~P. 2015, MNRAS, 449, 1819

\bibitem[{{Clarke} {et~al.}(2017){Clarke}, {Whitworth}, {Duarte-Cabral}, \&
  {Hubber}}]{clarke17}
{Clarke}, S.~D., {Whitworth}, A.~P., {Duarte-Cabral}, A., \& {Hubber}, D.~A.
  2017, MNRAS, 468, 2489

\bibitem[{{Clarke} {et~al.}(2016){Clarke}, {Whitworth}, \& {Hubber}}]{clarke16}
{Clarke}, S.~D., {Whitworth}, A.~P., \& {Hubber}, D.~A. 2016, MNRAS, 458, 319

\bibitem[{{Cox} {et~al.}(2016){Cox}, {Arzoumanian}, {Andr{\'e}}, {Rygl},
  {Prusti}, {Men'shchikov}, {Royer}, {K{\'o}sp{\'a}l}, {Palmeirim}, {Ribas},
  {K{\"o}nyves}, {Bernard}, {Schneider}, {Bontemps}, {Merin}, {Vavrek}, {Alves
  de Oliveira}, {Didelon}, {Pilbratt}, \& {Waelkens}}]{cox16}
{Cox}, N.~L.~J., {Arzoumanian}, D., {Andr{\'e}}, P., {et~al.} 2016, \aap, 590,
  A110

\bibitem[{{Dirienzo} {et~al.}(2015){Dirienzo}, {Brogan}, {Indebetouw},
  {Chandler}, {Friesen}, \& {Devine}}]{dirienzo15}
{Dirienzo}, W.~J., {Brogan}, C., {Indebetouw}, R., {et~al.} 2015, ApJ, 150, 159

\bibitem[{{Duarte-Cabral} {et~al.}(2012){Duarte-Cabral}, {Chrysostomou},
  {Peretto}, {Fuller}, {Matthews}, {Schieven}, \& {Davis}}]{duarte-cabral12}
{Duarte-Cabral}, A., {Chrysostomou}, A., {Peretto}, N., {et~al.} 2012, \aap,
  543, A140

\bibitem[{{Egan} {et~al.}(1998){Egan}, {Shipman}, {Price}, {Carey}, {Clark}, \&
  {Cohen}}]{egan98}
{Egan}, M.~P., {Shipman}, R.~F., {Price}, S.~D., {et~al.} 1998, ApJL, 494, L199

\bibitem[{{Federrath} {et~al.}(2016){Federrath}, {Rathborne}, {Longmore},
  {Kruijssen}, {Bally}, {Contreras}, {Crocker}, {Garay}, {Jackson}, {Testi}, \&
  {Walsh}}]{federrath16}
{Federrath}, C., {Rathborne}, J.~M., {Longmore}, S.~N., {et~al.} 2016, ApJ,
  832, 143

\bibitem[{{Fern{\'a}ndez-L{\'o}pez} {et~al.}(2014){Fern{\'a}ndez-L{\'o}pez},
  {Arce}, {Looney}, {Mundy}, {Storm}, {Teuben}, {Lee}, {Segura-Cox}, {Isella},
  {Tobin}, {Rosolowsky}, {Plunkett}, {Kwon}, {Kauffmann}, {Ostriker}, {Tassis},
  {Shirley}, \& {Pound}}]{fernandez-lopez14}
{Fern{\'a}ndez-L{\'o}pez}, M., {Arce}, H.~G., {Looney}, L., {et~al.} 2014,
  \apjl, 790, L19

\bibitem[{{Fiege} \& {Pudritz}(2000)}]{fiege00}
{Fiege}, J.~D. \& {Pudritz}, R.~E. 2000, \mnras, 311, 85

\bibitem[{{Fuller} \& {Myers}(1992)}]{fuller92}
{Fuller}, G.~A. \& {Myers}, P.~C. 1992, \apj, 384, 523

\bibitem[{{Girichidis} {et~al.}(2011){Girichidis}, {Federrath}, {Banerjee}, \&
  {Klessen}}]{girichidis11}
{Girichidis}, P., {Federrath}, C., {Banerjee}, R., \& {Klessen}, R.~S. 2011,
  MNRAS, 413, 2741

\bibitem[{{Goldsmith} \& {Langer}(1999)}]{goldsmithlanger99}
{Goldsmith}, P.~F. \& {Langer}, W.~D. 1999, ApJ, 517, 209

\bibitem[{{Goodman} {et~al.}(1998){Goodman}, {Barranco}, {Wilner}, \&
  {Heyer}}]{goodman98}
{Goodman}, A.~A., {Barranco}, J.~A., {Wilner}, D.~J., \& {Heyer}, M.~H. 1998,
  \apj, 504, 223

\bibitem[{{Gunther-Mohr}(1954)}]{gunther}
{Gunther-Mohr}, G.~R. 1954, PhD thesis, COLUMBIA UNIVERSITY.

\bibitem[{{Hacar} {et~al.}(2016){Hacar}, {Kainulainen}, {Tafalla}, {Beuther},
  \& {Alves}}]{hacar16}
{Hacar}, A., {Kainulainen}, J., {Tafalla}, M., {Beuther}, H., \& {Alves}, J.
  2016, \aap, 587, A97

\bibitem[{{Hacar} \& {Tafalla}(2011)}]{hacar11}
{Hacar}, A. \& {Tafalla}, M. 2011, \aap, 533, A34

\bibitem[{{Harju} {et~al.}(1993){Harju}, {Walmsley}, \& {Wouterloot}}]{harju93}
{Harju}, J., {Walmsley}, C.~M., \& {Wouterloot}, J.~G.~A. 1993, \aaps, 98, 51

\bibitem[{{Heitsch}(2013)}]{heitsch13}
{Heitsch}, F. 2013, ApJ, 769, 115

\bibitem[{{Henshaw} {et~al.}(2016){Henshaw}, {Caselli}, {Fontani},
  {Jim{\'e}nez-Serra}, {Tan}, {Longmore}, {Pineda}, {Parker}, \&
  {Barnes}}]{henshaw16}
{Henshaw}, J.~D., {Caselli}, P., {Fontani}, F., {et~al.} 2016, MNRAS, 463, 146

\bibitem[{{Ho} \& {Townes}(1983)}]{HoTownes83}
{Ho}, P.~T.~P. \& {Townes}, C.~H. 1983, ARAA, 21, 239

\bibitem[{{Inutsuka} \& {Miyama}(1992)}]{inutsuka92}
{Inutsuka}, S.-I. \& {Miyama}, S.~M. 1992, ApJ, 388, 392

\bibitem[{{Kirk} {et~al.}(2013){Kirk}, {Myers}, {Bourke}, {Gutermuth},
  {Hedden}, \& {Wilson}}]{kirk13}
{Kirk}, H., {Myers}, P.~C., {Bourke}, T.~L., {et~al.} 2013, \apj, 766, 115

\bibitem[{{Klessen} {et~al.}(2005){Klessen}, {Ballesteros-Paredes},
  {V{\'a}zquez-Semadeni}, \& {Dur{\'a}n-Rojas}}]{klessen05}
{Klessen}, R.~S., {Ballesteros-Paredes}, J., {V{\'a}zquez-Semadeni}, E., \&
  {Dur{\'a}n-Rojas}, C. 2005, ApJ, 620, 786

\bibitem[{{Klessen} \& {Hennebelle}(2010)}]{klessen10}
{Klessen}, R.~S. \& {Hennebelle}, P. 2010, \aap, 520, A17

\bibitem[{{K{\"o}nyves} {et~al.}(2015){K{\"o}nyves}, {Andr{\'e}},
  {Men'shchikov}, {Palmeirim}, {Arzoumanian}, {Schneider}, {Roy}, {Didelon},
  {Maury}, {Shimajiri}, {Di Francesco}, {Bontemps}, {Peretto}, {Benedettini},
  {Bernard}, {Elia}, {Griffin}, {Hill}, {Kirk}, {Ladjelate}, {Marsh}, {Martin},
  {Motte}, {Nguy{\^e}n Luong}, {Pezzuto}, {Roussel}, {Rygl}, {Sadavoy},
  {Schisano}, {Spinoglio}, {Ward-Thompson}, \& {White}}]{konyves15}
{K{\"o}nyves}, V., {Andr{\'e}}, P., {Men'shchikov}, A., {et~al.} 2015, \aap,
  584, A91

\bibitem[{{Kukolich}(1967)}]{kuko}
{Kukolich}, S.~G. 1967, Physical Review, 156, 83

\bibitem[{{Li} {et~al.}(2003){Li}, {Goldsmith}, \& {Menten}}]{li03}
{Li}, D., {Goldsmith}, P.~F., \& {Menten}, K. 2003, ApJ, 587, 262

\bibitem[{{Liu} {et~al.}(2013){Liu}, {Ho}, {Wright}, {Su}, {Hsieh}, {Sun},
  {Kim}, \& {Minh}}]{liu13}
{Liu}, H.~B., {Ho}, P.~T.~P., {Wright}, M.~C.~H., {et~al.} 2013, ApJ, 770, 44

\bibitem[{Lu {et~al.}(2014)Lu, Zhang, Liu, Wang, \& Gu}]{lu14}
Lu, X., Zhang, Q., Liu, H.~B., Wang, J., \& Gu, Q. 2014, ApJ, 790, 84

\bibitem[{{Maret} {et~al.}(2009){Maret}, {Faure}, {Scifoni}, \&
  {Wiesenfeld}}]{maret09}
{Maret}, S., {Faure}, A., {Scifoni}, E., \& {Wiesenfeld}, L. 2009, MNRAS, 399,
  425

\bibitem[{{Marsh} {et~al.}(2016){Marsh}, {Kirk}, {Andr{\'e}}, {Griffin},
  {K{\"o}nyves}, {Palmeirim}, {Men'shchikov}, {Ward-Thompson}, {Benedettini},
  {Bresnahan}, {Francesco}, {Elia}, {Motte}, {Peretto}, {Pezzuto}, {Roy},
  {Sadavoy}, {Schneider}, {Spinoglio}, \& {White}}]{marsh16}
{Marsh}, K.~A., {Kirk}, J.~M., {Andr{\'e}}, P., {et~al.} 2016, MNRAS, 459, 342

\bibitem[{{Masters} {et~al.}(2011){Masters}, {Garwood}, {Langston}, \&
  {Shelton}}]{masters}
{Masters}, J., {Garwood}, B., {Langston}, G., \& {Shelton}, A. 2011, in
  Astronomical Society of the Pacific Conference Series, Vol. 442, Astronomical
  Data Analysis Software and Systems XX, ed. I.~N. {Evans}, A.~{Accomazzi},
  D.~J. {Mink}, \& A.~H. {Rots}, 127

\bibitem[{{McGuire} {et~al.}(2016){McGuire}, {Fuller}, {Peretto}, {Zhang},
  {Traficante}, {Avison}, \& {Jimenez-Serra}}]{mcguire16}
{McGuire}, C., {Fuller}, G.~A., {Peretto}, N., {et~al.} 2016, \aap, 594, A118

\bibitem[{{McMullin} {et~al.}(1992){McMullin}, {Waters}, {Schiebel}, {Young},
  \& {Golap}}]{mcmullin}
{McMullin}, J.~P., {Waters}, B., {Schiebel}, D., {Young}, W., \& {Golap}, K.
  1992, (ASP Conf. Ser. 376), 127

\bibitem[{{Miyama} {et~al.}(1994){Miyama}, {Nakamoto}, {Kikuchi}, {Inutsuka},
  {Kobayashi}, \& {Takeuchi}}]{miyama94}
{Miyama}, S.~M., {Nakamoto}, T., {Kikuchi}, N., {et~al.} 1994, {The stability
  of circumstellar disks.}, ed. J.~{Franco}, S.~{Lizano}, L.~{Aguilar}, \&
  E.~{Daltabuit}

\bibitem[{{Molinari} {et~al.}(2010){Molinari}, {Swinyard}, {Bally}, {Barlow},
  {Bernard}, {Martin}, {Moore}, {Noriega-Crespo}, {Plume}, {Testi}, {Zavagno},
  {Abergel}, {Ali}, {Andr{\'e}}, {Baluteau}, {Benedettini}, {Bern{\'e}},
  {Billot}, {Blommaert}, {Bontemps}, {Boulanger}, {Brand}, {Brunt}, {Burton},
  {Campeggio}, {Carey}, {Caselli}, {Cesaroni}, {Cernicharo}, {Chakrabarti},
  {Chrysostomou}, {Codella}, {Cohen}, {Compiegne}, {Davis}, {de Bernardis}, {de
  Gasperis}, {Di Francesco}, {di Giorgio}, {Elia}, {Faustini}, {Fischera},
  {Fukui}, {Fuller}, {Ganga}, {Garcia-Lario}, {Giard}, {Giardino}, {Glenn},
  {Goldsmith}, {Griffin}, {Hoare}, {Huang}, {Jiang}, {Joblin}, {Joncas},
  {Juvela}, {Kirk}, {Lagache}, {Li}, {Lim}, {Lord}, {Lucas}, {Maiolo},
  {Marengo}, {Marshall}, {Masi}, {Massi}, {Matsuura}, {Meny}, {Minier},
  {Miville-Desch{\^e}nes}, {Montier}, {Motte}, {M{\"u}ller}, {Natoli}, {Neves},
  {Olmi}, {Paladini}, {Paradis}, {Pestalozzi}, {Pezzuto}, {Piacentini},
  {Pomar{\`e}s}, {Popescu}, {Reach}, {Richer}, {Ristorcelli}, {Roy}, {Royer},
  {Russeil}, {Saraceno}, {Sauvage}, {Schilke}, {Schneider-Bontemps},
  {Schuller}, {Schultz}, {Shepherd}, {Sibthorpe}, {Smith}, {Smith},
  {Spinoglio}, {Stamatellos}, {Strafella}, {Stringfellow}, {Sturm}, {Taylor},
  {Thompson}, {Tuffs}, {Umana}, {Valenziano}, {Vavrek}, {Viti}, {Waelkens},
  {Ward-Thompson}, {White}, {Wyrowski}, {Yorke}, \& {Zhang}}]{molinari10}
{Molinari}, S., {Swinyard}, B., {Bally}, J., {et~al.} 2010, PASP, 122, 314

\bibitem[{{Morgan} {et~al.}(2013){Morgan}, {Moore}, {Allsopp}, \&
  {Eden}}]{morgan13}
{Morgan}, L.~K., {Moore}, T.~J.~T., {Allsopp}, J., \& {Eden}, D.~J. 2013,
  MNRAS, 428, 1160

\bibitem[{{Myers}(1983)}]{myers83}
{Myers}, P.~C. 1983, \apj, 270, 105

\bibitem[{{Myers}(2009)}]{myers09}
{Myers}, P.~C. 2009, ApJ, 700, 1609

\bibitem[{{Ostriker}(1964)}]{ostriker64}
{Ostriker}, J. 1964, ApJ, 140, 1056

\bibitem[{{Padoan} {et~al.}(2001){Padoan}, {Nordlund}, {R{\"o}gnvaldsson}, \&
  {Goodman}}]{padoan01}
{Padoan}, P., {Nordlund}, {\AA}., {R{\"o}gnvaldsson}, {\"O}.~E., \& {Goodman},
  A. 2001, in Astronomical Society of the Pacific Conference Series, Vol. 243,
  From Darkness to Light: Origin and Evolution of Young Stellar Clusters, ed.
  T.~{Montmerle} \& P.~{Andr{\'e}}, 279

\bibitem[{{Palmeirim} {et~al.}(2013){Palmeirim}, {Andr{\'e}}, {Kirk},
  {Ward-Thompson}, {Arzoumanian}, {K{\"o}nyves}, {Didelon}, {Schneider},
  {Benedettini}, {Bontemps}, {Di Francesco}, {Elia}, {Griffin}, {Hennemann},
  {Hill}, {Martin}, {Men'shchikov}, {Molinari}, {Motte}, {Nguyen Luong},
  {Nutter}, {Peretto}, {Pezzuto}, {Roy}, {Rygl}, {Spinoglio}, \&
  {White}}]{palmeirim13}
{Palmeirim}, P., {Andr{\'e}}, P., {Kirk}, J., {et~al.} 2013, \aap, 550, A38

\bibitem[{{Panopoulou} {et~al.}(2017){Panopoulou}, {Psaradaki}, {Skalidis},
  {Tassis}, \& {Andrews}}]{panopoulou17}
{Panopoulou}, G.~V., {Psaradaki}, I., {Skalidis}, R., {Tassis}, K., \&
  {Andrews}, J.~J. 2017, MNRAS, 466, 2529

\bibitem[{{Panopoulou} {et~al.}(2014){Panopoulou}, {Tassis}, {Goldsmith}, \&
  {Heyer}}]{panopoulou14}
{Panopoulou}, G.~V., {Tassis}, K., {Goldsmith}, P.~F., \& {Heyer}, M.~H. 2014,
  MNRAS, 444, 2507

\bibitem[{{Perault} {et~al.}(1996){Perault}, {Omont}, {Simon}, {Seguin},
  {Ojha}, {Blommaert}, {Felli}, {Gilmore}, {Guglielmo}, {Habing}, {Price},
  {Robin}, {de Batz}, {Cesarsky}, {Elbaz}, {Epchtein}, {Fouque}, {Guest},
  {Levine}, {Pollock}, {Prusti}, {Siebenmorgen}, {Testi}, \&
  {Tiphene}}]{perault96}
{Perault}, M., {Omont}, A., {Simon}, G., {et~al.} 1996, \aap, 315, L165

\bibitem[{{Peretto} {et~al.}(2012){Peretto}, {Andr{\'e}}, {K{\"o}nyves},
  {Schneider}, {Arzoumanian}, {Palmeirim}, {Didelon}, {Attard}, {Bernard}, {Di
  Francesco}, {Elia}, {Hennemann}, {Hill}, {Kirk}, {Men'shchikov}, {Motte},
  {Nguyen Luong}, {Roussel}, {Sousbie}, {Testi}, {Ward-Thompson}, {White}, \&
  {Zavagno}}]{peretto12}
{Peretto}, N., {Andr{\'e}}, P., {K{\"o}nyves}, V., {et~al.} 2012, \aap, 541,
  A63

\bibitem[{{Peretto} \& {Fuller}(2009)}]{peretto09}
{Peretto}, N. \& {Fuller}, G.~A. 2009, \aap, 505, 405

\bibitem[{{Peretto} {et~al.}(2014){Peretto}, {Fuller}, {Andr{\'e}},
  {Arzoumanian}, {Rivilla}, {Bardeau}, {Duarte Puertas}, {Guzman Fernandez},
  {Lenfestey}, {Li}, {Olguin}, {R{\"o}ck}, {de Villiers}, \&
  {Williams}}]{peretto14}
{Peretto}, N., {Fuller}, G.~A., {Andr{\'e}}, P., {et~al.} 2014, \aap, 561, A83

\bibitem[{{Peretto} {et~al.}(2013){Peretto}, {Fuller}, {Duarte-Cabral},
  {Avison}, {Hennebelle}, {Pineda}, {Andr{\'e}}, {Bontemps}, {Motte},
  {Schneider}, \& {Molinari}}]{peretto13}
{Peretto}, N., {Fuller}, G.~A., {Duarte-Cabral}, A., {et~al.} 2013, \aap, 555,
  A112

\bibitem[{{Peretto} {et~al.}(2010){Peretto}, {Fuller}, {Plume}, {Anderson},
  {Bally}, {Battersby}, {Beltran}, {Bernard}, {Calzoletti}, {Digiorgio},
  {Faustini}, {Kirk}, {Lenfestey}, {Marshall}, {Martin}, {Molinari}, {Montier},
  {Motte}, {Ristorcelli}, {Rod{\'o}n}, {Smith}, {Traficante}, {Veneziani},
  {Ward-Thompson}, \& {Wilcock}}]{peretto10}
{Peretto}, N., {Fuller}, G.~A., {Plume}, R., {et~al.} 2010, \aap, 518, L98

\bibitem[{{Pillai} {et~al.}(2011){Pillai}, {Kauffmann}, {Wyrowski}, {Hatchell},
  {Gibb}, \& {Thompson}}]{pillai11}
{Pillai}, T., {Kauffmann}, J., {Wyrowski}, F., {et~al.} 2011, \aap, 530, A118

\bibitem[{{Pillai} {et~al.}(2006){Pillai}, {Wyrowski}, {Carey}, \&
  {Menten}}]{pillai06}
{Pillai}, T., {Wyrowski}, F., {Carey}, S.~J., \& {Menten}, K.~M. 2006, \aap,
  450, 569

\bibitem[{{Pineda} {et~al.}(2010){Pineda}, {Goodman}, {Arce}, {Caselli},
  {Foster}, {Myers}, \& {Rosolowsky}}]{pineda10}
{Pineda}, J.~E., {Goodman}, A.~A., {Arce}, H.~G., {et~al.} 2010, ApJL, 712,
  L116

\bibitem[{{Pineda} {et~al.}(2015){Pineda}, {Offner}, {Parker}, {Arce},
  {Goodman}, {Caselli}, {Fuller}, {Bourke}, \& {Corder}}]{pineda15}
{Pineda}, J.~E., {Offner}, S.~S.~R., {Parker}, R.~J., {et~al.} 2015, \nat, 518,
  213

\bibitem[{{Polychroni} {et~al.}(2013){Polychroni}, {Schisano}, {Elia}, {Roy},
  {Molinari}, {Martin}, {Andr{\'e}}, {Turrini}, {Rygl}, {Di Francesco},
  {Benedettini}, {Busquet}, {di Giorgio}, {Pestalozzi}, {Pezzuto},
  {Arzoumanian}, {Bontemps}, {Hennemann}, {Hill}, {K{\"o}nyves},
  {Men'shchikov}, {Motte}, {Nguyen-Luong}, {Peretto}, {Schneider}, \&
  {White}}]{polychroni13}
{Polychroni}, D., {Schisano}, E., {Elia}, D., {et~al.} 2013, ApJL, 777, L33

\bibitem[{{Pon} {et~al.}(2012){Pon}, {Toal{\'a}}, {Johnstone},
  {V{\'a}zquez-Semadeni}, {Heitsch}, \& {G{\'o}mez}}]{pon12}
{Pon}, A., {Toal{\'a}}, J.~A., {Johnstone}, D., {et~al.} 2012, ApJ, 756, 145

\bibitem[{{Purcell} {et~al.}(2012){Purcell}, {Longmore}, {Walsh}, {Whiting},
  {Breen}, {Britton}, {Brooks}, {Burton}, {Cunningham}, {Green},
  {Harvey-Smith}, {Hindson}, {Hoare}, {Indermuehle}, {Jones}, {Lo}, {Lowe},
  {Phillips}, {Thompson}, {Urquhart}, {Voronkov}, \& {White}}]{purcell12}
{Purcell}, C.~R., {Longmore}, S.~N., {Walsh}, A.~J., {et~al.} 2012, MNRAS, 426,
  1972

\bibitem[{{Ragan} {et~al.}(2011){Ragan}, {Bergin}, \& {Wilner}}]{ragan11}
{Ragan}, S.~E., {Bergin}, E.~A., \& {Wilner}, D. 2011, ApJ, 736, 163

\bibitem[{{Rathborne} {et~al.}(2006){Rathborne}, {Jackson}, \&
  {Simon}}]{rathjack06}
{Rathborne}, J.~M., {Jackson}, J.~M., \& {Simon}, R. 2006, ApJ, 641, 389

\bibitem[{{Recchi} {et~al.}(2014){Recchi}, {Hacar}, \& {Palestini}}]{recchi14}
{Recchi}, S., {Hacar}, A., \& {Palestini}, A. 2014, MNRAS, 444, 1775

\bibitem[{{Rosolowsky} {et~al.}(2008){Rosolowsky}, {Pineda}, {Kauffmann}, \&
  {Goodman}}]{rosolowsky08dendro}
{Rosolowsky}, E.~W., {Pineda}, J.~E., {Kauffmann}, J., \& {Goodman}, A.~A.
  2008, ApJ, 679, 1338

\bibitem[{{Salji} {et~al.}(2015){Salji}, {Richer}, {Buckle}, {Francesco},
  {Hatchell}, {Hogerheijde}, {Johnstone}, {Kirk}, {Ward-Thompson}, \& {JCMT GBS
  Consortium}}]{salji15}
{Salji}, C.~J., {Richer}, J.~S., {Buckle}, J.~V., {et~al.} 2015, MNRAS, 449,
  1782

\bibitem[{{Schisano} {et~al.}(2014){Schisano}, {Rygl}, {Molinari}, {Busquet},
  {Elia}, {Pestalozzi}, {Polychroni}, {Billot}, {Carey}, {Paladini},
  {Noriega-Crespo}, {Moore}, {Plume}, {Glover}, \&
  {V{\'a}zquez-Semadeni}}]{schisano14}
{Schisano}, E., {Rygl}, K.~L.~J., {Molinari}, S., {et~al.} 2014, ApJ, 791, 27

\bibitem[{{Schneider} {et~al.}(2012){Schneider}, {Csengeri}, {Hennemann},
  {Motte}, {Didelon}, {Federrath}, {Bontemps}, {Di Francesco}, {Arzoumanian},
  {Minier}, {Andr{\'e}}, {Hill}, {Zavagno}, {Nguyen-Luong}, {Attard},
  {Bernard}, {Elia}, {Fallscheer}, {Griffin}, {Kirk}, {Klessen}, {K{\"o}nyves},
  {Martin}, {Men'shchikov}, {Palmeirim}, {Peretto}, {Pestalozzi}, {Russeil},
  {Sadavoy}, {Sousbie}, {Testi}, {Tremblin}, {Ward-Thompson}, \&
  {White}}]{schneider12}
{Schneider}, N., {Csengeri}, T., {Hennemann}, M., {et~al.} 2012, \aap, 540, L11

\bibitem[{{Schneider} \& {Elmegreen}(1979)}]{schneider_elmegreen79}
{Schneider}, S. \& {Elmegreen}, B.~G. 1979, ApJS, 41, 87

\bibitem[{{Seifried} \& {Walch}(2015)}]{seifried15}
{Seifried}, D. \& {Walch}, S. 2015, \mnras, 452, 2410

\bibitem[{{Shirley}(2015)}]{shirley15}
{Shirley}, Y.~L. 2015, \pasp, 127, 299

\bibitem[{{Simon} {et~al.}(2006){Simon}, {Rathborne}, {Shah}, {Jackson}, \&
  {Chambers}}]{simon06}
{Simon}, R., {Rathborne}, J.~M., {Shah}, R.~Y., {Jackson}, J.~M., \&
  {Chambers}, E.~T. 2006, ApJ, 653, 1325

\bibitem[{{Smith} {et~al.}(2016){Smith}, {Glover}, {Klessen}, \&
  {Fuller}}]{smith16}
{Smith}, R.~J., {Glover}, S.~C.~O., {Klessen}, R.~S., \& {Fuller}, G.~A. 2016,
  MNRAS, 455, 3640

\bibitem[{{Teixeira} {et~al.}(2016){Teixeira}, {Takahashi}, {Zapata}, \&
  {Ho}}]{teixeira16}
{Teixeira}, P.~S., {Takahashi}, S., {Zapata}, L.~A., \& {Ho}, P.~T.~P. 2016,
  \aap, 587, A47

\bibitem[{{Toal{\'a}} {et~al.}(2012){Toal{\'a}}, {V{\'a}zquez-Semadeni}, \&
  {G{\'o}mez}}]{toala12}
{Toal{\'a}}, J.~A., {V{\'a}zquez-Semadeni}, E., \& {G{\'o}mez}, G.~C. 2012,
  ApJ, 744, 190

\bibitem[{{Traficante} {et~al.}(2018){Traficante}, {Fuller}, {Smith}, {Billot},
  {Duarte-Cabral}, {Peretto}, {Molinari}, \& {Pineda}}]{traficante18}
{Traficante}, A., {Fuller}, G.~A., {Smith}, R.~J., {et~al.} 2018, \mnras, 473,
  4975

\bibitem[{{Tremblin} {et~al.}(2014){Tremblin}, {Schneider}, {Minier},
  {Didelon}, {Hill}, {Anderson}, {Motte}, {Zavagno}, {Andr{\'e}},
  {Arzoumanian}, {Audit}, {Benedettini}, {Bontemps}, {Csengeri}, {Di
  Francesco}, {Giannini}, {Hennemann}, {Nguyen Luong}, {Marston}, {Peretto},
  {Rivera-Ingraham}, {Russeil}, {Rygl}, {Spinoglio}, \& {White}}]{tremblin14}
{Tremblin}, P., {Schneider}, N., {Minier}, V., {et~al.} 2014, \aap, 564, A106

\bibitem[{{Ungerechts} {et~al.}(1986){Ungerechts}, {Winnewisser}, \&
  {Walmsley}}]{ungerechts86}
{Ungerechts}, H., {Winnewisser}, G., \& {Walmsley}, C.~M. 1986, \aap, 157, 207

\bibitem[{{V{\'a}zquez-Semadeni} {et~al.}(2009){V{\'a}zquez-Semadeni},
  {G{\'o}mez}, {Jappsen}, {Ballesteros-Paredes}, \&
  {Klessen}}]{vazquez-semadeni09}
{V{\'a}zquez-Semadeni}, E., {G{\'o}mez}, G.~C., {Jappsen}, A.-K.,
  {Ballesteros-Paredes}, J., \& {Klessen}, R.~S. 2009, ApJ, 707, 1023

\bibitem[{{V{\'a}zquez-Semadeni} {et~al.}(2017){V{\'a}zquez-Semadeni},
  {Gonz{\'a}lez-Samaniego}, \& {Col{\'{\i}}n}}]{vazquez-semadeni17}
{V{\'a}zquez-Semadeni}, E., {Gonz{\'a}lez-Samaniego}, A., \& {Col{\'{\i}}n}, P.
  2017, MNRAS, 467, 1313

\bibitem[{{Walsh} {et~al.}(2008){Walsh}, {Lo}, {Burton}, {White}, {Purcell},
  {Longmore}, {Phillips}, \& {Brooks}}]{walsh08}
{Walsh}, A.~J., {Lo}, N., {Burton}, M.~G., {et~al.} 2008, PASA, 25, 105

\bibitem[{{Whitworth} \& {Ward-Thompson}(2001)}]{whitworth01}
{Whitworth}, A.~P. \& {Ward-Thompson}, D. 2001, ApJ, 547, 317

\bibitem[{{Zhang} {et~al.}(2009){Zhang}, {Wang}, {Pillai}, \&
  {Rathborne}}]{zhang09}
{Zhang}, Q., {Wang}, Y., {Pillai}, T., \& {Rathborne}, J. 2009, ApJ, 696, 268

\end{thebibliography}

\begin{appendix}

\section{Channel map} \label{appendix-channelmap}

Here we show the channel map of the main NH$_{3}$(1,1) emission in SDC13.

\begin{figure*}[!t]
\centering
\includegraphics[scale=.75,center]{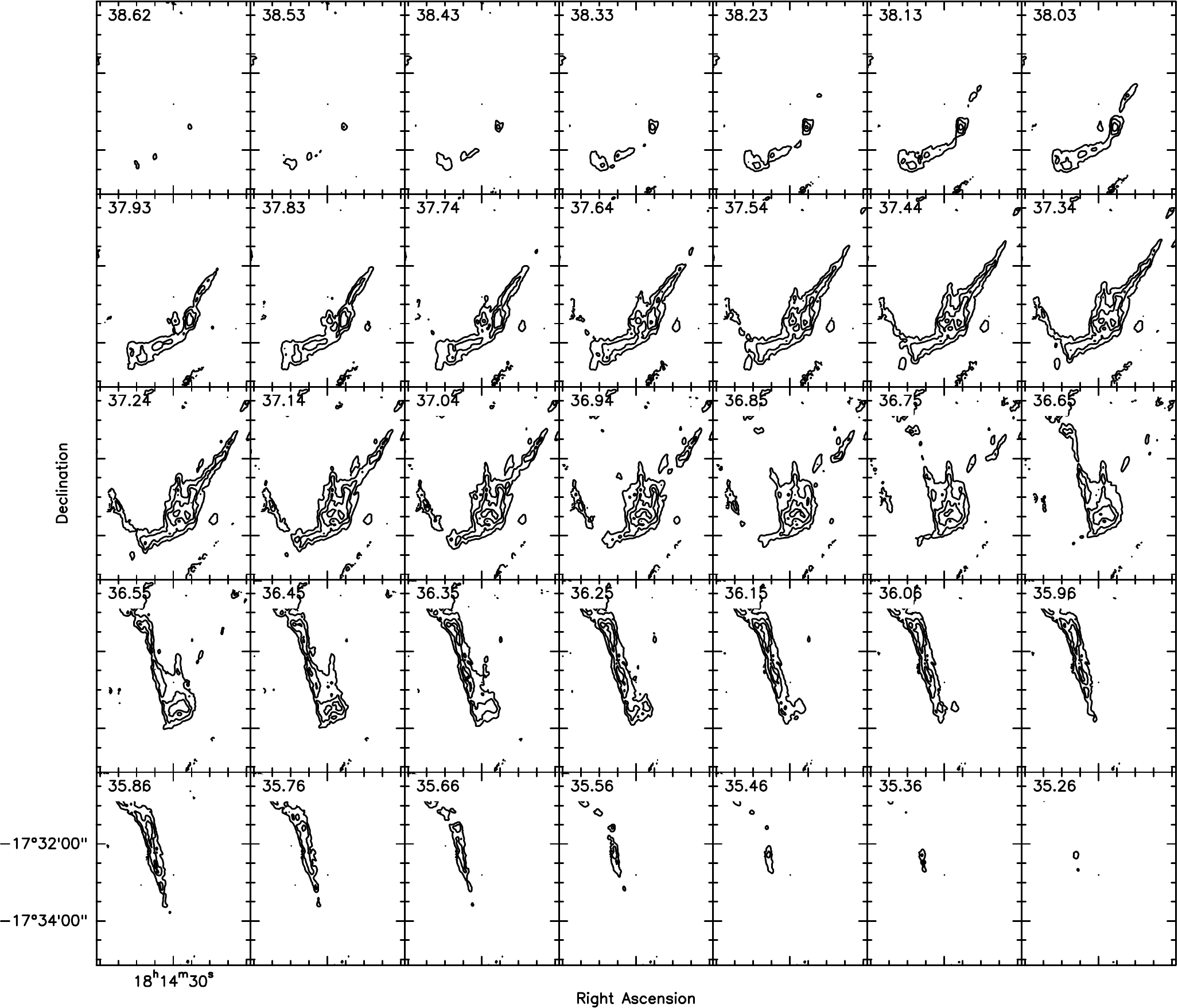} 
\caption{Channel map of the main transition of NH$_{3}$(1,1) in the model combined data cube. Contour levels are placed at 0.05~Jy/beam steps, from 0.05~Jy/beam to 0.2~Jy/beam. The channel velocity in kms$^{-1}$ is plotted in the top left corner of each panel. Right Ascension and Declination are in the J2000 epoch.}
\label{fig:channel}
\end{figure*}

\section{Opacity, temperature and column density} \label{appendix-A}

Here we present a more in depth discussion on the calculation of the opacity, temperature and column density directly from the relative intensities of the NH$_{3}$(1,1) and NH$_{3}$(2,2) lines.

\subsection{Opacity} \label{sec:opacity}

The optical depth defines the extent to which the NH$_{3}$ gas is opaque or transparent, and was solved numerically \citep{barrett77},
    \begin{equation}
        \frac{\Delta T_{a}(1,1,m)}{\Delta T_{a}(1,1,s)} = \frac{1 -e^{-\tau (1,1,m)} }{1 - e^{-\mathrm{a}\tau(1,1,m)} } \,,
        \label{eq:opac1}
    \end{equation}
\noindent where the parameters $\Delta T_{a}(1,1,m)$ and $\Delta T_{a}(1,1,s)$ refer to the observed peak spectra of the NH$_{3}$(1,1) main (denoted by the \textit{m}) and satellite (denoted by the \textit{s}) components, which we take from the model hyperfine structure fit.  The expected relative intensities between the main and inner satellite lines is denoted by \textit{a}.  The opacity of the main NH$_{3}$(1,1) line is denoted by $\tau (1,1,m)$.  This opacity is converted to the total opacity, $\tau(1,1,t)$, by dividing by the relative strength of 0.502 \citep{li03}.

The opacity of NH$_{3}$(2,2) is calculated similarly \citep{barrett77},
    \begin{equation}
        \frac{\Delta T_{a}(1,1,m)}{\Delta T_{a}(2,2,m)} = \frac{1 -e^{-\tau (1,1,m)} }{1 - e^{-\tau(2,2,m)} } \, ,
        \label{eq:opac2}
    \end{equation}

\noindent where $\Delta T_{a}(2,2,m)$ and $\tau(2,2,m)$ are the observed peak of the spectra and the opacity of the NH$_{3}$(2,2) main component, respectively.  This opacity is converted to the total opacity, $\tau(2,2,t)$ of NH$_{3}$(2,2) by dividing by the relative strength of 0.796 \citep{li03}.

\subsection{Temperature} \label{sec:trot}

\begin{figure}
    \centering
    \begin{minipage}{.5\linewidth}
    \subfloat{\label{fig:trotmap}\includegraphics[trim={-1.25cm 0 0 0},scale=.42,center]{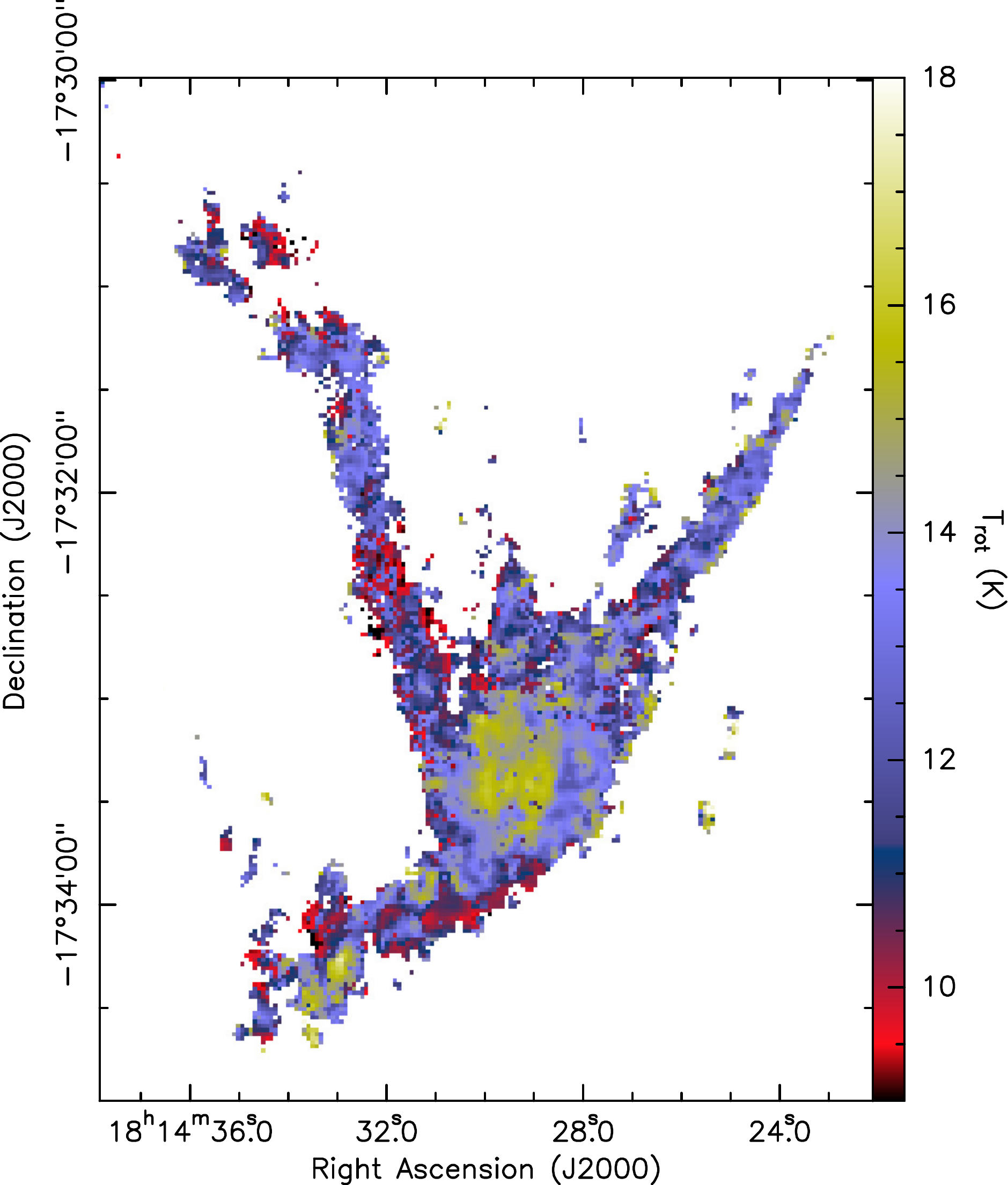}}
\end{minipage}\par\medskip
\begin{minipage}{.5\linewidth}
    \subfloat{\label{fig:trothist}\includegraphics[scale=.42,center]{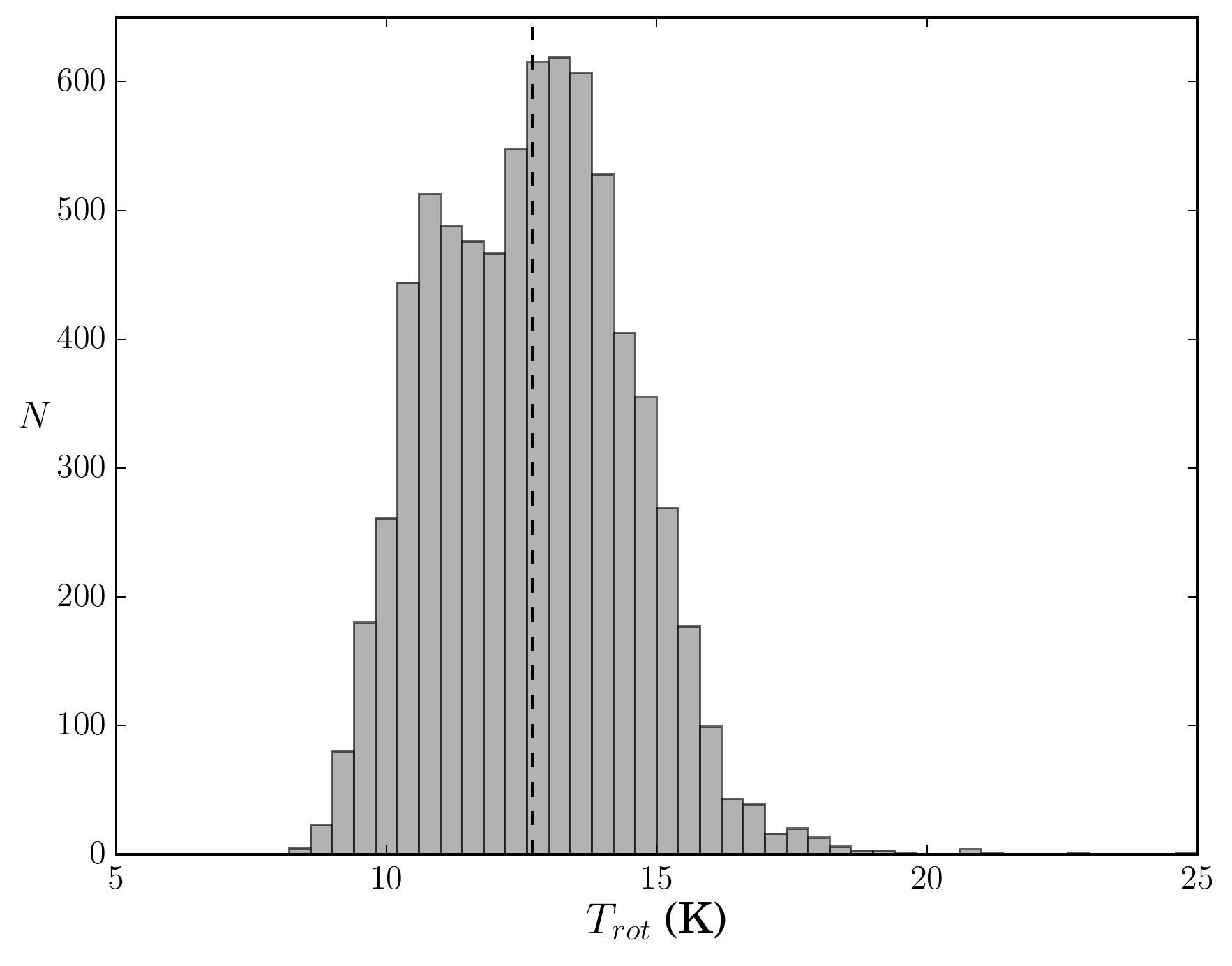}} 
\end{minipage}
\caption{\textit{Top}: Map of the rotational temperature calculated using equation \ref{eq:trot}.  It is clear that the coverage of the image is limited to that of the weaker NH$_{3}$(2,2) line.  No correlation is found between temperature and core position. \textit{Bottom}: Histogram of the $T_{rot}$ map in the top panel. Vertical dashed line indicates the median value of 12.7\,K. The standard deviation is 1.8\,K.}
\label{fig:trot}
\end{figure}

The rotational temperature, which describes the relative populations of the NH$_{3}$(1,1) and NH$_{3}$(2,2) levels, was found using the following equation \citep{ungerechts86},
    \begin{equation}
        T_{rot} = -\mathrm{T_{0}} \,/\, \mathrm{ln} \left( \frac{9}{20} \frac{\tau(2,2,t)}{\tau(1,1,t)} \left[ \frac{\Delta v_{2,2}}{\Delta v_{1,1}} \right] \right) \, ,
        \label{eq:trot}
    \end{equation}

\noindent where $\tau(2,2,t)$ and $\tau(1,1,t)$ are the total optical depths, and $\Delta \mathrm{v}_{2,2}$ and $\Delta \mathrm{v}_{1,1}$ are the velocity widths of the NH$_{3}$(2,2) and NH$_{3}$(1,1) lines respectively, and $\mathrm{T_{0}}$\,=\,$(E_{2,2}$\,$-$\,$E_{1,1})/\mathrm{k}$\,=\,41\,K, where $E_{1,1}$ and $E_{2,2}$ are the energies of the NH$_{3}$(1,1) and NH$_{3}$(2,2) lines respectively.  The resulting $T_{rot}$ map is shown in the top panel of Figure \ref{fig:trot}, with its corresponding histogram in the bottom panel.  We find no correlation between temperature and core positions, and only notice a very mild warming of material at the filament hub junctions.  


As outlined in Section \ref{sec:structure}, the NH$_{3}$(2,2) emission is approximately 6 times weaker than that of NH$_{3}$(1,1), resulting in 42\% less coverage in the NH$_{3}$(2,2) maps, and hence, the temperature maps. This limits the calculation of the column density of the cloud to this lesser coverage.  To overcome this hindrance, we took a median $T_{rot}$ across the entire cloud of 12.7\,K \citep[e.g.][]{morgan13}.  We discounted using the full temperature map whilst setting the median in the NH$_{3}$(1,1) regions of emission only, as this caused edge effects in the resulting column density.




\subsection{Column density} \label{sec:column-density}

With opacity and temperature, we calculated the total column density of NH$_{3}$ and H$_{2}$.  Firstly, the column density of the NH$_{3}$(1,1) upper level only was calculated \citep{goldsmithlanger99},
    \begin{equation}
         N[\mathrm{NH}_{3}(1,1)] = \frac{8\mathrm{k\pi}\nu^{2}_{1,1}}{\mathrm{A}_{\mathrm{u,l}}\mathrm{hc}^{3}} \int_{-\infty}^{\infty}T_{b}\mathrm{d}v\left  ( \frac{\tau(1,1,t)}{1-e^{-\tau(1,1,t)}} \right ) \,,
         \label{eq:n11}
    \end{equation}
\noindent where k is the Boltzmann's constant, $\nu_{1,1}$ is the rest frequency of the NH$_{3}$(1,1) line, $\mathrm{A}_{ul}$ is the Einstein Coefficient for spontaneous emission, h is the Planck constant, c is the speed of light, $\tau(1,1,t)$ is the total opacity of the NH$_{3}$(1,1) line, and $\int_{-\infty}^{\infty}T_{b}$d$v$ is the integrated intensity of NH$_{3}$(1,1) in Figure \ref{fig:integ11}.  Beam information has been cancelled out under the assumption that the source fills the beam.

Assuming local thermodynamic equilibrium (LTE), where the rotational temperature is assumed to equal the excitation temperature, we converted $N[\mathrm{NH}_{3}(1,1)]$ into the total column density of Ammonia, $N(\mathrm{NH}_{3})$, using \citep{goldsmithlanger99},
    \begin{equation}
        N(\mathrm{NH}_{3}) =\frac{N[\mathrm{NH}_{3}(1,1)]}{g_{1,1}} e^{E_{1,1}/{\mathrm{k}T_{rot}}}Q(T_{rot}) \, ,
        \label{eq:Nnh31}
    \end{equation}

\noindent where $g_{1,1}$ is the statistical weight, $E_{1,1}$ is the energy of the NH$_{3}$(1,1) line, k is the Boltzmann constant, \textit{T$_{rot}$} is the rotational temperature, and $Q(T_{rot})$ is the partition function.  Expanding the partition function, $Q(T_{rot}) = \sum_{i}g_{i}e^{-E_{i}/kT_{rot}}$ over J and K energy levels, where J and K are the transition quantum numbers \citep{ungerechts86},
    \begin{equation}
        N(\mathrm{NH}_{3}) = N[\mathrm{NH}_{3}(1,1)]\left [ \frac{1}{3}e^{+23.26/T_{rot}}+1+\frac{5}{3}e^{-41.18/T_{rot}}\cdots \right ] \, ,
        \label{eq:Nnh32}
    \end{equation}
\noindent  where \textit{T$_{rot}$} is the median rotational temperature of 12.7\,K, and $N[\mathrm{NH}_{3}(1,1)]$ follows from Equation \ref{eq:n11}.  

We calculated the H$_{2}$ column density by converting from NH$_{3}$ using an abundance of [NH$_{3}$]/[H$_{2}$]$\sim$ $3\times10^{-8}$ \citep{harju93}.  The final H$_{2}$ column density map is shown in Figure \ref{fig:n:nh2}.

\section{Deprojected filaments} \label{appendix-B}

\begin{figure}
    \centering
    \begin{minipage}{.5\linewidth}
    \subfloat{\label{fig:deprojse}\includegraphics[scale=.38,center]{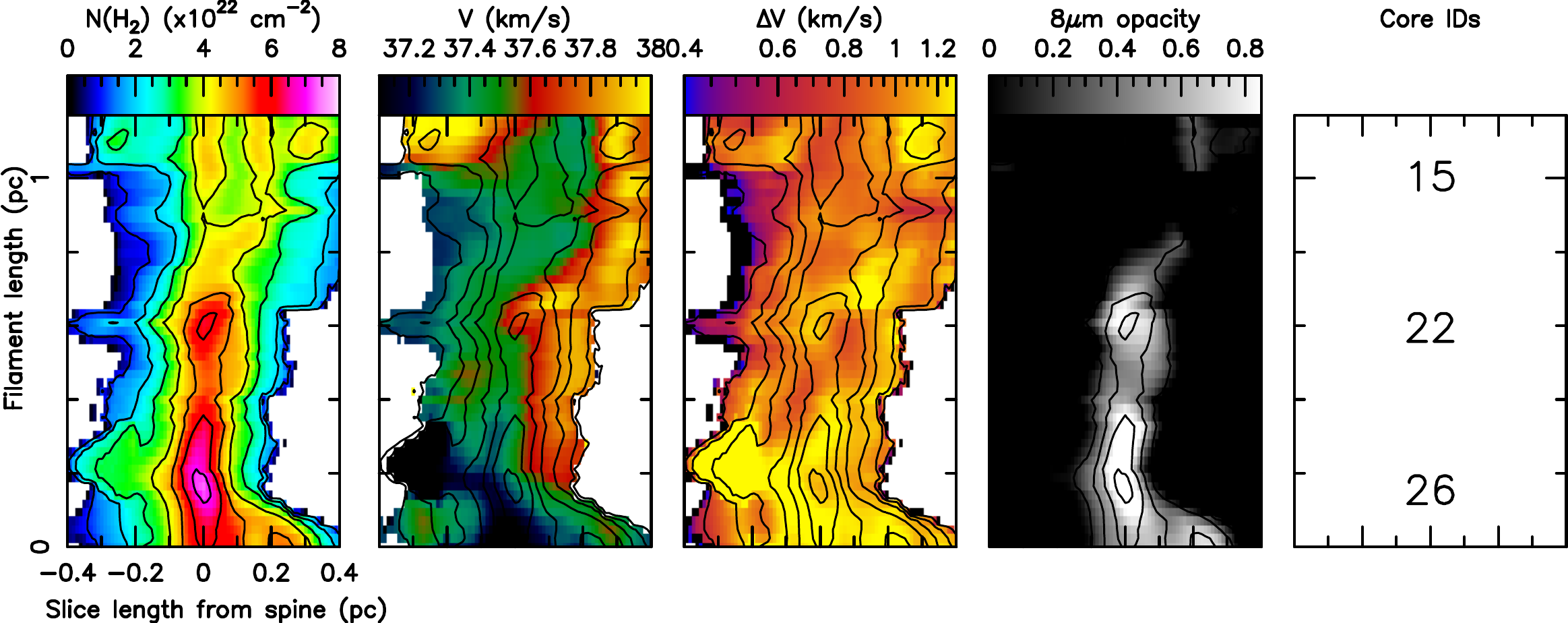}}
\end{minipage}\par\medskip
\begin{minipage}{.5\linewidth}
    \subfloat{\label{fig:deprojn}\includegraphics[scale=.54,center]{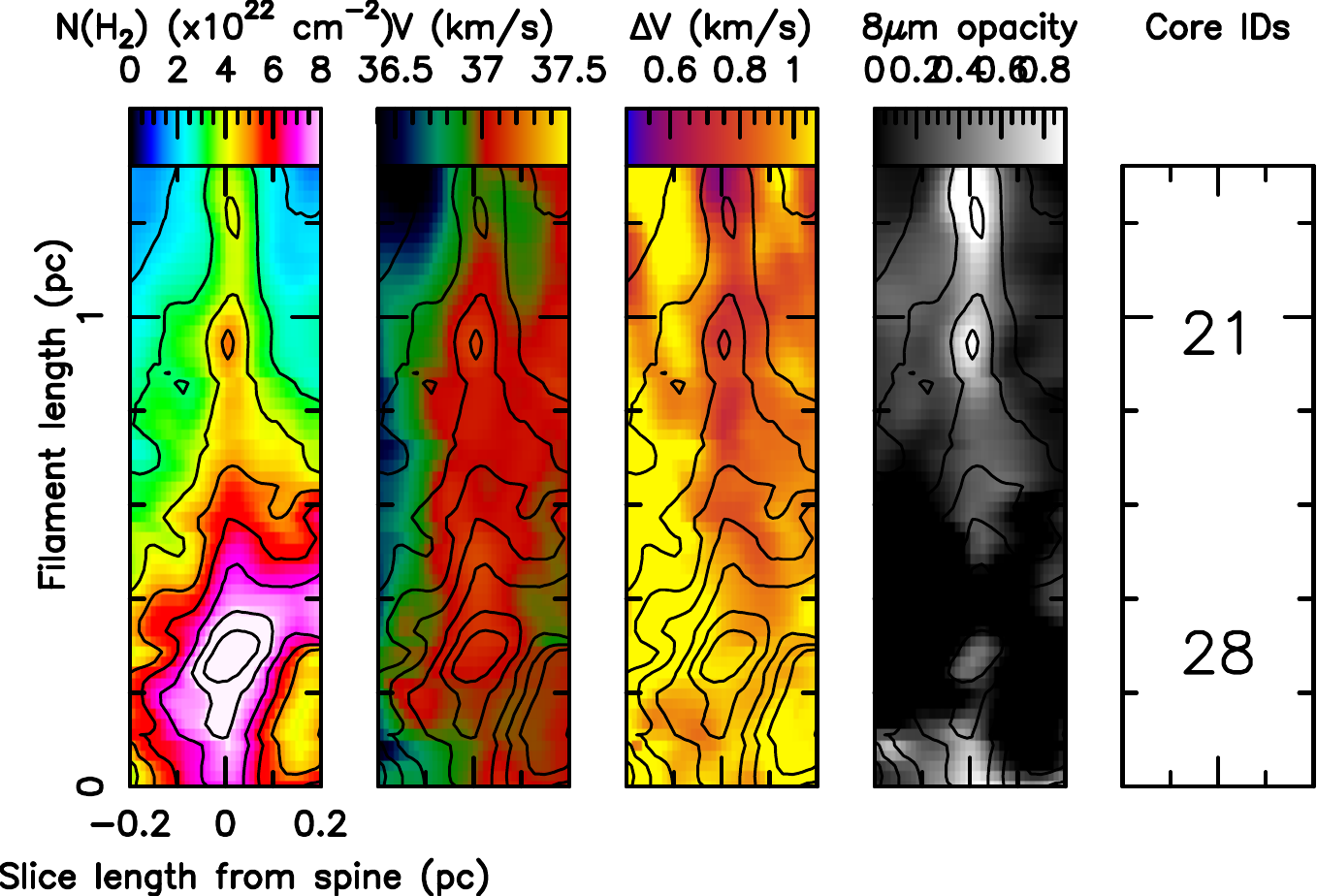}}
\end{minipage}
\caption{Deprojected views of the South-East (top) and North (bottom) filaments. The North filament was restricted to 0.2\,pc either side of the spine due to contamination from the North-West and North-East filaments. The filament length is plotted on the y-axis, while the length of the radial slices from the central spine pixel in plotted on the x-axis. In both sub-figures (from the left), the first panel shows the H$_{2}$ column density in units of 10$^{22}$\,cm$^{-2}$, the second panel shows the line-of-sight velocity in km/s, the third panel shows the velocity width in km\,s$^{-1}$, the fourth panel shows the opacity derived from the 8\,$\mu$m Spitzer emission, whilst the final panel labels the ID number of the identified cores. Contours in each panel are of the column density, from 1$\times$10$^{22}$\,cm$^{-2}$ to 11$\times$10$^{22}$\,cm$^{-2}$, spaced by 1$\times$10$^{22}$\,cm$^{-2}$.}
\label{fig:deprojected-app}
\end{figure}

In Figure \ref{fig:deprojected-app} we show the deprojected views of the South-East and North filaments (in the top and bottom panels respectively). It is clear that there are radial velocity gradients present in both, stronger in the South-East.

\section{GBT data} \label{appendix-gbt}

\begin{figure*}[!htbp]
\centering
\begin{minipage}{.345\linewidth}
\subfloat{\label{fig:integgbt11}\includegraphics[trim={0.5cm 0 0 0},scale=.39,left]{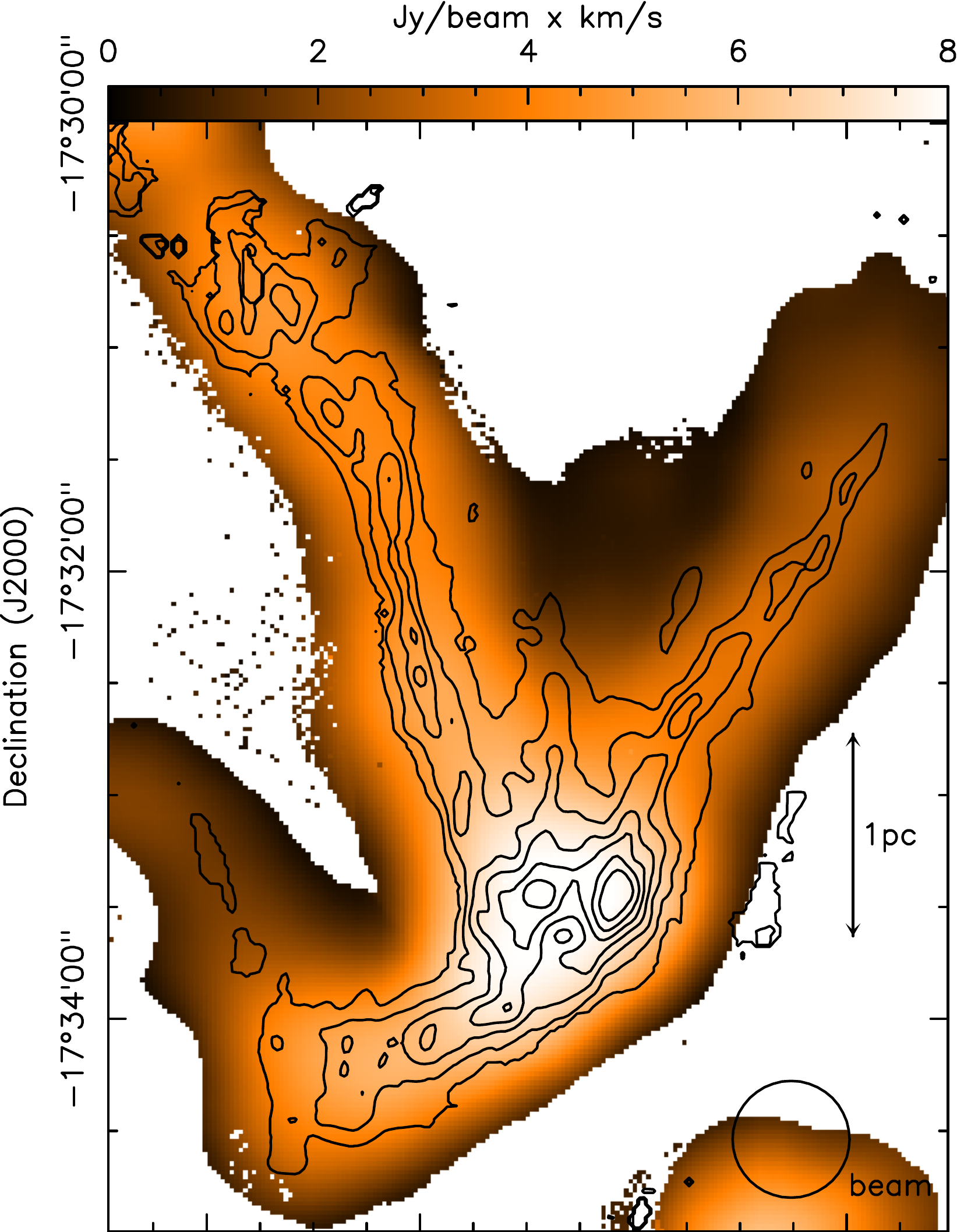}}
\end{minipage}%
\begin{minipage}{.315\linewidth}
\subfloat{\label{fig:velogbt11}\includegraphics[trim={0.4cm 0 0 0},scale=.39,center]{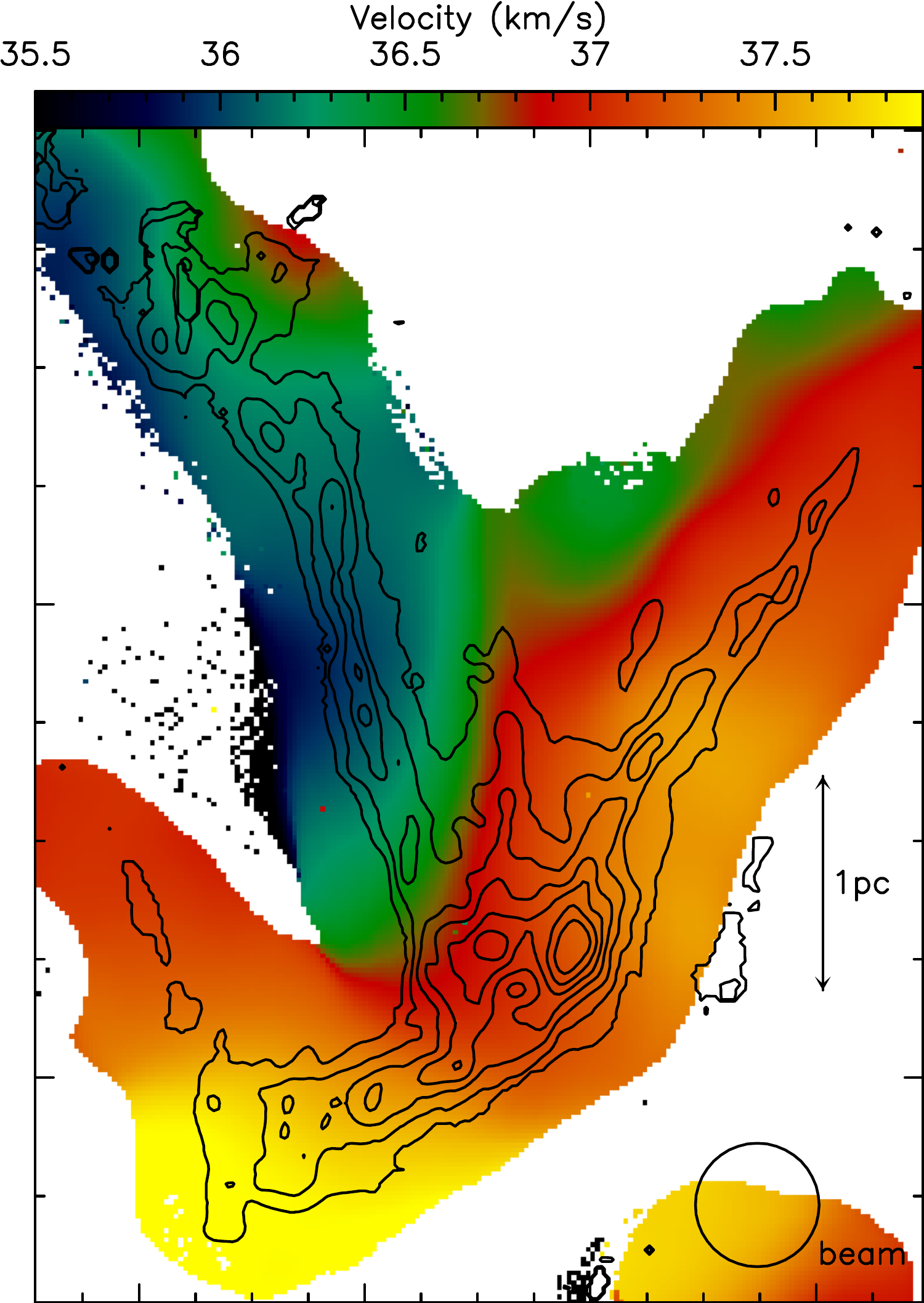}}
\end{minipage}%
\centering
\begin{minipage}{.315\linewidth}
\subfloat{\label{fig:widthgbt11}\includegraphics[trim={0 0 0.5cm 0},scale=.39,right]{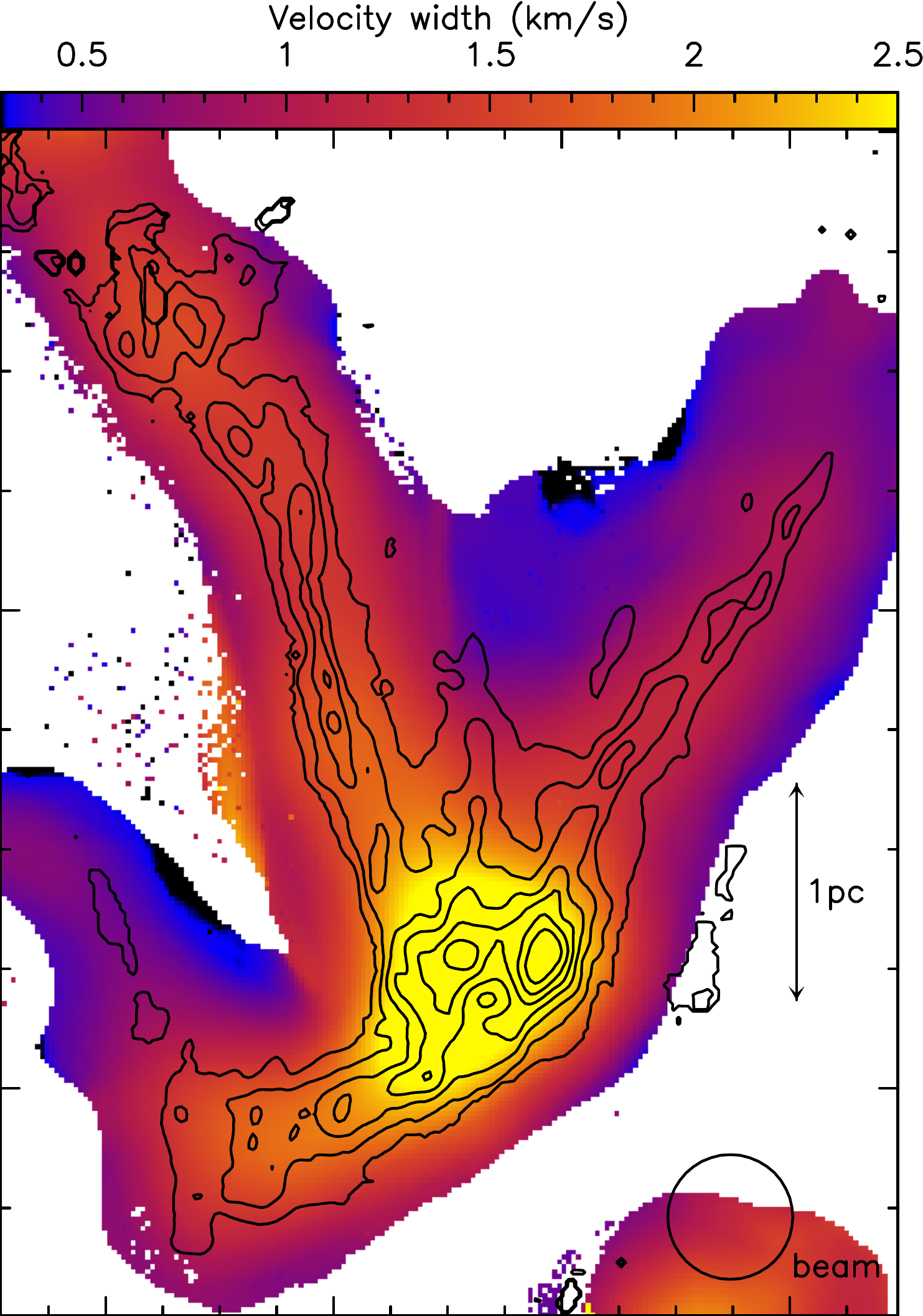}}
\end{minipage}\par\bigskip
\begin{minipage}{.345\linewidth}
\subfloat{\label{fig:integgbt22}\includegraphics[trim={0.5cm 0 0 0},scale=.39,left]{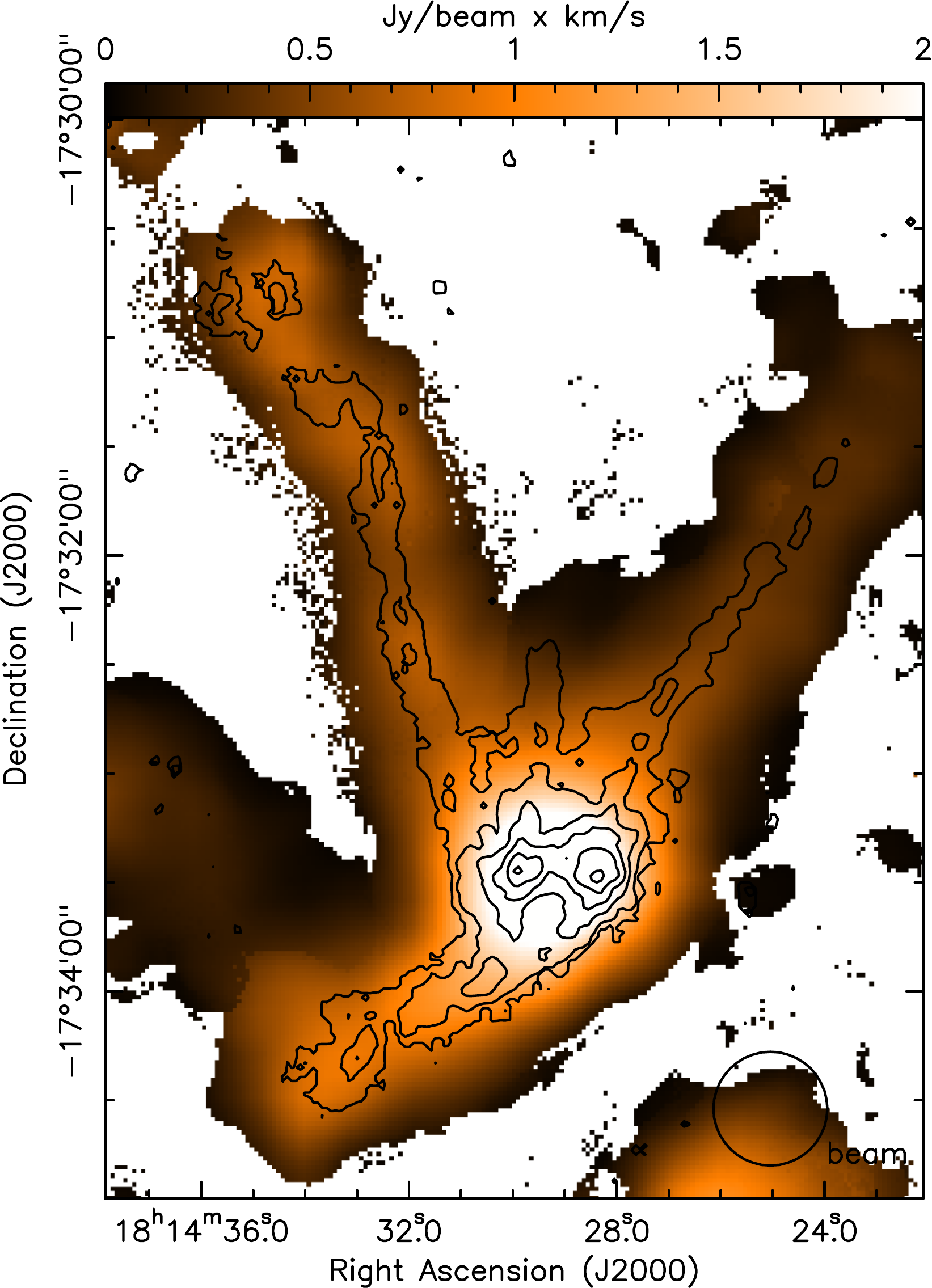}}
\end{minipage}%
\centering
\begin{minipage}{.315\linewidth}
\subfloat{\label{fig:velogbt22}\includegraphics[trim={0.1cm 0 0 0},scale=.39,center]{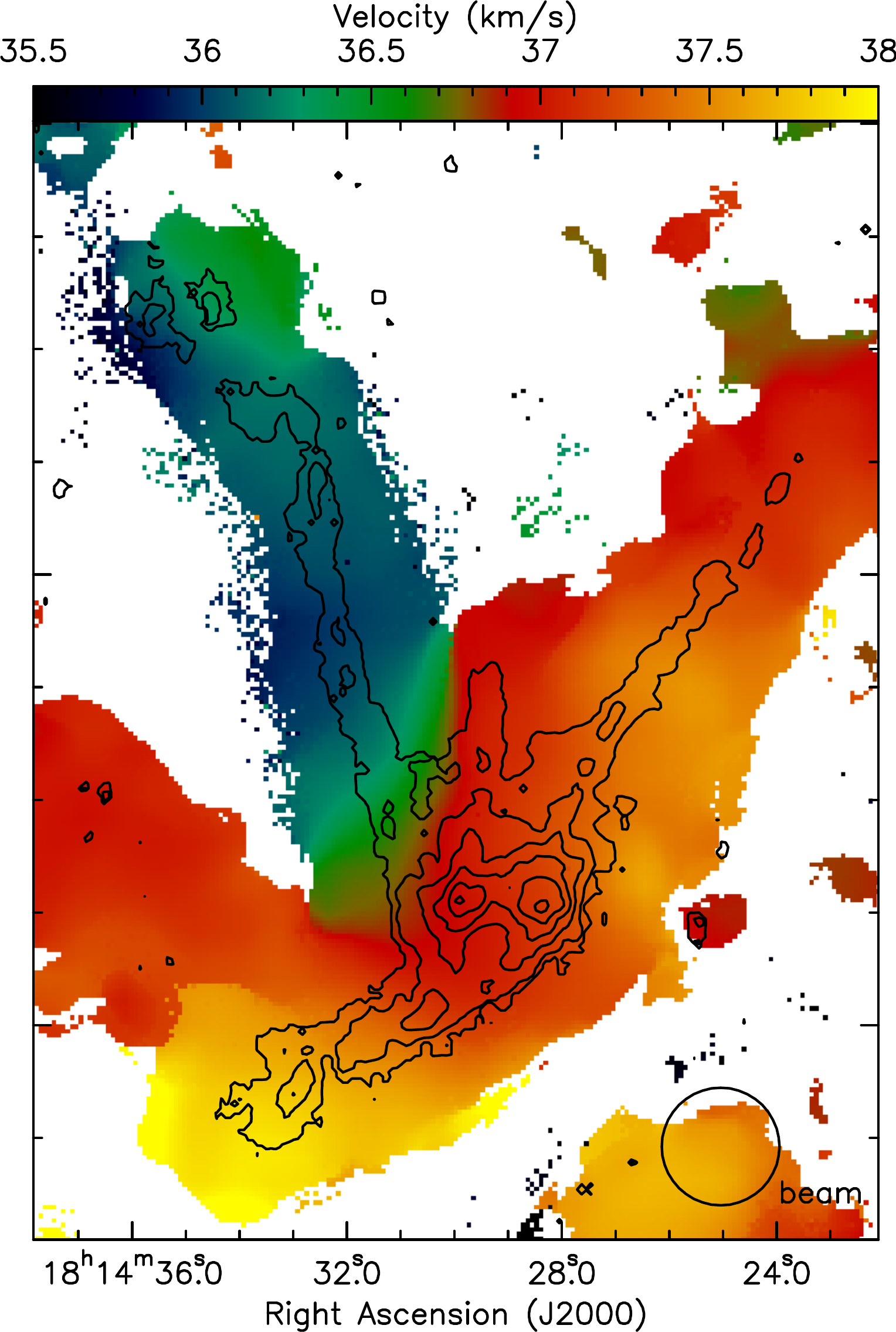}}
\end{minipage}%
\begin{minipage}{.315\linewidth}
\subfloat{\label{fig:widthgbt22}\includegraphics[trim={0 0 0.5cm 0},scale=.39,right]{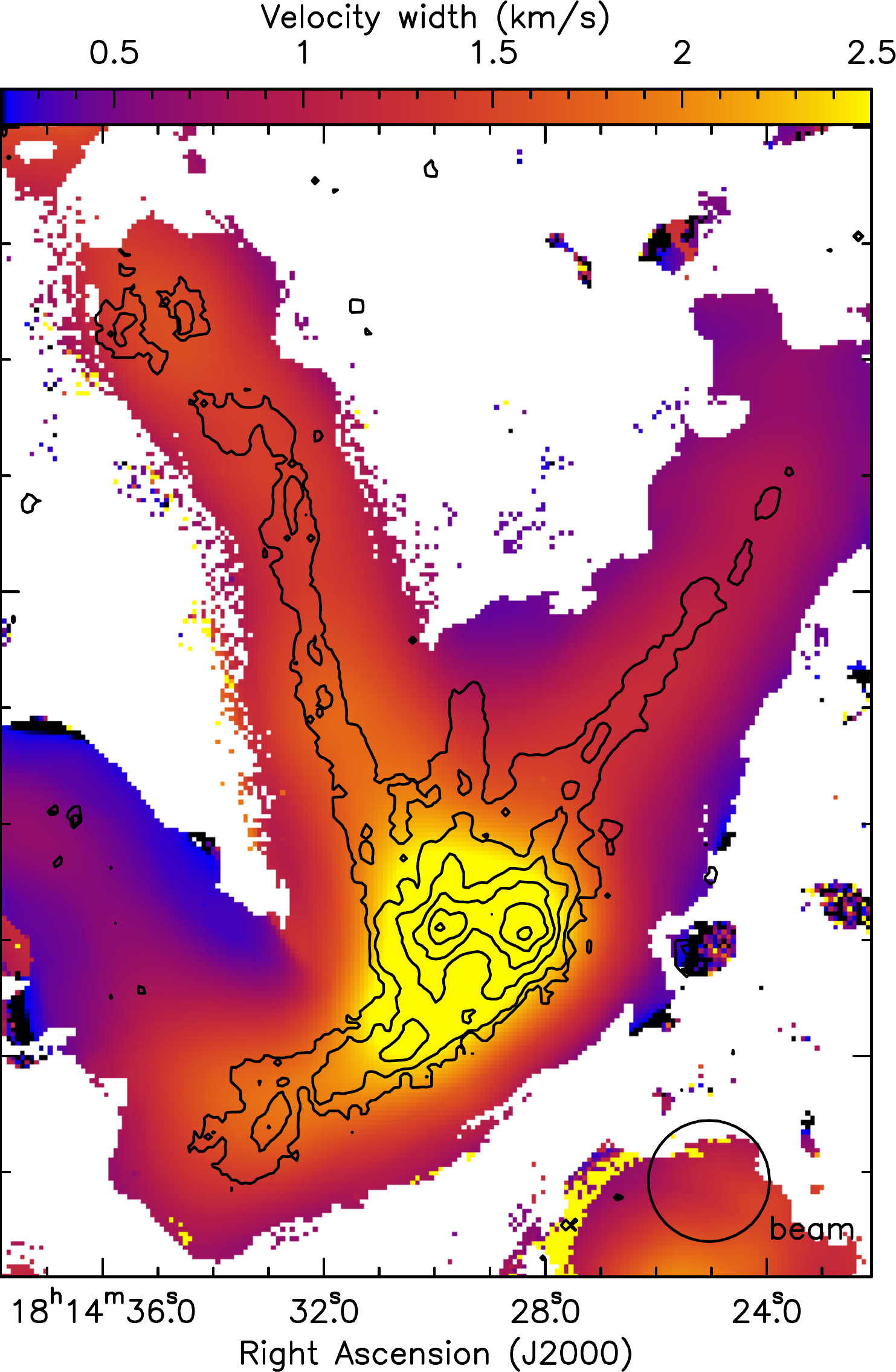}}
\end{minipage}
\caption{Integrated intensity (first column), centroid velocity (middle column) and velocity dispersion (right column) of the NH$_{3}$(1,1) transition (top row) and  NH$_{3}$(2,2) transition (bottom row) of the GBT data. The data has been regridded to the JVLA pixel size.  The NH$_{3}$(1,1) data was masked to 3$\sigma$, while the weaker NH$_{3}$(2,2) data was masked to 2$\sigma$. Contours in the top row correspond to the integrated intensity of the combined JVLA and GBT NH$_{3}$(1,1) data in steps of 0.04\,Jy/beam\,$\times$\,km/s, from 0.05\,Jy/beam\,$\times$\,km/s to 0.29\,Jy/beam\,$\times$\,km/s, while contours in the bottom row correspond to the integrated intensity of the combined NH$_{3}$(2,2) data in steps of 0.008\,Jy/beam\,$\times$\,km/s, from 0.008\,Jy/beam\,$\times$\,km/s to 0.096\,Jy/beam\,$\times$\,km/s. Beam information is plotted in the bottom right corner of each panel, while the scale of 1\,pc is plotted in each panel of the top row.}
\label{fig:GBTmaps}
\end{figure*}

In Figure \ref{fig:GBTmaps} we show the NH$_{3}$(1,1) and NH$_{3}$(2,2) integrated intensity, centroid line-of-sight velocity and velocity width of the GBT data.

\section{Characterising radial column density profiles} \label{app:radial_nh2}

\begin{figure*}[!t]
\centering
\includegraphics[scale=.37,left]{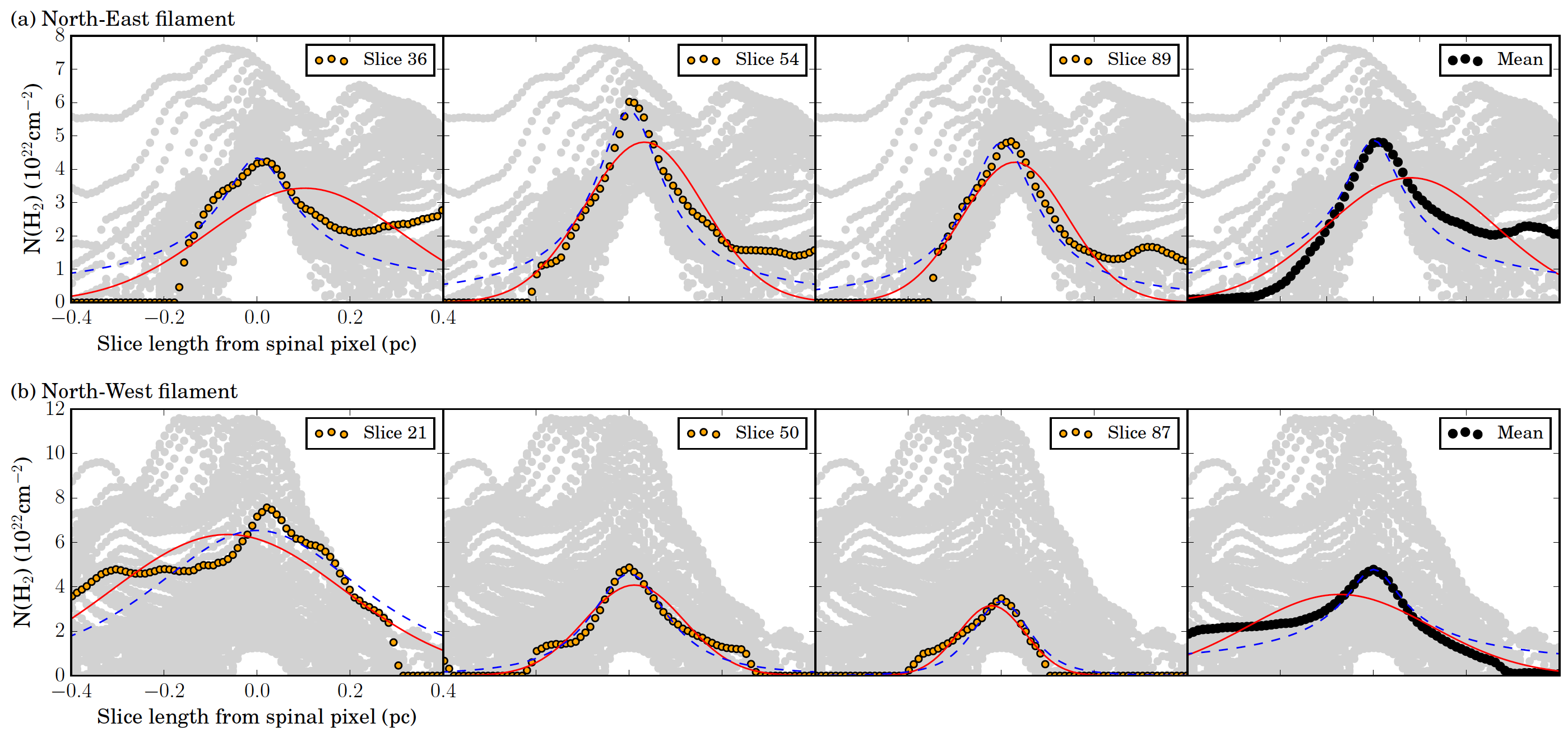}
\caption{Plots showing the results of fitting Gaussian (red line) and Plummer (blue dashed line) profiles to \textit{individual} radial column density slices (orange circles in the first, second and third panels of each row) and the \textit{mean} radial column density profile (black circles in the fourth panel of each row) of (a) the North-East filament (top row) and (b) the North-West filament (bottom row). The grey points in the background of each panel plots all of the radial column density slices simultaneously.  Plummer profiles may on occasion well describe the inner regions of the filaments only, whilst Gaussian profiles fail to fit any of the distributions.  Gaussian profiles are only found to reasonably fit some of the data once the fitting range is restricted to the inner portion of the filament.}
\label{fig:radial_nh2_fits}
\end{figure*}

To characterise the SDC13 filaments, we proceeded to fit each radial column density slice (found in Figure \ref{fig:deprojected}) with Gaussian and Plummer \citep{whitworth01} profiles, and intended to find the function that best described the observed structure.  Examples of such fits are shown in Figure~\ref{fig:radial_nh2_fits}.  Looking at the mean profiles of the North-East and North-West filaments (the fourth panels of Figure~\ref{fig:radial_nh2_fits}), it is clear that they are not well described by neither Plummer nor Gaussian fits.  This trend continues when considering individual column density slices along the filaments (shown in the first three panels of Figure~\ref{fig:radial_nh2_fits}).  Although a Plummer may reasonably well describe slice 50 in the North-West filament (and perhaps even the centre-most portion of slice 54 in the North-East filament) the extended emission in the majority of slices renders fitting them with Gaussian or Plummer profiles inappropriate.  We have evaluated further the impact of constraining the Gaussian fit parameters on the goodness of the fit itself.  While setting the peak position of the Gaussian profiles to lie on the spine (i.e. $x=0$) does not improve the produced fits, restricting the fitting to the inner portion of the filaments does result in better Gaussian fits of the central portion of the filament.  However, filament properties (width and line mass) derived from such fits are very much dependent on the exact fitting range used. Such properties calculated from Trapezoidal integration of the asymmetrical profiles (as shown in Table~\ref{tab:fil-properties} and Figure~\ref{fig:width_mline}) are less affected by such restrictions, and therefore more robust.  
Characterising filaments is fraught with difficulty, and caution must be observed whilst deriving global properties from their mean profiles \citep[e.g.][]{panopoulou17}.

\section{Derivation of the analytical expression of the virial ratio} \label{app:virial}

Consider a uniform density filament fragmenting at the length scale equal to the separation of a core, $\lambda_{core}$.  In an idealised scenario, energy is conserved during collapse whilst fragmentation is happening.  However, there is a fraction of energy converted from gravitational potential energy ($E_{g}$) into kinetic energy ($E_{k}$) at the time the filament fragmentation starts.  This conversion may be expressed in the virial equation as an efficiency $\epsilon$, where if no energy is converted, $\epsilon = 0$.  If energy conversion does occur, this can be expressed as a fractional increase in $E_{k}$, expressed as $E_{k} = E_{k,0} + \Delta E_{k}$, where $E_{k,0}$ is the initial kinetic energy of the fragmenting core before collapse.  This fractional increase in kinetic energy can be expressed as $\Delta E_{k} = \epsilon \left( E_{g} - E_{g,0}\right)$, where $E_{g,0}$ is the initial gravitational potential energy.  Substitution into the standard virial ratio of $\alpha_{vir} = 2E_{k}/E_{g}$ gives:
    \begin{equation}
        \alpha_{vir}=\frac{2 E_{k,0} + 2\epsilon\left(E_{g,0} - E_{g}\right) }{E_{g}}\, ,
    \end{equation}
 \noindent and simplifying gives:
    \begin{equation}
        \alpha_{vir}=2\frac{E_{k,0}}{|E_g|}+2\epsilon\left(1-\frac{|E_{g,0}|}{|E_g|}\right) \, .
    \end{equation}

The energy terms may be expressed as:
    \begin{equation}
        E_{k,0} = \frac{3}{2}Ma_{eff,0}^{2} \, ,
    \end{equation}
\noindent where $a_{eff,0}$ is the 1D velocity dispersion, and:
    \begin{equation}
        E_{g} = - \beta\frac{GM^{2}}{R} \, ,
    \end{equation}
\noindent where $\beta = \frac{3 - k_{\rho}}{5 - 2k_{\rho}}$ is a factor depending on the power law index $k_{\rho}$ of the density profile of collapsing cores, and:
    \begin{equation}
        E_{g,0} = - \beta_{0}\frac{GM^{2}}{R_{0}} \, ,
    \end{equation}
\noindent where $\beta_{0}$ and $R_{0}$ are the initial values of $\beta$ and $R$ before the onset of collapse.  Making these substitutions gives:
    \begin{equation}
        \alpha_{vir}=\frac{3}{\beta G}\frac{Ra_{eff,0}^2}{M}+2\epsilon\left(1-\frac{\beta_{0}}{\beta}\frac{R}{R_{0}}\right) \, .
    \end{equation}
One may rewrite $R_{o}$ as $\lambda_{core} / 2$ under the assumption that all cores are initially their separation in size, and $\beta_{0} = 3/5$ under the assumption that the filament is initially uniform in density (i.e.  $k_{\rho} = 0$) before the onset of collapse.  Making these assumptions simplifies the equation to:
    \begin{equation}
        \alpha_{vir}=\frac{3}{\beta G}\frac{Ra_{eff,0}^2}{M}+2\epsilon\left(1-\frac{6}{5\beta}\frac{R}{\lambda_{core}}\right) \, .
    \end{equation}


\end{appendix}

\end{document}